\documentclass[prd,preprint,tightenlines,showpacs,nofootinbib,eqsecnum,amsfonts,amssymb,amsmath]{revtex4}
\usepackage{graphicx}   
\usepackage{dcolumn}      
\usepackage{bm}           
\usepackage{color}
\newif\ifbidon\bidontrue
\ifbidon \else \fi
\newcommand{\etal}{{\it et al.}}
\def\kref{k_{\rm ref}}
\def\dd{{\rm d}}

\def\HH{\mathcal{H}}
\def\bk{{\bf k}}
\def\bx{{\bf x}}

\def\sva{\sigma_{_{\rm V}a}}
\def\svb{\sigma_{_{\rm V}b}}
\def\stl{\sigma_{_{\rm T}\lambda}}
\def\spar{\sigma_{_\parallel}}
\def\stplus{\sigma_{_{\rm T}+}}
\def\stcross{\sigma_{_{\rm T}\times}}
\newcommand{\urel}[1]{{\tt #1}}
\begin{document}

\title{Predictions from an anisotropic inflationary era}

\author{Cyril Pitrou}
 \email{pitrou@iap.fr}
 \affiliation{
             Institut d'Astrophysique de Paris,
             Universit\'e Pierre~\&~Marie Curie - Paris VI,
             CNRS-UMR 7095, 98 bis, Bd Arago, 75014 Paris, France,}

\author{Thiago S. Pereira}
 \email{thiago@if.usp.br}
 \affiliation{
             Instituto de F\'isica,
             Universidade de S\~ao Paulo
             CP 66318, 05315-970 S\~ao Paulo, Brazil,}

\author{Jean-Philippe Uzan}
 \email{uzan@iap.fr}
 \affiliation{
             Institut d'Astrophysique de Paris,
             Universit\'e Pierre~\&~Marie Curie - Paris VI,
             CNRS-UMR 7095, 98 bis, Bd Arago, 75014 Paris, France.}

\begin{abstract}
This article investigates the predictions of an inflationary phase
starting from a homogeneous and anisotropic universe of the
Bianchi~$I$ type. After discussing the evolution of the background
spacetime, focusing on the number of $e$-folds and the
isotropization, we solve the perturbation equations and predict
the power spectra of the curvature perturbations and gravity waves
at the end of inflation.

The main features of the early anisotropic phase is (1) a
dependence of the spectra on the direction of the modes, (2) a
coupling between curvature perturbations and gravity waves, and
(3) the fact that the two gravity waves polarisations do not share
the same spectrum on large scales. All these effects are
significant only on  large scales and die out on small scales
where isotropy is recovered. They depend on a characteristic scale that can,
but a priori must not, be tuned to some observable scale.

To fix the initial conditions, we propose a procedure that
generalises the one standardly used in inflation but that takes
into account the fact that the WKB regime is violated at early
times when the shear dominates. We stress that there exist modes
that do not satisfy the WKB condition during the shear-dominated 
regime and for which the amplitude at the end of inflation depends 
on unknown initial conditions. On such scales, inflation loses 
its predictability.

This study paves the way to the determination of the cosmological
signature of a primordial shear, whatever the Bianchi~$I$
spacetime. It thus stresses the importance of the WKB regime to draw
inflationary predictions and demonstrates that when the number of
$e$-folds is large enough, the predictions converge toward those
of inflation in a Friedmann-Lema\^{\i}tre spacetime but that they
are less robust in the case of an inflationary era with a small
number of $e$-folds.
\end{abstract}

 \date{18 novembre 2007\\}

 \maketitle
 \tableofcontents

\section{Introduction}

Inflation~\cite{lindebook,pubook} (see Ref.~\cite{linde2007} for a
recent review of its status) is now one of the cornerstones of the
standard cosmological model. In its simplest form, inflation has
very definite predictions: the existence of adiabatic initial
scalar perturbations and gravitational waves, both with Gaussian
statistics and an almost scale invariant power
spectrum~\cite{ref:inf,mbf}. Various extensions, which in general
involve more fields, allow e.g. for isocurvature
perturbations~\cite{iso}, non-Gaussianity~\cite{ng}, and modulated
fluctuations~\cite{bku}. All these features let us hope that
future data will shed some light on the details (and physics) of
this primordial phase of the universe.

Almost the entire literature on inflation assumes that the
universe is homogeneous and isotropic while homogeneity and
isotropy are what inflation is supposed to explain. Indeed, the
dynamics of anisotropic inflationary universes has been widely
discussed~\cite{olive}. It was demonstrated that under a large
variety of conditions, inflation occurs even if the spacetime is
initially anisotropic~\cite{infanisogen}, regardless of whether it
is dominated by a pure cosmological constant or a slow-rolling
scalar field. The isotropization of the universe was even
generalized to Bianchi braneworld models~\cite{braneinf} recently.
It should be emphasized, however, that a deviation from 
isotropy~\cite{infanisogen} or flatness~\cite{inflK} may have a 
strong effect on the dynamics of inflation, and in particular 
on the number of $e$-folds.\\

The study of the perturbations during the isotropization phase has
been overlooked, mainly because in the past decades one was
mostly focused on large field inflationary models, which
generically give very long inflationary phases. This was also
backed up by the ideas of chaotic inflation and eternal
inflation~\cite{linde2007}. In such cases, it is thus an excellent
approximation to describe the universe by a
Friedmann-Lema\^{\i}tre (FL) spacetime when focusing on the modes
observable today since they exited the Hubble radius
approximatly in the last 60 $e$-folds. In this case, the origin
of the density perturbations is understood as the amplification of
vacuum quantum fluctuations of the inflaton. In particular, the
degrees of freedom that should be quantized deep in the
inflationary phase, and known as the Mukhanov-Sasaki
variables~\cite{MSvar}, were identified~\cite{mbf}, which
completely fixes the initial conditions and makes inflation a very
predictive theory.

In the context of string theory, constructing a string
compactification whose low energy effective Lagrangian is able to
produce inflation is challenging (see Ref.~\cite{Allister}). In
particular, it has proved to be difficult to build large field
models~\cite{bauman} (in which the inflaton moves over large
distances compared to the Planck scale in field space). This has led
to the idea that, in this framework, an inflationary phase with a
small number of $e$-folds is favored (see however
Ref.~\cite{roulette}). If so, the predictions of inflation are
expected to be sensitive to the initial conditions, and in
particular the classical inhomogeneities are not expected to be
exponentially suppressed, which makes the search for large scale
deviations from homogeneity and isotropy much more motivated, as well 
as the possibility that the inflaton has not reached the
inflationary attractor. More important, it is far from obvious (as
we shall discuss in details later) that all observable modes can be
assumed to be in their Bunch-Davies vacuum initially, and more
puzzling, that for a given mode modulus the possibility of setting
the initial conditions will depend on its direction. If so, then
the initial conditions for these modes would have to be set in the
stringy phase, an open issue at the time.

The theory of cosmological perturbations in a Bianchi universe was
roughed out in Ref.~\cite{TomitaDen,Dunsby,pitrou07} (see also
Ref.~\cite{NohHwang} and Ref.~\cite{Abbotetal} for the case of
higher-dimensional Kaluza-Klein models and Ref.~\cite{Qaniso} for
the quantization of test fields and particle production in an anisotropic spacetime). 
Recently, we performed a full analysis of
the cosmological perturbations in an arbitrary Bianchi~$I$
universe~\cite{ppu1}. It was soon followed by an
analysis~\cite{GCP} that focused on Bianchi~$I$ universe with a
planar symmetry. As we shall see in this work, the case of a
planar symmetric spacetime is not generic (both for the dynamics
of the background and the evolution of the perturbations).\\

From a more observationally oriented perspective, the primordial
anisotropy can imprint a preferred direction in the primordial
power spectra. This could be related to the possible large scale
statistical anomalies~\cite{lowQ} of the cosmic microwave
background (CMB) anisotropies. Many possible explanations have
been proposed, including foregrounds~\cite{prunet}, non-trivial
spatial topology~\cite{topologie} (which implies a violation of
global isotropy~\cite{modetopo}), the breakdown of local isotropy
due to multiple scalar fields~\cite{picon}, the presence of
spinors~\cite{spinor} or dynamical vectors \cite{dulaney}, the effect of the spatial gradient of the
inflaton~\cite{donoghue} or a late time violation of
isotropy~\cite{Jaffe2005}. In the case of universes with planar
symmetry, the signatures on the CMB were derived by many
authors~\cite{GCP,CMB_bianchi,acker}. In this case, a large
primordial shear is indeed necessary, which is not in contradiction
with the constraints obtained from the CMB~\cite{LimitShear} or
from the big-bang nucleosynthesis~\cite{bbn}. This brings a secondary
motivation to our analysis: can the CMB anisotropy be related to an
anisotropic primordial phase or, on the other hand, can it constrain the
primordial shear and what are the
exact predictions of a primordial  anisotropic phase? \\

This is however not our primary motivation. In the first place, we
are interested in understanding the genericity of the predictions
of inflation, and in particular with respect to the symmetries of
the background spacetime. As we shall see, the simple extension
considered in this article drives a lot of questions, 
concerning both the initial conditions in inflation and, more
generally, quantum field theory in curved spacetime.

In this article, we build on our previous work~\cite{ppu1} to
investigate the dynamics and predictions of one field inflation
starting from a generic Bianchi~$I$ universe. After discussing the
dynamics of the background in \S~\ref{sec1}, we focus on the
evolution of the perturbations in \S~\ref{sec2}. In particular, we
will need to understand the procedure for the quantization during
inflation. As we shall see in~\S~\ref{sec2.2}, this procedure
deviates from the one standardly used in a Friedmann-Lema\^{\i}tre
spacetime, and even in a planar symmetric spacetime as discussed
in Ref.~\cite{GCP}. The main reason for such an extension lies in
the fact that there always exist modes that were not in a WKB
regime during the shear-dominated inflationary era. It follows that, 
while our procedure leads to similar predictions to the standard one on
small scales, it appears that there is a lack of predictibility on
large scales. Indeed, we do not want to push our description
beyond the Planck or string scale where extensions of general
relativity have to be considered. This may give a description of
both the early phase of the inflationary era and also a procedure
to fix the initial conditions without any ambiguity. In
\S~\ref{sec3}, we explicitly compute the primordial spectra, for
both scalar modes and gravity waves. We will describe in details
the effect of the anisotropy and show how the isotropic
predictions are recovered on small scales.

\section{Background dynamics}\label{sec1}

We first set our notations in \S~\ref{sec1.1} and describe the
dynamics of the background, focusing on its general solutions
in~\S~\ref{sec1.2}. We then turn to the slow-roll regime and to
the case of a massive free field, that we shall use as our working
example, respectively in \S~\ref{sec1.3} and~\S~\ref{sec1.4}.

\subsection{Definitions and notations}\label{sec1.1}

Bianchi spacetimes~\cite{bianchimath} are spatially homogeneous
and those of type~$I$ have Euclidean hypersurfaces of homogeneity.
In comoving coordinates, and using cosmic time, their metric takes
the general form
\begin{equation}\label{metric1}
 \dd s^2 = -\dd t^2+ \sum_{i=1}^3 X_i^2(t)\,\left(\dd x^i\right)^2\ ,
\end{equation}
which includes three {\it a priori} different scale factors. It
includes the Friedmann-Lema\^{\i}tre spacetimes as a subcase when the
three scale factors are equal, and the extensively studied planar
symmetric universes when only two of the three scale factors are
different. The average scale factor, defined by
\begin{equation}\label{def_a}
 S(t) \equiv \left[X_1(t)X_2(t)X_3(t)\right]^{1/3}\,,
\end{equation}
characterises the volume expansion of the universe. The
metric~(\ref{metric1}) can then be recast under the equivalent
form
\begin{equation}\label{metric2}
 \dd s^2 = -\dd t^2+ S^2(t)\gamma_{ij}(t)\dd x^i\dd x^j\, ,
\end{equation}
where the ``spatial metric'', $\gamma_{ij}$, is the metric on
constant time hypersurfaces. It can be decomposed as
\begin{equation}\label{metricdec}
 \gamma_{ij}= \hbox{exp}\left[{2\beta_i(t)}\right]\delta_{ij}\,,
\end{equation}
where the functions $\beta_i$ must satisfy the constraint
\begin{equation}\label{beta}
 \sum_{i=1}^{3}\beta_{i}=0\,.
\end{equation}

Let us introduce some useful definitions. First, we consider the
scale factors
\begin{equation}
 a_i \equiv \hbox{e}^{\beta_i(t)}\ ,
 \qquad
 X_i = S a_i\, .
\end{equation}
They are associated to the following Hubble parameters
\begin{equation}
 H\equiv \frac{\dot S}{S}\ ,\qquad
 h_i = \frac{\dot X_i}{X_i}\ ,\qquad
 \dot\beta_i =\frac{\dot a_i}{a_i}\ ,
\end{equation}
which are trivially related by
\begin{equation}
 h_i = H + \dot\beta_i\ ,\qquad
 H=\frac{1}{3}\sum_{i=1}^3 h_i\ ,
\end{equation}
where the dot refers to a derivative with respect to physical
time. We define the shear as
\begin{equation}
 \hat\sigma_{ij}\equiv\frac{1}{2}\dot\gamma_{ij}
\end{equation}
and introduce the scalar shear by
\begin{equation}\label{e:210}
 \hat\sigma^2\equiv\hat\sigma_{ij}\hat\sigma^{ij} = \sum_i\dot\beta_i^2\ .
\end{equation}
(See appendix~A of Ref.~\cite{ppu1} to see the relation with
the shear usually defined in the $1+3$ formalism.) To finish, we
define the conformal time by $\dd t \equiv S\dd\eta$, in terms of
which the metric~(\ref{metric2}) is recast as
\begin{equation}\label{metricconforme}
 \dd s^2 = S^2(\eta)\left[-\dd \eta^2+ \gamma_{ij}(\eta)\dd x^i\dd x^j\right].
\end{equation}
We define the comoving Hubble parameter by $\mathcal{H}\equiv
S'/S$, where a prime refers to a derivative with respect to the
conformal time. The shear tensor, now defined as
\begin{equation}\label{defsigmaij}
 \sigma_{ij} \equiv \frac{1}{2}\gamma_{ij}'\ ,
\end{equation}
is clearly related to $\hat\sigma_{ij}$ by
$\sigma_{ij}=S\hat\sigma_{ij}$ so that $\sigma^2\equiv
\sigma_{ij}\sigma^{ij}$ is explicitly given by
$\sigma^2=\sum_{i=1}^3(\beta_i')^2$ and is related to its cosmic
time analogous by $\sigma=S\hat\sigma$.

\subsection{Friedmann equations and their general solutions}\label{sec1.2}

\subsubsection{Friedmann equations}

In cosmic time, assuming a general fluid as matter source with
stress-energy tensor
\begin{equation}\label{eq:a8}
 T_{\mu\nu}=  \rho u_\mu u_\nu + P(g_{\mu\nu}+u_\mu u_\nu) + \pi_{\mu\nu},
\end{equation}
where $\rho$ is the energy density, $P$ the isotropic pressure and
$\pi_{\mu\nu}$ the anisotropic stress ($\pi_{\mu\nu} u^\mu=0$ and
$\pi_\mu^\mu=0$), the Einstein equations take the form
\begin{eqnarray}
 3 H^2 &=& \kappa\rho + \frac{1}{2}\hat\sigma^2\,,\label{e:fried1C}\\
 \frac{\ddot S}{S}&=& -\frac{\kappa}{6}(\rho+3P) -\frac{1}{3}\hat\sigma^2\,,\label{e:fried2C}\\
 (\hat\sigma^i_j)^. &=& - 3H \hat\sigma^i_j +\kappa \tilde\pi^i_j\,,\label{e:fried3C}
\end{eqnarray}
and the conservation equation for the matter reads
\begin{equation}\label{e:cons1C}
 \dot\rho+ 3 H (\rho + P) + \hat \sigma_{ij} \tilde{\pi}^{ij}
 =0\,,
\end{equation}
where the $ij$-component of $\pi_{\mu\nu}$ has been defined as
$S^2 \tilde{\pi}_{ij}$ (so that
$\tilde{\pi}^i_{j}=\gamma^{ik}\tilde{\pi}_{kj}$).

In this work, we focus on inflation and assume that the matter
content of the universe is described by a single scalar field so that
\begin{equation}
 T_{\mu\nu}=\partial_\mu\varphi\partial_\nu\varphi-\left(\frac{1}{2}\partial_\alpha\varphi
 \partial^\alpha\varphi + V\right)g_{\mu\nu}\,.
\end{equation}
This implies that
\begin{eqnarray}
 3 H^2 &=& \kappa\left[\frac{1}{2}\dot\varphi^2+V(\varphi)\right]
           + \frac{1}{2}\hat\sigma^2\,,\label{e:fried1Cb}\\
 \frac{\ddot S}{S}&=&
 -\frac{\kappa}{3}\left[\dot\varphi^2-V(\varphi)\right]
  -\frac{1}{3}\hat\sigma^2\,,\label{e:fried2Cb}\\
 (\hat\sigma^i_j)^. &=& - 3H \hat\sigma^i_j\ .\label{e:fried3Cb}
\end{eqnarray}
The last of these equations is easily integrated and gives
\begin{equation}
 \hat\sigma^i_j=\frac{{\cal K}^i_j}{S^3}
\end{equation}
where ${\cal K}^i_j$ is a constant tensor, $({\cal K}^i_j)^{\cdot}=0$.
This implies that
\begin{equation}\label{e:223}
 \hat\sigma^2=\frac{{\cal K}^2}{S^6}\,,
\end{equation}
with ${\cal K}^2\equiv {\cal K}^i_j {\cal K}^j_i$, from which we
deduce that
\begin{equation}\label{SigKprim}
 \dot{\hat\sigma}=-3H\hat\sigma\,.
\end{equation}
The conservation equation reduces to the Klein-Gordon equation,
which keeps its Friedmannian form,
\begin{equation}\label{e:KGgen}
 \ddot\varphi+3H\dot\varphi+V_\varphi=0.
\end{equation}

\subsubsection{General solutions}

Let us concentrate on the particular case in which
$\pi_{\mu\nu}=0$ (relevant for scalar fields) and first set
\begin{equation}
 \beta_i = B_i W(t)\,,
\end{equation}
where $B_i$ are constants yet to be determined.
Equations~(\ref{e:210}) and (\ref{e:223}) then imply that
$$
 \left(\sum B_i^2\right)\dot W^2(t) = \frac{{\cal K}^2}{S^6}
 \, ,
$$
from which we deduce that
\begin{equation}\label{W}
 W(t)=\int\frac{\dd t}{S^3}
\end{equation}
The constraints~(\ref{beta}) and~(\ref{e:210}) imply that the
$B_i$ must satisfy
\begin{equation}
 \sum_{i=1}^3 B_i=0,\quad
 \sum_{i=1}^3 B_i^2 = {\cal K}^2\, ,
\end{equation}
which can be trivially satisfied by setting
\begin{equation}
 B_i = \sqrt{\frac{2}{3}}{\cal K}\sin\alpha_i,\quad {\rm with}\quad
 \alpha_i=\alpha+\frac{2\pi}{3}i,\quad i\in\{1,2,3\}.
\end{equation}
Thus, the general solution is of the form
\begin{equation}
 \beta_i(t) =
\sqrt{\frac{2}{3}} {\cal K}
\sin\left(\alpha+\frac{2\pi}{3}i\right)\times W(t),
\end{equation}
where $S$ is solution of
\begin{equation}
 3H^2=\kappa\rho +\frac{1}{2}\frac{{\cal K}^2}{S^6}.
\end{equation}
Once an equation of state is specified, the conservation equation
gives $\rho[S]$ and we can solve for $S(t)$.

As we shall see, it is convenient to introduce the reduced
shear
\begin{equation}\label{e232}
 x \equiv \frac{1}{\sqrt{6}}\frac{\hat\sigma}{H}=\frac{1}{\sqrt{6}}\frac{\sigma}{\HH}
\end{equation}
in terms of which the Friedmann equation takes the form
$$
 (1-x^2)H^2=\frac{\kappa}{3}\rho\ ,
$$
so that the local positivity of the energy density implies that
$x^2<1$.

\subsubsection{Particular case of a cosmological constant}\label{Sec_Pure_Lambda}

First, let us consider the case of a pure cosmological constant,
$V={\rm constant}$ and $\dot\varphi=0$. This case is relevant for the
initial stage of the inflationary period since we expect first the
shear to dominate and the field energy density to be dominated by
its potential energy. If not, then the field is fast rolling and
its energy density behaves as $S^{-6}$ exactly as the square of the
shear.

The Friedmann equation now takes the form
$$
 H^2 =  V_0 \left[ 1 + \left(\frac{S_*}{S}\right)^6 \right],
$$
with\footnote{Note the dimensions: $\rho\sim M^4$, $G\sim M^{-2}$,
$H\sim M$, $\hat\sigma\sim M$, $B_i\sim M$, $W\sim M^{-1}$,
$V_0\sim M^{2}$, $S_*\sim M^0$ where $M$ is a mass scale.}
$V_0\equiv \kappa V/3$ and $S_*\equiv ({\cal K}^2/6V_0)^{1/6}$. It
can be easily integrated to get
\begin{equation}\label{Eq_S_purelambda}
 S(t) = S_*\left[\sinh\left(t/\tau_* \right) \right]^{1/3}
\end{equation}
where we have introduced the characteristic time
\begin{equation}
 \tau_*^{-1}=3\sqrt{V_0}\ .
\end{equation}
Thus, we obtain from Eq.~(\ref{W})
\begin{equation}\label{Eq_Wdet}
 W(t) = W_0 + \frac{\tau_*}{S_*^3}\log\left[\tanh\left(\frac{t}{2\tau_*} \right) \right]\ ,
\end{equation}
where the constant $W_0$ can be set to zero (corresponding to the
choice of the origin of time). It follows, using that
$\sqrt{3/2}{\mathcal K}=S_*^3/\tau_*$, that the directional scale
factors behave as
\begin{eqnarray}
 X_i &=& S_* \left[\sinh\left(\frac{t}{\tau_*} \right) \right]^{1/3}
 \left[\tanh\left(\frac{t}{2\tau_*}\right)
 \right]^{\frac{2}{3}\sin\alpha_i}.
\end{eqnarray}
From this expression, we deduce that the directional Hubble
parameters evolve as
\begin{eqnarray}
 h_i &=& \frac{1}{3\tau_*}\frac{1}{\sinh(t/\tau_*)}\left[2\sin\alpha_i+\cosh(t/\tau_*)\ \right]
\end{eqnarray}
and further that the average Hubble parameter and reduced shear are given by
\begin{eqnarray}\label{Eq_H_purelambda}
 H &=& \frac{1}{3\tau_*}\frac{1}{\tanh(t/\tau_*)}\,,\qquad x=\frac{1}{\cosh(t/\tau_*)}\ .
\end{eqnarray}

\begin{figure}[htb]
 \includegraphics[]{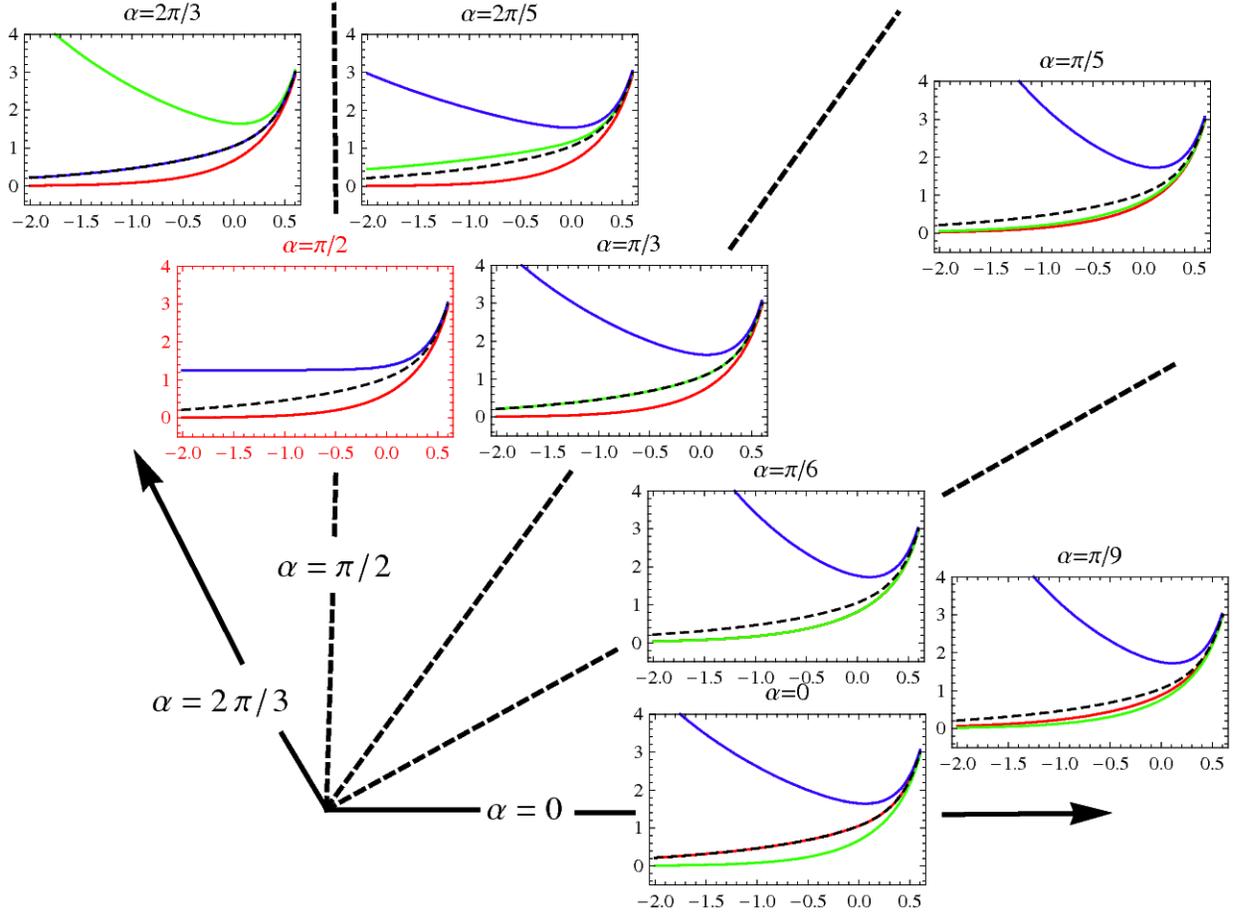}
 \caption{Evolution of the scale factors according to the value of
 the parameter $\alpha$. We depict the three
 directional scale factors and the average scale factor $S$
 (dashed line), all in units of $S_*$. The three
 directional scale factors are permuted when $\alpha$ is changed
 by $2\pi/3$. Note that there exist two particular cases in which
 the spacetime has an extra rotational symmetry when
 $\alpha=\pi/6$ or $\alpha=\pi/2$. The latter case is even more
 peculiar since this is the only Bianchi~$I$ universe for which
 none of the direction is bouncing.
}\label{fig1}
\end{figure}

The behaviours of the directional scale factors are depicted on
Fig.~\ref{fig1} for various values of the parameter $\alpha$. We
restrict to $\alpha\in[0,2\pi/3]$ and it is clear that for
$\alpha<\pi/2$ the $i=3$ direction is bouncing. At $\alpha=\pi/2$
none of the direction is contracting and then it switches to
direction $i=2$, when $\alpha>\pi/2$. It is thus clear that there is always one
bouncing direction, except in the particular case in which
$\alpha=\pi/2$.

\begin{figure}[htb]
 \includegraphics[width=6cm]{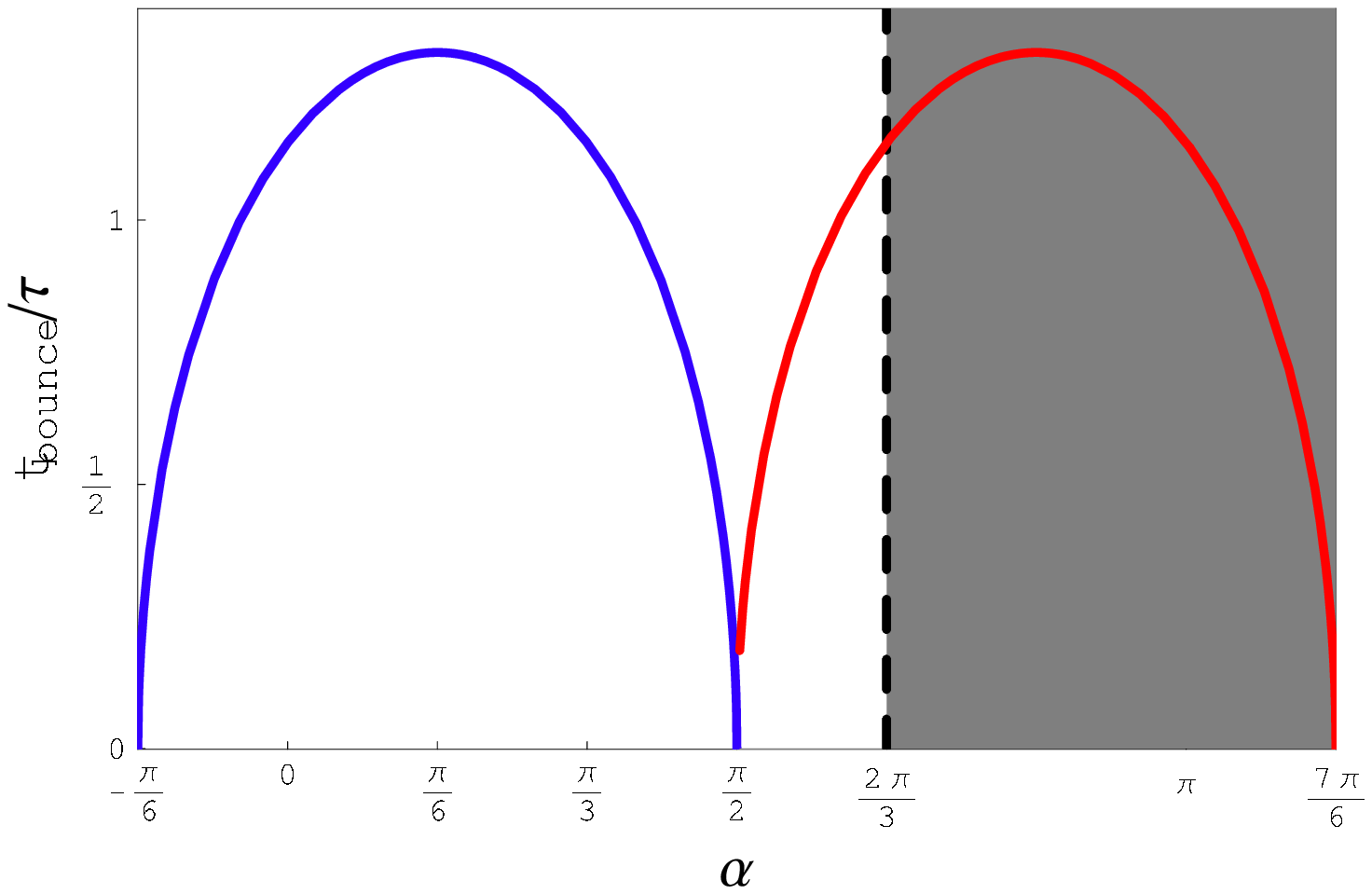}
 \includegraphics[width=7cm]{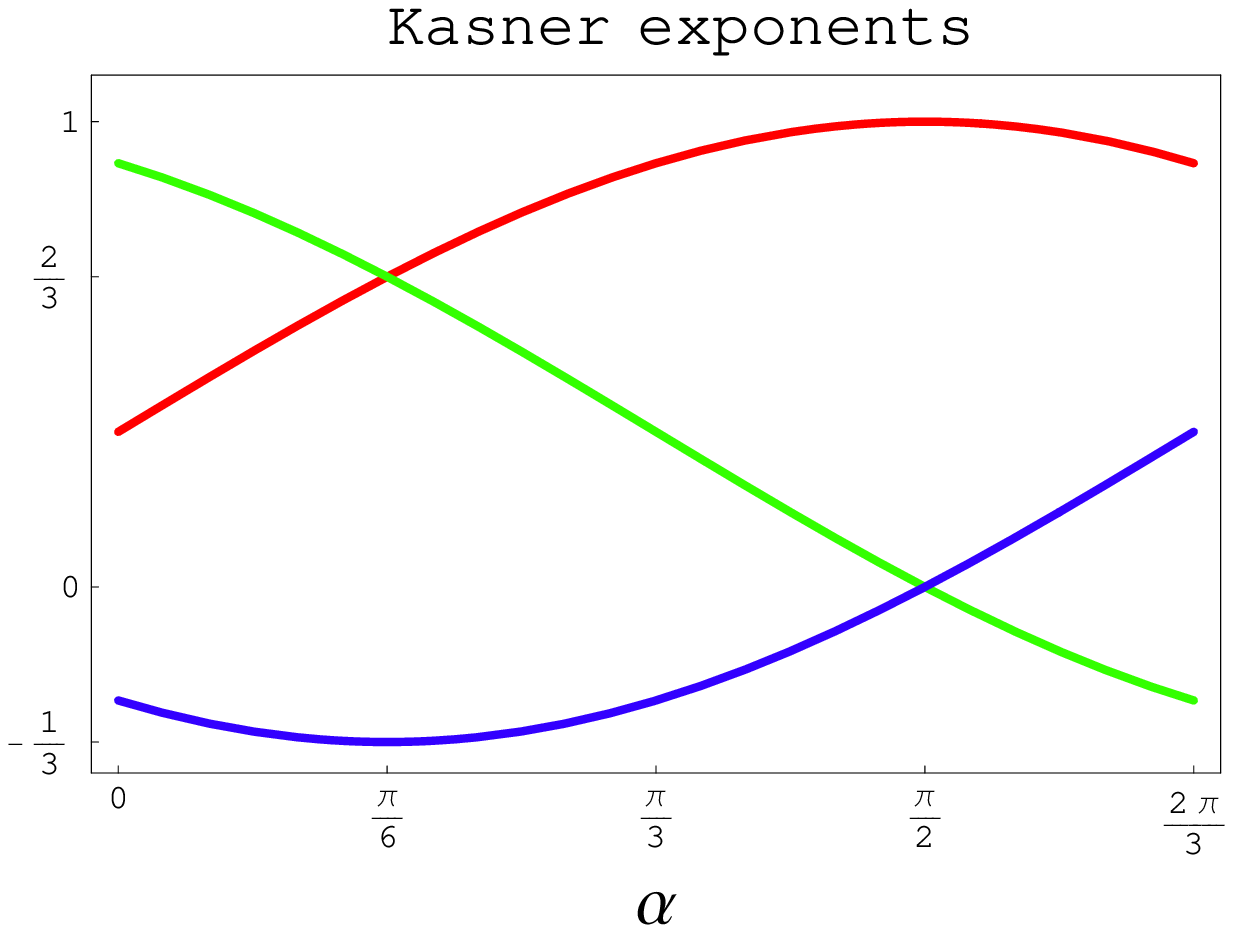}
 \caption{(left) Time at which the direction $i$ bounces (blue: $i=3$), (red: $i=2$).
 At $\pi/2$, none of the direction is contracting and a change of
 the contracting direction occurs. (right) Value of the Kasner exponents as
 a function of the parameter $\alpha$ characterising the Bianchi~$I$ model close to the singularity.
}\label{fig0}
\end{figure}

As a first conclusion, let us compute the time at which the
contracting direction bounces. Fig~\ref{fig0} (left) depicts the
value of the time of the bounce as a function of $\alpha$ and we
conclude that it is always smaller than $1.4\tau_*$.

\subsubsection{Behaviour close to the singularity}

Whatever the potential chosen for the inflaton, the Friedmann
equation will be dominated by the shear close to the singularity,
so that we can use the solutions obtained in the case of a pure
cosmological constant for the sake of the discussion. From the
previous analysis, we obtain that
\begin{eqnarray}
 X_i &=& S_*
 \left(\frac{t}{2\tau_*}\right)^{\frac{2}{3}\sin\alpha_i+\frac{1}{3}}\left[1
 +\frac{1}{18}(1-\sin\alpha_i)\left(\frac{t}{\tau_*}\right)^2+\mathcal{O}\left
 (\left(\frac{t}{\tau_*}\right)^4\right)\right].
\end{eqnarray}
The metric can thus be expanded around a Kasner solution of the
form
\begin{equation}
 \dd s^2_{_{\rm Kasner}}=-\dd t^2 + S_*^2\sum_{i=1}^3 \left(\frac{t}{2\tau_*}\right)^{2p_i}(\dd x^i)^2
\end{equation}
with the indices
\begin{equation}
 p_i(\alpha) = \frac{2}{3}\sin\alpha_i+\frac{1}{3}\ ,
\end{equation}
that clearly satisfy $\sum p_i=\sum p_i^2=1$ [see Fig.~\ref{fig0}
(right)]. We thus have
\begin{equation}
 \dd s^2\simeq-\dd t^2 + S_*^2\sum_{i=1}^3 \left(\frac{t}{2\tau_*}\right)^{2p_i}\left[
 1+\frac{1}{12}(1-p_i)\left(\frac{t}{\tau_*}\right)^2\right](\dd x^i)^2
\end{equation}
up to terms of order $(t/\tau_*)^4$.

The invariants of the metric behave as
\begin{eqnarray}
 R&=&\frac{4}{3 \tau_*^2}\ ,\\
 R_{\mu\nu}R^{\mu\nu}&=&\frac{4}{9\tau_*^2}\ ,\\
 R_{\mu\nu\rho\sigma}R^{\mu\nu\rho\sigma}&=&\frac{1}{27 \tau_*^2}\left\{ 8 +
  \frac{4}{\cosh^4\left[t/(2\tau_*)\right]}\, +
  \right. \nonumber\\
 &&\qquad\quad\left.
  +\frac{32
    \cosh\left(t/\tau_*\right)}{\sinh^4\left(t/\tau_*\right)}\left[3 \cos(\alpha)^2
     \sin( \alpha) - \sin( \alpha)^3 +1 \right]\right\}\ .
\end{eqnarray}
Clearly, we see that $R$ and $R_{\mu\nu}R^{\mu\nu}$ are regular at
the singularity, which is expected for a cosmological constant
dominated universe since $R_{\mu\nu}=-\Lambda g_{\mu\nu}$. The
third invariant $R_{\mu\nu\rho\sigma}R^{\mu\nu\rho\sigma}$
diverges when we are approaching the singularity, the only
exception being the case where $\alpha=\pi/2$ which corresponds to
the positive branch considered in Ref.~\cite{GCP}. In this
particular case, the Kasner metric has exponents $(1,0,0)$. We
emphasize that the spacetime $\alpha=\pi/2$ is a singular point in
the set of Bianchi spacetimes since there is no uniform
convergence of the invariants of the metric evaluated on the
singularity when $\alpha\rightarrow\pi/2$.

\subsection{Slow roll parameters}\label{sec1.3}

In order to discuss our results, we introduce the slow-roll
parameters in the usual way by
\begin{equation}\label{sr}
 \epsilon\equiv3\frac{\varphi^{\prime2}}{\varphi^{\prime2}+2S^2V}\,,
 \qquad
 \delta\equiv 1-\frac{\varphi''}{\HH\varphi'}=-\frac{\ddot\varphi}{H\dot\varphi}\,\, .
\end{equation}

With these definitions, the Friedmann and Klein-Gordon equations
take the form
\begin{equation}\label{srF}
 (1-x^2)\HH^2=\frac{\kappa}{3-\epsilon}VS^2\,,
 \qquad
 (3-\delta)\HH\varphi'+V_\varphi S^2=0\,,
\end{equation}
and we deduce that
\begin{equation}\label{Hprimeenslowroll}
\frac{\HH'}{\HH^2}=(1-\epsilon)+\left(\epsilon-3\right)x^2\,,
\end{equation}
so that
\begin{equation}
\frac{S^{\prime\prime}}{S}=\mathcal{H}^{2}\left[2-\epsilon+\left(\epsilon-3\right)x^{2}\right]
\end{equation}
and
\begin{equation}\label{Eq_epsilon_prime}
\epsilon'=2 \HH \epsilon(\epsilon-\delta)\,.
\end{equation}
Interestingly, it is easy to check that
$$
 x'= -\HH x(1-x^2)(3-\varepsilon)\ ,
$$
from which we deduce that
\begin{equation}
\delta'=\HH\left[ -9 x^2 +\frac{S^2 V_{, \varphi\varphi}}{\HH^2} -
(1-x^2) (3\epsilon + 3 \delta) + \delta(\delta +
\epsilon(1-x^2))\right]\ .
\end{equation}

From the definition~(\ref{sr}) and making use of the second
equation of Eqs.~(\ref{srF}), we deduce that the slow-roll
$\epsilon$ parameter takes the form
$$
\epsilon = \frac{(1-x^2)}{2 \kappa}
 \left(\frac{V_{,\varphi}}{V}\right)^2
 \left(\frac{3-\epsilon}{3-\delta}\right)^2\,.
$$
Once we use the relation $\dot{\varphi}^2= 2\epsilon
V/(3-\epsilon)$, we deduce that
\begin{equation}\label{phiprime}
\dot\varphi =
 -\frac{1}{\sqrt{\kappa}}
  \sqrt{\frac{(3-\epsilon)(1-x^2)}{(3-\delta)^2}}
 \frac{V_{,\varphi}}{\sqrt{V}}\ .
\end{equation}
As long as $\dot\varphi^2\ll V(\varphi)$ and in the limit
$x\rightarrow 1$, which corresponds to the shear dominated period
prior to the inflationary phase, we have that
\begin{equation}\label{Eq_slowroll_asymptotique}
 \epsilon\rightarrow 0\ ,
 \qquad
 \delta\rightarrow -3\ ,
\end{equation}
but we still have $\delta'\rightarrow0$. It follows that
initially, even if the shear decreases rapidly, $\varphi$ remains
almost constant and $\delta$ remains close to $-3$. This solution
converges when $t \rightarrow 0$ to the pure cosmological constant
solution of \S \ref{Sec_Pure_Lambda}. Note that this conclusion
differs from the statement of Ref.~\cite{lidsey}.

Before, we reach the slow-roll attractor, we may be in a
transitory regime in which $\dot \varphi^2 \gg V$. Then, this
implies that
$$
 \epsilon\rightarrow 3,\quad \delta\rightarrow 3,\quad \dot
\varphi\simeq \frac{\varphi_0}{t}\ .
$$
The field velocity decreases so that this solution converges
rapidly toward the slow-rolling attractor. These general results on 
the limiting behaviours will be important to understand the dynamics 
of the inflaton.

\subsection{Numerical integration for a massive scalar field}\label{sec1.4}

In this article, we will consider the explicit example of chaotic
inflation with a potential
\begin{equation}\label{def:Vpot}
 V=\frac{1}{2}m^2\varphi^2\, .
\end{equation}
The dynamical equations~(\ref{e:fried1Cb}) and~(\ref{e:KGgen}) can
be rescaled as
\begin{eqnarray}\label{sys-dimensionless1}
 && h^2=\frac{1}{6}\left[\frac{1}{2}\dot\psi^2 + \psi^2
  +\left(\frac{S_*}{S}\right)^6 \right]\\\label{sys-dimensionless2}
 &&\ddot\psi+3h\dot\psi+\psi=0\ ,
\end{eqnarray}
where we use $\tau=mt$ as a time variable (so that $h=H/m$),
$\psi=\varphi/M_p$ (with $M_p^{-2}=8\pi G=\kappa$) as the field
variable, and where $S_*=({\cal K}/m)^{1/3}$. Under this form, it
is clear that the various solutions of the system
(\ref{sys-dimensionless1}-\ref{sys-dimensionless2}) are, in
general, characterised by the three numbers
$\lbrace\psi(t_0),\dot\psi(t_0),S_*\rbrace$. As we will now see,
this extra dependence on the parameter $S_*$ causes our dynamical
system to behave differently from its analogous in
Friedmann-Lema\^{\i}tre spacetimes\footnote{There is another way
to see this: the term in $S^{-6}$ in
Eq.~(\ref{sys-dimensionless1}) is equivalent to the contribution
of a massless scalar field, $\chi$ say, that would satisfy the
Klein-Gordon equation $\ddot\chi+3 h\dot\chi=0$, implying
$\dot\chi\propto S^{-3}$. One requires initial conditions for this
extra-degree of freedom as well.}. Under this form, we also see clearly
that one cannot continuously go from a Bianchi to a FL spacetime
(because either $S_*=0$ or $S_*\neq0$), even though Bianchi spacetime
isotropizes.

It is well known that the dynamics of the inflationary stage in a
Friedmann-Lema\^{\i}tre spacetime is characterised by attractor
solutions, clearly seen in the phase space. For a large set of
initial conditions, the solutions converge to the slow-roll stage
defined by an almost constant $\dot \varphi$ and $\varphi$
decreasing accordingly. It is thus given by a roughly horizontal
line in the phase portrait which ends with the oscillations of the
scalar field at the bottom of its potential. As previously
mentioned, the slow-roll inflationary stage in Bianchi~$I$
spacetimes possesses the same attractor behaviour, although it is
quite different during the initial shear-dominated phase. When the
shear dominates, the solutions are rapidly attracted to the point
$\dot \varphi \simeq 0$ and $\varphi$  nearly constant. This is
the attractor whose solution is given by
Eqs.~(\ref{Eq_dev_H_B1}-\ref{Eq_dev_epsilon_B1}). It then
converges towards the Friedmann-Lema\^{\i}tre behaviour when the
shear becomes negligible (see Fig.~\ref{fig:phase3D}).

In conclusion, we have a double attraction mechanism, namely of
the field dynamics toward the slow-roll attractor and of the
Bianchi spacetime toward a Friedmann-Lema\^{\i}tre solution.

\begin{figure}[htb]
 \includegraphics[width=14cm]{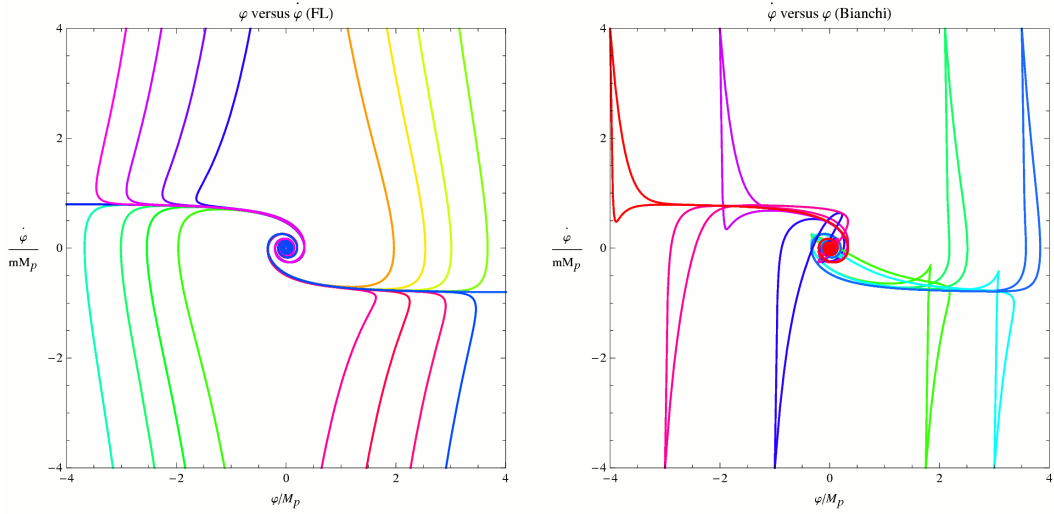}
 \caption{Comparison of the phase portraits of a Friedmann-Lema\^{\i}tre (left) and a
 Bianchi~$I$ inflationary phases. The curves with same colour
 corresponds to same initial conditions for the scalar field but
 starting with different initial shear (see Fig.~\ref{fig:phase3D} for a 3-dimensional
 representation).
}\label{fig:dynaFLvB}
\end{figure}

\begin{figure}[htb]
 \includegraphics[width=8cm]{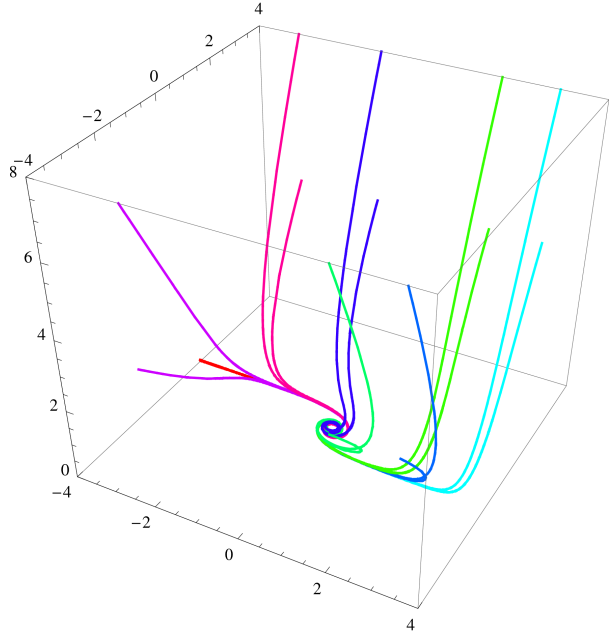}
 \caption{Phase portrait of a
 Bianchi~$I$ inflationary phases in the space
 $\{\varphi,\dot\varphi,\sigma\}$. The plane $\sigma=0$ corresponds
 to the FL-limit (see Fig.~\ref{fig:dynaFLvB}). This illustrates the
 double attraction mechanism of the spacetime toward FL and of the
 solution toward the slow-roll attractor.}\label{fig:phase3D}
\end{figure}

\begin{figure}[htb]
 \includegraphics[width=8cm]{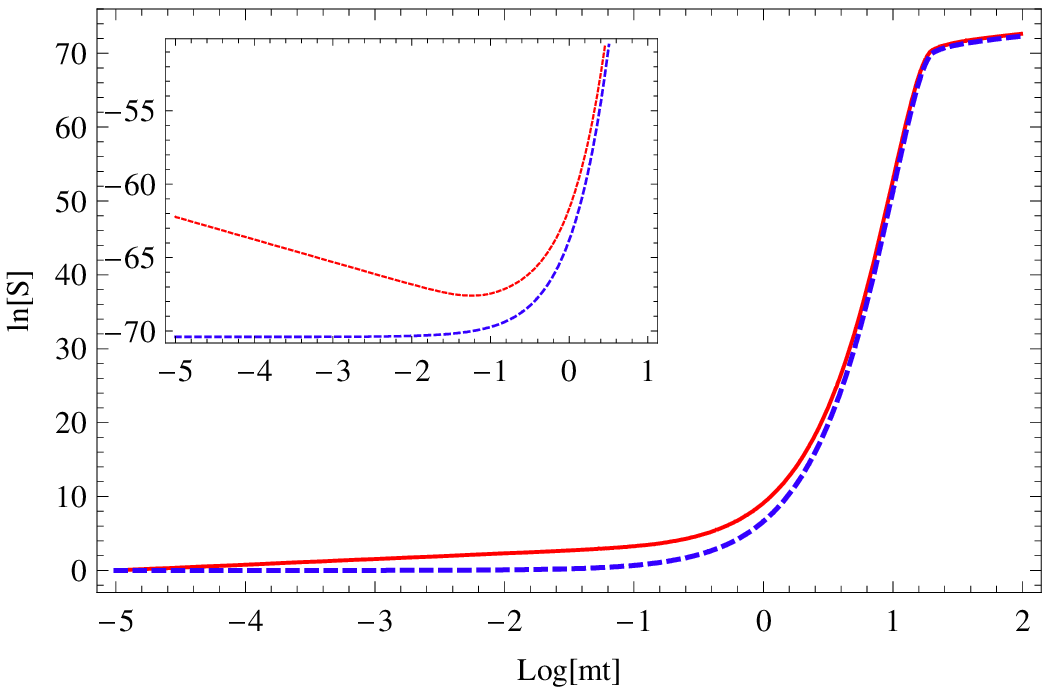}\includegraphics[width=8cm]{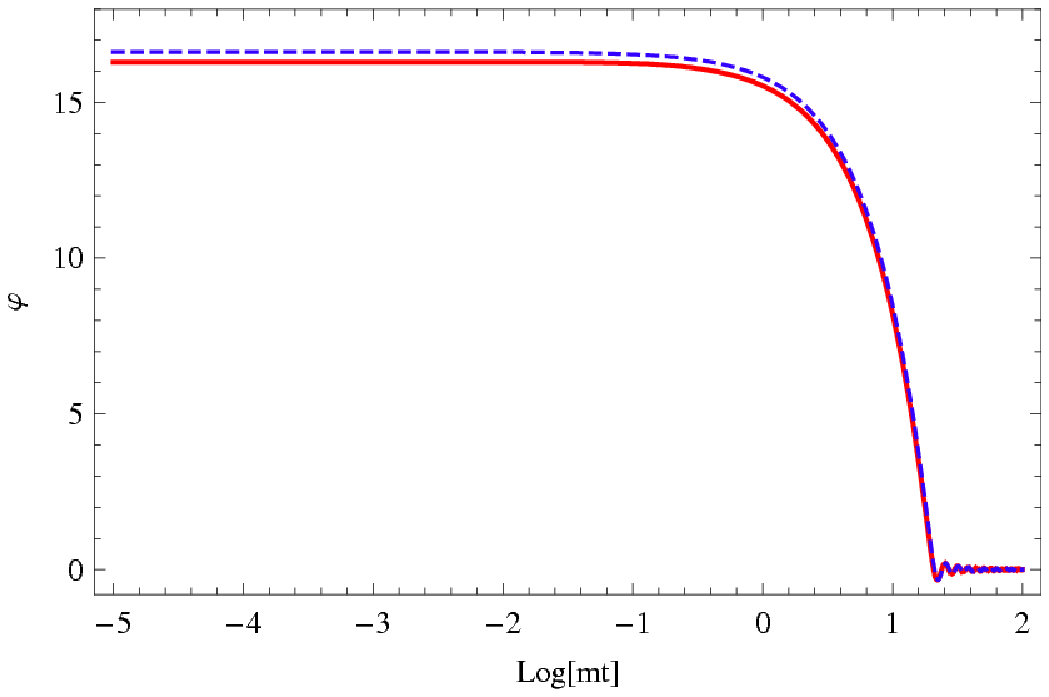}
 \caption{Evolution of the average scale factor (left) and scalar field (right)
 as a function of time for both the Bianchi (red, solid line) and FL cases (blue, dashed line).
 The Bianchi  solution is characterised by $\varphi_0=16 M_p$. We have normalized the
 solutions such that the scale factors have the same value at the
 end of the inflation when the shear is negligible. The inner left figure shows the
 logarithmic evolution of the velocity of the expansion in both cases, the minimum
 of which indicates the time at which $\ddot S=0$ for the Bianchi
 case.}\label{fig:dyn-inflation}
\end{figure}

In Fig.~\ref{fig:dyn-inflation}, we compare the dynamics of the
inflationary phase of a Bianchi~$I$ and a Friedmann-Lema\^{\i}tre
spacetimes. While in both cases the (average) scale factors grow,
the effect of a shear-dominated phase is to initially decrease the
velocity of the expansion of the Bianchi universe. On the other
hand, the dynamics of the scalar field is barely affected by the
presence of the shear as long as we have reached the slow-roll
attractor. In the following we will assume that we have reached
this attractor. If the number of $e$-folds is small, it is not
clear that this attractor has been reached, regardless of whether 
or not we assume a Friedmann-Lema\^{\i}tre spacetime. In such a case, one
has to rely on families of trajectories to draw the observational
predictions (see e.g. Ref.~\cite{levprivate}).

Let us now consider the behaviour close to the singularity and
introduce the characteristic time $\tau_*$, explicitly given by
\begin{equation}\label{valtau}
 \tau_*= \sqrt{\frac{2}{3}} \frac{M_p}{m \varphi_0}\ .
\end{equation}
By developing all the previous equations in powers of $t/\tau_*$,
we obtain that
\begin{eqnarray}
H(t) &=& \frac{1}{3t}\left[1+ \frac{1}{3}\frac{t^2}{\tau_*^2}+
     \mathcal{O}\left(\frac{t^4}{\tau_*^4}\right)\right]\ ,\label{Eq_dev_H_B1}\\
x(t) &=& 1 - \frac{1}{2}\frac{t^2}{\tau_*^2}+
     \mathcal{O}\left(\frac{t^4}{\tau_*^4}\right)\ ,\\
\varphi(t) &=& \varphi_0\left[1- \frac{1}{6}
\left(\frac{M_p}{\varphi_0}\right)^2
 \frac{t^2}{\tau_*^2}+
     \mathcal{O}\left(\frac{t^4}{\tau_*^4}\right)\right]\ ,\\
\delta(t) &=&-3 \left[1 - \frac{1}{2}\frac{t^2}{\tau_*^2}+
     \mathcal{O}\left(\frac{t^4}{\tau_*^4}\right)\right]\ ,\\
\epsilon(t) &=& \frac{1}{3} \left(\frac{M_p}{\varphi_0}\right)^2
 \frac{t^2}{\tau_*^2}+
     \mathcal{O}\left(\frac{t^4}{\tau_*^4}\right)\ ,\label{Eq_dev_epsilon_B1}
\end{eqnarray}
in complete agreement with the expansions obtained in
Ref.~\cite{GCP}. In this limit, we understand why
$\delta\rightarrow-3$ close to the singularity. It simply reflects the fact that the field decreases as $t^2$. This is different
from the case of chaotic inflation in a Friedmann-Lema\^{\i}tre
spacetime where the field varies linearly with time during the
slow-roll regime since in that case
\begin{eqnarray}
 \varphi = \varphi_i\left[1 - \left(\frac{M_p}{\varphi_i}\right)^2\frac{t}{\tau_*}\right]\
 , \qquad
 S=S_i\, \exp\left\lbrace{\frac{1}{M_p^2}[\varphi_i^2-\varphi^2(t)]}\right\rbrace \ , \nonumber
\end{eqnarray}
and the slow-roll parameters are explicitly given by
$$
 \epsilon = 2\frac{M_p^2}{\varphi^2}\ ,
 \qquad
 \delta= 0 \ .
$$

\begin{figure}[htb]
 \includegraphics[width=16.5cm,height=6.5cm]{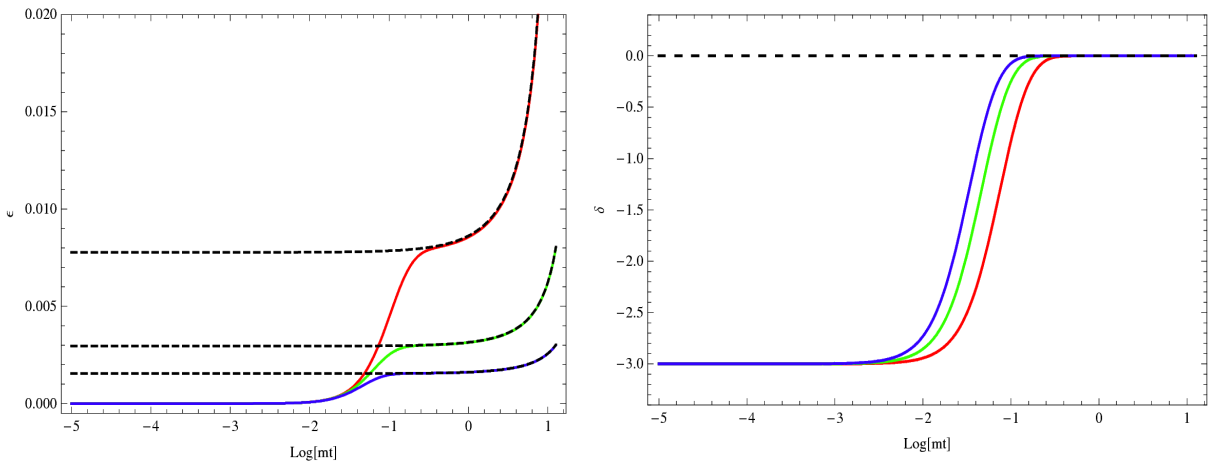}
 \caption{Evolution of the slow-roll parameters $\epsilon$ (left) and $\delta$ (right) during the
 inflationary phase for a Bianchi~$I$ model (solid lines) with $\alpha=\pi/4$
 compared to the case of a Friedmanian model (dashed lines). We assume a potential
 of the form~(\ref{def:Vpot}) and initial values $\varphi_0=(16,26,36)M_p$ corresponding respectively
to the red, green and blue lines.
}\label{fig:srpara}
\end{figure}

Let us come back on the slow-roll parameters defined in
\S~\ref{sec1.3}. In the particular case of a quadratic potential,
we have
\begin{equation}
\delta'= \HH (3-\delta)\left[\frac{\epsilon^2-3
\delta}{3-\epsilon} -x^2(3-\epsilon) \right]\ ,
\end{equation}
since $S^2 V_{,\varphi\varphi} = \epsilon {\HH^2}
(3-\delta)^2/(3-\epsilon)$. During the shear-dominated phase,
$\delta\sim-3$ and the previous equation tells us that $\delta$
remains constant. It follows from Eq.~(\ref{Eq_epsilon_prime})
that $\epsilon'$ is of the same order than $\epsilon$ (note the
difference between the standard case in which $\epsilon'$ is
second order). It follows that the variation of $\epsilon$ cannot
be neglected until $\delta$ has converged toward $0$. Then
$\epsilon$ can be considered as constant until the end of
inflation. While the universe isotropizes, both $\delta$ and
$\epsilon$ converged toward their Friedmann-Lema\^{\i}tre value.
Fig.~\ref{fig:srpara} illustrates the evolution of the two
slow-roll parameters and compare them to their values in a
Friedmann-Lema\^{\i}tre universe.

\begin{figure}[htb]
 \includegraphics[width=8cm,height=6.5cm]{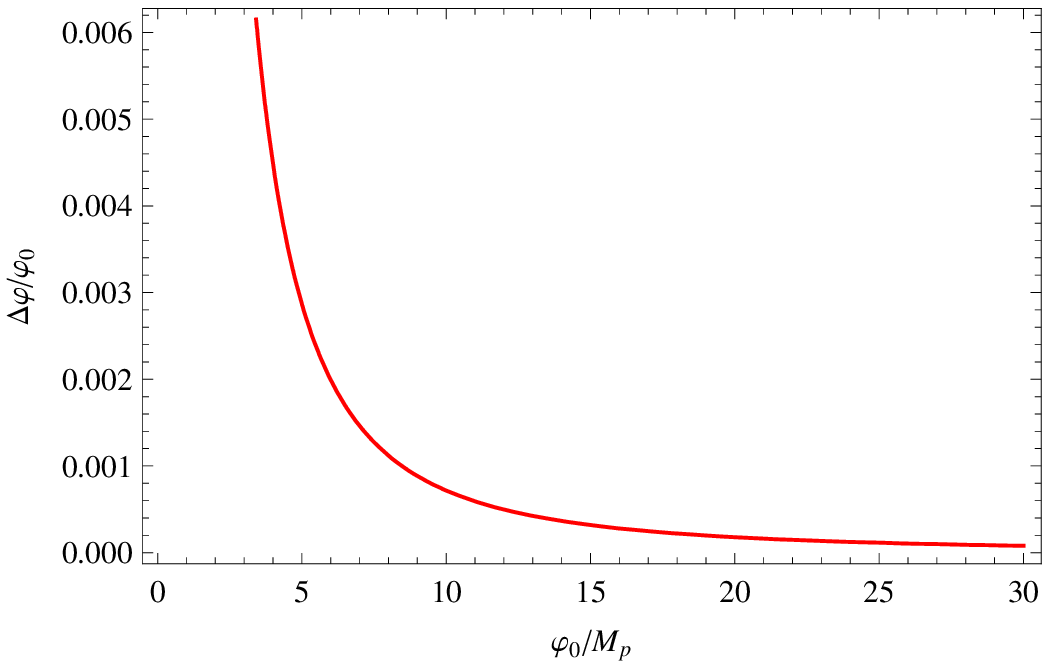}
 \includegraphics[width=8cm,height=6.5cm]{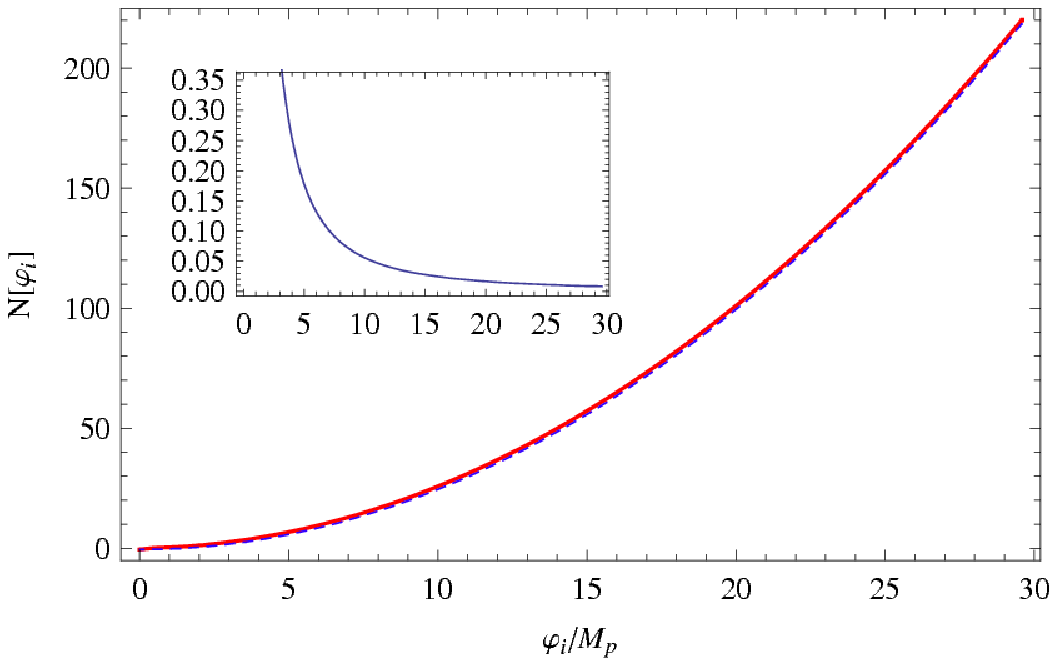}
 \caption{(left) Relative variation of the scalar field during the shear dominated era in
 function of its initial value.  (right) Comparison of the
number of $e$-folds for a Bianchi (solid, red line) and Friedmann-Lema\^itre (blue, dashed line)
 inflationary period as a function of $\varphi_i$. The inside plot is the
 relative difference between the number of $e$-folds of the Bianchi and
 Friedmann-Lema\^itre inflationary period as a function of $\varphi_i$.}\label{fig:efold}
\end{figure}

For any initial value of the scalar field, $\varphi_0$, we define
$\varphi_i$ as the value of the field when the universe starts to
inflate, that is at the time when $\ddot S=0$. Then, the number of
$e$-folds of the inflationary period is defined as
\begin{equation}\label{Def_efolds}
 N[\varphi_i] \equiv \ln\frac{S(\varphi_f)}{S(\varphi_i)}
   = \int_{\varphi_i}^{\varphi_f}\frac{H}{\dot\varphi}\dd\varphi\ ,
\end{equation}
where $\varphi_f$ is the value of the field at which $\epsilon=1$.
Since at that time the shear is negligible, $\varphi_f$ is given
by
\begin{equation}
 \varphi_f = \sqrt{2} M_p\ .
\end{equation}
It follows that for any initial value of the field, we can
characterise the shear-dominated phase by
\begin{equation}\label{deltaphi}
 \Delta\varphi[\varphi_0]=\varphi_0 - \varphi_i[\varphi_0]\ ,
\end{equation}
which indicates by how much the field has moved prior to
inflation. Then, the duration of the inflationary phase is given
\begin{equation}
 N[\varphi_0]
   = \int_{\varphi_i(\varphi_0)}^{\sqrt{2} M_p}\frac{H}{\dot\varphi}\dd\varphi\
   .
\end{equation}
This has to be compared with the number of $e$-folds in the FL
case
$$
 N_{\rm  FL}[\varphi_i] =
 \frac{1}{4}\left(\frac{\varphi_i}{M_p}\right)^2 - \frac{1}{2}\ .
$$
In Fig.~\ref{fig:efold} (left), we depict the fractional duration
of the shear-dominated phase [see Eq. (\ref{deltaphi})] as a
function of $\varphi_0$. We see that the larger the initial value
of the field, the smaller the impact of the shear. Also shown in
Fig.~\ref{fig:efold} is the number of $e$-folds for the
Friedmann-Lema\^itre and Bianchi spacetimes. The presence of the
shear slightly increases the number of $e$-folds (solid red line in
Fig.~\ref{fig:efold}).

\subsection{Discussion}

This general study of the background shows that generically there
is always one bouncing direction, except in the particular case
$\alpha=\pi/2$ considered in Ref.~\cite{GCP} (positive branch).
Regarding the dynamics of the universe, the expansions
(\ref{Eq_dev_H_B1}-\ref{Eq_dev_epsilon_B1}) at lowest order
exactly reproduce the exact (numerical) solutions.

This analysis actually shows that the shear is effective typically until the
characteristic time $\tau_*$. Going backward in time, the universe
goes rapidly to an initial singularity and is thus past incomplete
(in fact as other inflationary models~\cite{borde}). Indeed, as we
shall discuss later, we do not want to extrapolate such a model up
to the Planck or string times and we just assume that they are a
good description of the inflating universe after this time.

We have focused on the evolution of the ``slow-roll'' parameters
and concluded that the variation of $\epsilon$ cannot be neglected
until the shear has decayed. We have also shown that the scalar
field barely moves during the shear dominated era and that, with
same initial value for the scalar field, the number of $e$-folds
is almost not affected (even though slightly larger) by a
non-vanishing shear.

\section{Summary of the perturbation theory}\label{sec2}

In our previous work~\cite{ppu1}, we investigated the theory of
cosmological perturbations around a Bianchi~$I$ universe. We shall
now briefly summarize the main steps which are necessary to deduce
the dynamical system of equations that we want to solve in this
article.

\subsection{Mode decomposition}

First, we pick up a comoving coordinates system, $\{x^i\}$, on the
constant time hypersurfaces. Any scalar
function can then be decomposed in Fourier modes as
\begin{equation}\label{fft}
 f\left(x^j,\eta\right)= \int\frac{\dd^3k_i}{\left(2\pi\right)^{3/2}}
         \,\hat
         f\left(k_i,\eta\right)\,\hbox{e}^{\mathrm{i}k_ix^i}\,.
\end{equation}
In the Fourier space, the comoving wave co-vectors $k_i$ are
constant, $k_i'=0$. We now define $k^i\equiv\gamma^{ij}k_j$ that
is obviously a time-dependent quantity. Contrary to the standard
Friedmann-Lema\^{\i}tre case, we must be careful not to trivially
identify $k_i$ and $k^i$, since this does not commute with the
time evolution. Note however that $x_ik^i=x^ik_i$ remains
constant.

Besides, since $(k^i)'=-2\sigma^{ip}k_p$,  the modulus of the
comoving wave vector, $k^2 = k^{i}k_{i} = \gamma^{ij}k_{i}k_{j}$,
is now time-dependent and its rate of change is explicitly given
by
\begin{equation}
 \frac{k'}{k} = - \sigma^{ij}\hat k_{i}\hat k_{j}\ ,
\end{equation}
where we have defined the unit vector
\begin{equation}
 \hat k_i \equiv \frac{k_i}{k}\ .
\end{equation}
Now, we introduce the base $\{e^1,e^2\}$ of the subspace
perpendicular to $k^i$. By construction, it satisfies the
orthonormalisation conditions
$$
 e^a_i k_j \gamma^{ij}=0\ ,\qquad
 e^a_i e^b_j \gamma^{ij}= \delta^{ab}\ .
$$
Such a basis is indeed defined up to a rotation about the axis
$k^i$. These two basis vectors allow to define a projection
operator onto the subspace perpendicular to $k^i$ as
\begin{equation}\label{P}
 P_{ij} \equiv e_i^1e_j^1 + e_i^2e_j^2 =\gamma_{ij} -\hat k_i\hat k_j\ .
\end{equation}
It trivially satisfies $P^i_jP^j_k=P^i_k$, $P^i_jk^j=0$ and
$P^{ij}\gamma_{ij}=2$.

Any symmetric tensor $\bar V_{ij}$ that is transverse and
trace-free has only two independent components and can be
decomposed as
\begin{equation}\label{dec_tensor}
 \bar V_{ij}(k_i,\eta)=\sum_{\lambda=+,\times} V_\lambda(k^i,\eta)
 \, \varepsilon_{ij}^\lambda(\hat k_i)\ ,
\end{equation}
where the polarization tensors have been defined as
\begin{equation}\label{defespilonij}
 \varepsilon_{ij}^\lambda=\frac{e_i^1e_j^1 - e_i^2e_j^2}{\sqrt{2}}\delta^\lambda_+
 + \frac{e_i^1e_j^2 + e_i^2e_j^1}{\sqrt{2}}\delta^\lambda_\times\,.
\end{equation}
It can be checked that they are traceless
($\varepsilon_{ij}^\lambda \gamma^{ij}=0$), transverse
($\varepsilon_{ij}^\lambda k^i=0$), and that the two polarizations
are perpendicular ($\varepsilon_{ij}^\lambda
\varepsilon^{ij}_{\mu}=\delta^\lambda_{\mu}$). This defines the
two tensor degrees of freedom.

\subsection{Decomposition of the shear}

It is then fruitful to decompose the shear in a local basis
adapted to the mode that we are considering. The shear being a
symmetric trace-free tensor, it can be decomposed on the basis
$\{\hat k_i, e^1_i, e^2_j\}$ as
\begin{equation}\label{decshear}
 \sigma_{ij} = \frac{3}{2}\left(\hat k_i\hat k_j-\frac{1}{3}\gamma_{ij}\right)\spar
 + 2\sum_{a=1,2}\sva \,\hat k_{(i}e^a_{j)}
 + \sum_{\lambda=+,\times}\stl\,\varepsilon^\lambda_{ij}.
\end{equation}
This decomposition involves 5 independent components in a basis
adapted to the wavenumber $k_i$.  We must however stress that
$(\spar,\sva,\stl)$ do not have to be interpreted as the Fourier
components of the shear, even if they explicitly depend on $k_i$.
This dependence arises from the local anisotropy of space.

Using Eq.~(\ref{decshear}), it is easily worked out that
$\sigma_{ij}\gamma^{ij} = 0$, and that
\begin{equation}\label{Eq_dec_shear1}
 \sigma_{ij}\hat k^i = \spar\hat k_j + \sum_a\sva e^a_j\,,
 \qquad
 \sigma_{ij}\hat k^i\hat k^j = \spar\,,
\end{equation}
and
\begin{equation}\label{Eq_dec_shear2}
 \sigma_{ij}\varepsilon_\lambda^{ij}=\stl\,,\qquad
 \sigma_{ij}\hat k^i e^j_a = \sva\,.
\end{equation}
The scalar shear is explicitly given by
\begin{equation}
 \sigma^2 = \sigma_{ij}\sigma^{ij}
          = \frac{3}{2}\spar^2 + 2\sum_a\sva^2 + \sum_\lambda\stl^2\,,
\end{equation}
which is, by construction, independent of $k_i$. We emphasize that
the local positivity of the energy density of matter implies [see
Eq.~(\ref{e232})] that $\sigma^2/6 < \HH^2$ and thus
\begin{equation}\label{e:ssurH}
 \frac{1}{2}\spar \leq \frac{1}{\sqrt{6}}\sigma < \HH\,.
\end{equation}
This, in turn, implies that
\begin{equation}\label{e:ssurH2}
 {\spar}<2\HH\,,
\end{equation}
a property that shall turn to be very useful in the following of
our discussion. Analogously, we have that
\begin{equation}\label{e:ssurH3}
 {\stl}<\sqrt{6}\HH\,.
\end{equation}

\subsection{Gauge invariant variables}

We start from the most general metric of an almost Bianchi~$I$
spacetime,
\begin{equation}\label{dmet1}
  \dd s^{2}=S^2\left[-\left(1+2A\right)\dd\eta^{2}+2 B_{i}\dd x^{i}\dd\eta
    +\left(\gamma_{ij}+h_{ij}\right)\dd x^{i}\dd x^{j}\right].
\end{equation}
$B_{i}$ and $h_{ij}$ are further decomposed as
\begin{eqnarray}\label{tens-decomp}
B_{i} & = & \partial_{i}B+ \bar{B}_{i}\,, \\
h_{ij} & \equiv & 2C\left(\gamma_{ij}+\frac{\sigma_{ij}}{\HH}
\right)+2\partial_{i}\partial_{j}E+2\partial_{(i}E_{j)}+2
E_{ij}\,,
\end{eqnarray}
with
\begin{equation}
 \partial_i \bar{B}^{i}=0=\partial_i E^i, \quad
 E_i^i=0=\partial_iE^{ij}.
\end{equation}
We showed, that one can construct the following gauge invariant
variables
\begin{eqnarray}
  \Phi & \equiv & A+\frac{1}{S}\left\{
  S\left[B-\frac{\left(k^{2}E\right)'}{k^{2}}\right]\right\}'\,,\\
 \Psi & \equiv & -C-\mathcal{H}\left[B-\frac{\left(k^{2}E\right)'}{k^{2}}\right]\,.
\end{eqnarray}
for the scalar modes,
\begin{equation}
 \Phi_{i} \equiv \bar{B}_{i}-\gamma_{ij}\left(E^{j} \right)' +2\mathrm{i}k^j
 \sigma_{lj}P^{l}_{\,\,i} E,
\end{equation}
for the vector modes and that the tensor mode $E_{ij}$ is readily
gauge invariant.

Concerning the matter sector, one can introduce a single gauge
invariant variable associated with the scalar field perturbation,
\begin{equation}
 Q \equiv \delta\varphi - \frac{C}{\HH}\varphi'\ .
\end{equation}

\subsection{The Mukhanov-Sasaki variables and their evolution equations}

We established that the only degrees of freedom reduce to a scalar
mode and two tensor modes
\begin{equation}\label{e:defms}
 v\equiv SQ\,,\qquad
 \sqrt{\kappa}\mu_\lambda \equiv S E_\lambda\ .
\end{equation}
They evolve according to
\begin{eqnarray}\label{b1:v}
 && v'' + \omega^2_v(k_i,\eta) v =
 \sum_\lambda\aleph_\lambda(k_i,\eta)\mu_\lambda\ , \\
 && \mu_\lambda'' + \omega^2_\lambda(k_i,\eta) \mu_\lambda =
 \aleph_\lambda(k_i,\eta)v + \beth(k_i,\eta)\mu_{(1-\lambda)}\ ,\label{b1:mu}
\end{eqnarray}
where the pulsations are explicitly given by
\begin{equation}
 \omega^2_v(k_i,\eta) \equiv k^2 - \frac{z_{\rm s}''}{z_{\rm s}}\
 ,
 \qquad
 \omega^2_\lambda(k_i,\eta) \equiv k^2 -
 \frac{z_{\lambda}''}{z_\lambda}\ .
\end{equation}
The two functions $z_{\rm s}$ and $z_\lambda$ have been defined by
\begin{eqnarray}
 \frac{z_{\rm s}''}{z_{\rm s}}(\eta,k_i) &\equiv& \frac{S''}{S}-S^{2}V_{,\varphi\varphi}
 + \frac{1}{S^2}\left(\frac{2S^2\kappa
     \varphi^{\prime2}}{2\HH-\spar}\right)^{\prime}\ ,\\
 \frac{z_{\lambda}''}{z_\lambda}(\eta,k_i) &\equiv& \frac{S''}{S}
 +2\sigma_{_{\rm T} (1-\lambda)}^{2}+\frac{1}{S^2}\left(S^2\spar\right)'
 +\frac{1}{S^2}\left(\frac{2S^2\sigma_{_{\rm
T}\lambda}^2}{2\HH-\spar}\right)'\ ,
\end{eqnarray}
and we also need the coupling terms
\begin{eqnarray}
 \aleph_\lambda(\eta,k_i) &\equiv& \frac{1}{S^2}
     \sqrt{\kappa}\left(\frac{2S^2\varphi'\sigma_{_{\rm T}\lambda}}{2\HH-\spar}
           \right)'\ ,\\
 \beth(\eta,k_i) &\equiv& \frac{1}{S^2}\left(\frac{2^2 \stcross
    \stplus}{2\HH-\spar}\right)'-2 \stcross \stplus\
    .\label{systlast}
\end{eqnarray}

\section{Prescription for the initial conditions}\label{sec2.2}

The set of equations~(\ref{b1:v}-\ref{systlast}) completely
determines the evolution of the three degrees of freedom of our
problem. To be predictive, we must determine the initial
conditions.

In a FL spacetime, the procedure is well understood~\cite{mbf} and
relies on the quantization of the canonical variables on
sub-Hubble scales where it can be shown that they evolve
adiabatically.

We have to understand how far this procedure can be extended to a
Bianchi universe and how robust it is to the existence of a
non-vanishing primordial shear. We thus start, in \S~\ref{sec2.2.a} by 
a review of the standard FL procedure, to highlight its hypothesis. 
In \S~\ref{sec2.2.b}, we stress, and also quantify, the differences that appear in a
Bianchi universe. This will lead us (\S~\ref{sec2.2.c}) to propose
an extension of the quantization procedure. We shall finish in
\S~\ref{sec2.2.e} by critically discussing the limits and
weaknesses of our quantization procedure.

\subsection{Friedmann-Lema\^{\i}tre universes}\label{sec2.2.a}

For simplicity, let us consider the case of a pure de Sitter
phase. This represents no limitation to our following arguments and can 
be generalised to an almost-de Sitter phase.

\subsubsection{Quantization procedure}

In order to be quantized, the canonical variables are promoted to
the status of quantum operators~\cite{mbf} and are decomposed as
\begin{eqnarray}\label{e:modeexp}
 \hat v (\bx,\eta) &=& \int\frac{\dd^3\bk}{(2\pi)^{3/2}}\left[v_k(\eta)\hbox{e}^{i\bk.\bx}\hat a_\bk
 + v_k^*(\eta)\hbox{e}^{-i\bk.\bx}\hat
 a_\bk^\dag\right]\ ,\nonumber\\
 &\equiv&
 \int\frac{\dd^3\bk}{(2\pi)^{3/2}}\left[\hat v_\bk(\eta)\hbox{e}^{i\bk.\bx}
 + \hat v_\bk^\dag(\eta)\hbox{e}^{-i\bk.\bx}\right]\ ,
\end{eqnarray}
where the creation and annihilation operators satisfy the
commutation relations $[\hat a_\bk,\hat
a_{\bk'}^\dag]=\delta^{(3)}(\bk-\bk')$. The mode function,
$v_k(\eta)$, is solution of the classical Klein-Gordon equation
\begin{equation}\label{e:kgfl}
 v_k''+ \omega_v^2(k,\eta) v_k=0
 \qquad\hbox{with}\qquad
 \omega_v^2(k,\eta)=k^2-\frac{2}{\eta^2}\ ,
\end{equation}
which is the equation of motion for a harmonic oscillator with
time-dependent mass, which translates the fact that the field
lives in a time-dependent background spacetime. The general
solution of Eq.~(\ref{e:kgfl}) is
$$
 v_k(\eta) = \left[A(k) H_\nu^{(1)}(-k\eta) + B(k)
 H_\nu^{(2)}(-k\eta)\right]\sqrt{-\eta}
$$
where $H_\nu$ are the Hankel functions. In the
particular case of a de Sitter era considered here, $\nu=3/2$ so that
$$
 H_{3/2}^{(2)}(z) = \left[H_{3/2}^{(1)}(z)\right]^* = -\sqrt{\frac{2}{\pi
 z}}\hbox{e}^{-i z}\left(1+\frac{1}{iz}\right)\ .
$$
Canonical quantization consists in imposing the commutation rules
$[\hat v(\bx,\eta),\hat v(\bx',\eta)] = [\hat
\pi(\bx,\eta),\hat\pi(\bx',\eta)]= 0$ and $[\hat
v(\bx,\eta),\hat\pi(\bx',\eta)]= \delta^{(3)}(\bx-\bx')$ on
constant time hypersurfaces, $\hat\pi$ being the conjugate
momentum of $\hat v$. From Eq.~(\ref{e:modeexp}) and the
commutation rules of the annihilation and creation operators, this
implies that
\begin{equation}
 v_k v_k^{\prime *} - v_k^* v_k' = i\ ,
\end{equation}
which determines the normalisation of the Wronskian. The choice of
a specific mode function $v_k(\eta)$ corresponds to the choice of
a prescription for the physical vacuum $|0\rangle$, defined by
$$
 \hat a_\bk|0\rangle = 0\ .
$$
The most natural choice for the vacuum is to pick up the solution
that corresponds adiabatically to the usual Minkowsky vacuum so
that
$$
 v_k \rightarrow \frac{1}{\sqrt{2k}}\hbox{e}^{-i k\eta}
$$
when $k\eta\rightarrow-\infty$. This implies that the mode
function is
\begin{equation}
 v_k = \frac{1}{\sqrt{2k}}\left(1+\frac{1}{ik\eta} \right)\hbox{e}^{-i
 k\eta}\ .
\end{equation}
This choice is referred to as the Bunch-Davies vacuum.

\subsubsection{WKB approximation}

In more general cases, and for sure in the Bianchi case that
follows, we may not have exact solutions for the mode functions.
On can redo the previous construction by relying on a WKB approach
\cite{JeromeSchwartz} in which one introduces the WKB mode function
\begin{equation}
 v_k^{\rm WKB}(\eta) = \frac{1}{\sqrt{2\omega_v}}\hbox{e}^{\pm i \int\omega_v
 \dd\eta}\ .
\end{equation}
It is easily checked that it is solution of
$$
 {v_k^{\rm WKB}}'' + \left(\omega^2_v - Q_{\rm WKB}\right)v_k^{\rm WKB} = 0
$$
with
\begin{equation}\label{def:Qwkb}
 Q_{\rm WKB} = \frac{3}{4}\left(\frac{\omega_v'}{\omega_v} \right)^2 -
 \frac{1}{2}\frac{\omega_v''}{\omega_v}\ .
\end{equation}
For a function satisfying an equation such as Eq.~(\ref{e:kgfl}),
the WKB solution is thus a good approximation as long as the WKB
condition $|Q_{\rm WKB}/\omega_v^2|\ll1$ is satisfied. In the
example at hand, this condition reduces to $k\eta\rightarrow
-\infty$ so that on sub-Hubble scales, the mode function is
actually close to its WKB approximation. We see that the
quantization procedure thus relies on the fact that there exists
an adiabatic (WKB) solution on sub-Hubble scales.

\subsubsection{Primordial spectra on super-Hubble scales}

Once the initial conditions are fixed, $v_k$ is completely
determined and it can then be related to the scalar field
perturbation $Q$ (also promoted to the status of operator). After
the modes become super-Hubble, i.e. $k\eta\ll1$, the scalar field
perturbation in flat slicing gauge is given by
$$
 \hat Q \rightarrow \int\frac{\dd^3\bk}{(2\pi)^{3/2}}\,
 \hat Q_\bk\,\hbox{e}^{i\bk.\bx}=
 \int\frac{\dd^3\bk}{(2\pi)^{3/2}}\,
 \frac{H}{\sqrt{2k^3}}
 \left( \hat a_\bk+ \hat a_{-\bk}^\dag\right)\,\hbox{e}^{i\bk.\bx}\ ,
$$
where we have used that $S
(\eta)=-1/H\eta$ for a pure de Sitter
inflationary phase, $H$ being a constant in this case. All the
modes are proportional to $( \hat a_\bk+ \hat a_{-\bk}^\dag)$ so
that the variables $\hat Q_\bk$ commute. We thus deduce that $\hat
Q$ has actually the same statistical properties as a Gaussian
classical stochastic field. Effectively, we can replace our
quantum operators by stochastic fields with Gaussian statistics
and we introduce a unit Gaussian random variable, $e_v(\bk)$,
which satisfies
$$
 \langle e_v(\bk) \rangle = 0\ ,
 \qquad
 \langle e_v(\bk)e_v^*(\bk') \rangle = \delta^{(3)}(\bk-\bk')\ .
$$
In this description the mode operators are replaced by stochastic
variables according to $\hat v_\bk\rightarrow v_\bk =
v_k(\eta)e_v(\bk)$ and we identify the (quantum) average in the
vacuum, i.e. $\langle0|...|0\rangle$ by an ensemble (classical)
average, $\langle...\rangle$.

The correlation function of $v$ is defined as
$$
 \xi_v \equiv \langle0|\hat v(\bx,\eta)\hat v(\bx',\eta)|0\rangle\
 ,
$$
and takes the simple form
\begin{equation}
 \xi_v = \int\frac{\dd^3\bk}{(2\pi)^3} |v_k|^2 \hbox{e}^{i
 \bk.(\bx-\bx')}\ .
\end{equation}
Interestingly, in a Friedmann universe, isotropy implies that we
can integrate over the angle to get
\begin{equation}
 \xi_v = \int\frac{\dd k}{k}\, \frac{k^3}{2\pi^2}|v_k|^2\, \frac{\sin kr}{kr}\
 ,
\end{equation}
and it is , because of the symmetries of the background, a
function of $r=|\bx-\bx'|$ only. We thus define the power spectra
\begin{equation}
 P_v(k) = |v_k|^2\
 ,\qquad
 {\mathcal P}_v(k) = \frac{k^3}{2\pi^2}|v_k|^2 \ .
\end{equation}
In the stochastic picture, the correlator of $v_\bk$ is simply
given by
\begin{equation}
 \langle v_\bk v_{\bk'}^* \rangle = P_v(k)\, \delta^{(3)}(\bk-\bk')\ ,
\end{equation}
from which one easily deduces the power spectrum of the curvature
perturbation
$$
 P_{\mathcal R}(k) = \frac{2\pi^2}{k^3} {\mathcal P}_{\mathcal R}(k) = \frac{|v_k|^2}{z^2}\ .
$$

Indeed, we can proceed in the same way for gravity waves. Since
they are not coupled to scalar modes and since the two
polarisations are independent, we introduce two sets of creation
and annihilation of operators, $\hat b_{\bk,\lambda}$, one per
polarisation. On super-Hubble scales, the two modes can be
described by two independent Gaussian classical stochastic fields,
$\mu_{\bk,\lambda}=\mu_\bk e_\lambda(\bk)$ with
$$
 \langle e_\lambda(\bk)e_{\lambda'}^*(\bk')\rangle
 =\delta_{\lambda\lambda'}\delta^{(3)}(\bk-\bk')\ .
$$
The power spectra are thus given by
$$
 P_\lambda(k) = \frac{2\pi^2}{k^3} {\mathcal P}_\lambda(k) = |\mu_k|^2\ .
$$
Spatial isotropy implies that $P_+=P_\times$ so that the tensor
modes power spectrum is
\begin{equation}
 P_T(k)= 2\frac{\kappa}{S^2}P_+(k)\ .
\end{equation}

\subsubsection{Conclusion}

To conclude, this short review of the standard procedure
highlights (1) the importance of the WKB regime on sub-Hubble
scales which allows to construct a Bunch-Davies vacuum
adiabatically, (2)  the fact that on super-Hubble scales (where
one wants to draw the predictions for the initial power spectra)
the quantum operators can be conveniently replaced by stochastic
fields, (3) the importance of isotropy which implies that there
exist 3 independent stochastic directions (because modes are
decoupled) and (4) the fact that the two gravity wave
polarisations have the same power spectrum.

We shall now see which of these properties generalise to a Bianchi
universe.

\subsection{Generic Bianchi~$I$ universes}\label{sec2.2.b}

\subsubsection{Characteristic wavenumber}

In order to relate our predictions to observations, we introduce
the characteristic wavenumber $\kref$ by
\begin{equation}\label{Eq_defkref}
 \kref \equiv \left.SH\right|_{t=\tau_*} \ .
\end{equation}
We define ${\mathcal N}$ as the number of $e$-fold  with the quantity $S H$
rather than $S$ in the definition~(\ref{Def_efolds}). If we denote by
${\mathcal N}_{\text{ref}}$ the number of $e$-folds between
$t=\tau_*$ and the end of inflation, and ${\mathcal N}_0$ the number of
$e$-folds from the end of inflation until now, then we can relate
$\kref$ to the largest observable scale today, $k_0=S_0H_0$, by
\begin{equation}
\frac{\kref}{k_0} = e^{({\mathcal N}_0 - {\mathcal N}_{\text{ref}})} \,.
\end{equation}
Since during the inflationary era after $\tau_*$, $H$ is nearly
constant, then ${\mathcal N}_{\text{ref}}\simeq N_{\text{ref}}$. As for
${\mathcal N}_0$, it depends on the post-inflationary evolution and it can be
estimated~\cite{llN} by
\begin{eqnarray}
{\mathcal N}_0 \simeq 62 -\ln\left(\frac{10^{16}\,{\rm
GeV}}{V_{k_0}^{1/4}}\right)+\frac{1}{4}\ln\frac{V_{k_0}}{V_{\rm
end}}-\frac{1}{3}\ln\left(\frac{V_{\rm end}^{1/4}}{\rho_{\rm
reh}^{1/4}}\right)-\ln h,
\end{eqnarray}
$h$ being the Hubble parameter in units of 100~km.s$^{-1}$/Mpc.
Typically,  $V_{k_0}\sim V_{\rm end}$ as long as slow rolling holds.
The reheating temperature can be argued to be larger than
$\rho_{\rm reh}^{1/4}>10^{10}$~GeV to avoid the gravitino
problem~\cite{sarkar}. The amplitude of the cosmological
fluctuations (typically of order $2\times10^{-5}$ on Hubble
scales) roughly implies that $V_{\rm end}^{1/4}$ is smaller than a
few times $10^{16}$~GeV and, for the same reason as above, has to
be larger than $10^{10}$~GeV in the extreme case. This implies
that $N_0$ has approximately to lie between 50 and 70, which is
the order of magnitude also required to solve the horizon and
flatness problem.

$N_{\rm ref}$ can be computed from the background dynamics, and as
shown on Fig.~\ref{fig:efold}, it is almost equivalent to its
value in the FL case.

\subsubsection{Anisotropy}

The first obvious difference with the FL case arises from the
local spatial anisotropy.\\

First, it is clear from the set of
equations~(\ref{b1:v}-\ref{b1:mu}) that it implies that the
evolution of the mode functions shall depend on $k_i$ and not
simply on the modulus. This violation of isotropy will reflect
itself on the fact that
\begin{enumerate}
 \item the power spectra at the end of inflation will be functions of
$\bf k$ and not $k$, i.e. $P_v(k_i)$, $P_\lambda(k_i)$;
 \item because of the coupling between scalar and
gravity waves, there exists a cross-correlation between scalar and
tensor, i.e. $\langle v\mu_\lambda\rangle\not=0$;
 \item the two polarisations shall a priori
have two different power spectra, i.e. $P_+\not=P_\times$.
\end{enumerate}

\begin{figure}[htb]
 \includegraphics[width=9cm]{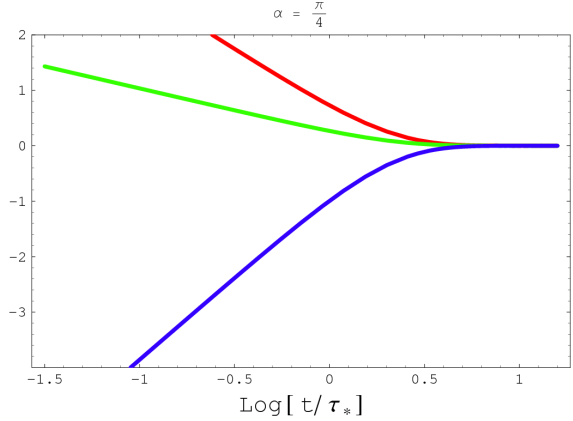}
 \caption{Logarithm of the ratio $k/\kref$ where $\kref$ is the modulus of the
 wavenumber when the shear is zero. We see that as soon as the shear grows, $k$ depends
 on the direction on which it is aligned (Each color corresponds to a
 principal axis of the Bianchi universe). We have considered a generic Bianchi universe with
 $\alpha=\pi/4$.}\label{fig2}
\end{figure}

A second related issue arises from the evolution of a comoving
wavenumber. Let us consider the different evolutions of a
wave-mode of modulus $k$ at the end of inflation according to its
orientation. As we see on Fig.~\ref{fig2}, depending on its
orientation, this mode has very different time evolutions before it
settles to a constant value.

\subsubsection{WKB regime}

Whatever the Bianchi universe we consider, there is a
shear-dominated phase prior to inflation. During this phase $\ddot
S<0$. Besides, in a generic Bianchi~$I$ spacetime (that is
$\alpha\not=\pi/2$) two of the scale factors go to zero while the
third is bouncing (in the $\alpha=\pi/2$ case, one scale factor
goes to zero while the two others remain constant).

For analysing the WKB regime we will consider, as usual, the ratio $k/SH$ to discuss whether a 
given mode is inside $(k/SH\gg 1)$ or outside $(k/SH\ll 1)$ the Hubble radius. According to the 
details shown in the Appendix~A, if we specify to only one direction we have $k\sim 1/a_i\sim S/X_i$, then 
$$ 
\frac{k}{SH}=\frac{1}{X_{i}H}\sim t^{2(1-\sin\alpha_{i})/3}
$$
during the shear-dominated regime. This shows that, except when $\alpha=\pi/2$, {\it any} given 
mode becomes super-Hubble when we approach the singularity $(t\rightarrow 0)$, and 
that this approach is faster (going backwards in time) for modes aligned with the 
bouncing direction.

\begin{figure}[htb]
 \includegraphics[width=8.1cm]{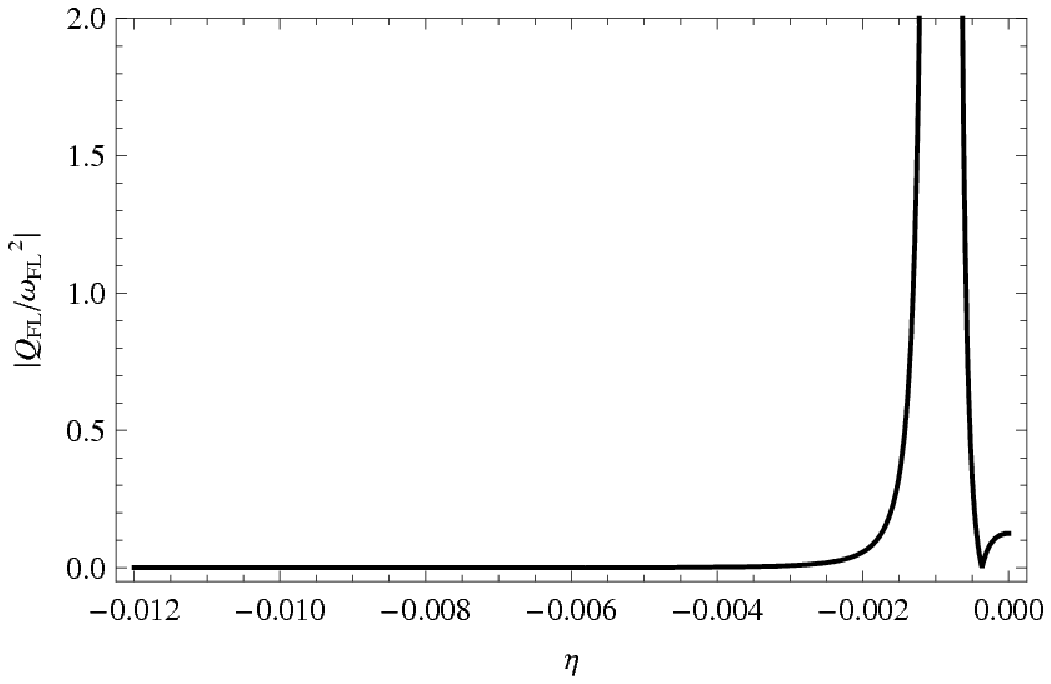}
 \includegraphics[width=8.1cm]{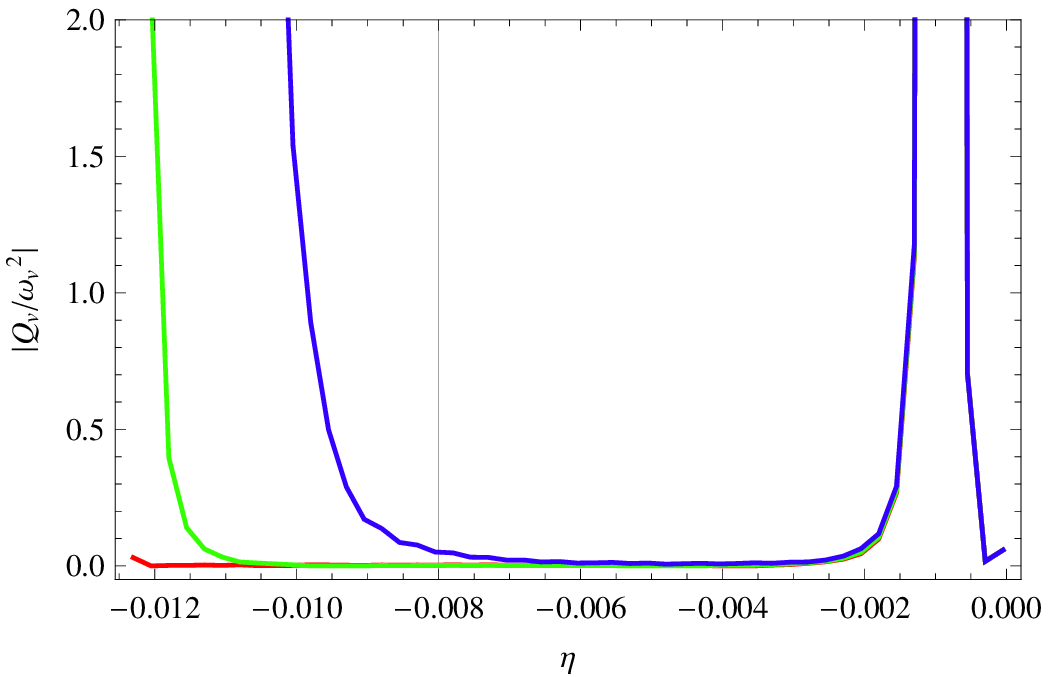}
 \caption{Left panel: Evolution of the quantity $|Q/\omega^2|$ for a FL universe.
Right panel: Evolution of $|Q_v^{\rm WKB}/\omega_v^2|$ for three
different modes, each of them aligned with one of the three
orthogonal directions (same color code as in Fig.~\ref{fig2}), and
with the same modulus $10k_{\rm ref}$ at the end of inflation. The vertical line correspond to 
the instant $\eta(\tau_*)$ and we have considered a generic Bianchi spacetime with
$\alpha=\pi/4$.}\label{fig9}
\end{figure}

In other words, {\it all} modes become super-Hubble in the past, and the mode aligned with the bouncing 
direction (blue line in Fig.~\ref{fig2}) becomes super-Hubble earlier (again, going backwards in time) 
than the ones with same $k$ at the end of inflation but aligned with a growing direction 
(blue line in Fig.~\ref{fig9}).

For these two reasons, we can doubt the existence of a
well-defined adiabatic vacuum for all modes through their early evolution, as happens in FL universes. 
However, we can still ask whether the WKB regime is reached in a short time interval when the shear is 
not complete negligible, and how long it lasts given the wavenumber.\\

Let us thus discuss quantitatively the validity of the
WKB approximation. First, we focus on the pulsation, we neglect
the effect of the couplings, and consider the WKB solutions
\begin{equation}\label{e:bwkb}
 v_\bk^{\rm WKB}(\eta) = \frac{1}{\sqrt{2\omega_v}}\hbox{e}^{\pm i \int\omega_v
 \dd\eta}\ ,
 \qquad
 \mu_{\bk,\lambda}^{\rm WKB}(\eta) = \frac{1}{\sqrt{2\omega_\lambda}}\hbox{e}^{\pm i
 \int\omega_\lambda\dd\eta}\ .
\end{equation}
They are good approximations of the solutions of
Eqs.~(\ref{b1:v}-\ref{b1:mu}) if $|Q_{v,\lambda}^{\rm WKB} /
\omega_{v,\lambda}^2|\ll 1$ where $Q_{\rm WKB}$ has been defined
in Eq.~(\ref{def:Qwkb}).

Fig.~\ref{fig9} illustrates the validity of the WKB approximation
for the scalar modes. It depicts the evolution of $|Q_v^{\rm
WKB}/\omega_v^2|$ as a function of time for three different modes
corresponding to the three principal axis of the Bianchi universe.
We see that, for a given comoving wavenumber $k$ at the end of
inflation, the WKB condition is always violated in the past and
that it is violated first in increasing order of the Kasner
coefficients (compare with Fig.~\ref{fig2}). It can be checked
that the larger the $k$ is, the longer the time during which the WKB
condition is restored. For long wavelength modes (typically of
order $1/\kref$), the WKB regime is never established.

\begin{figure}[htb]

 \includegraphics[width=8.1cm]{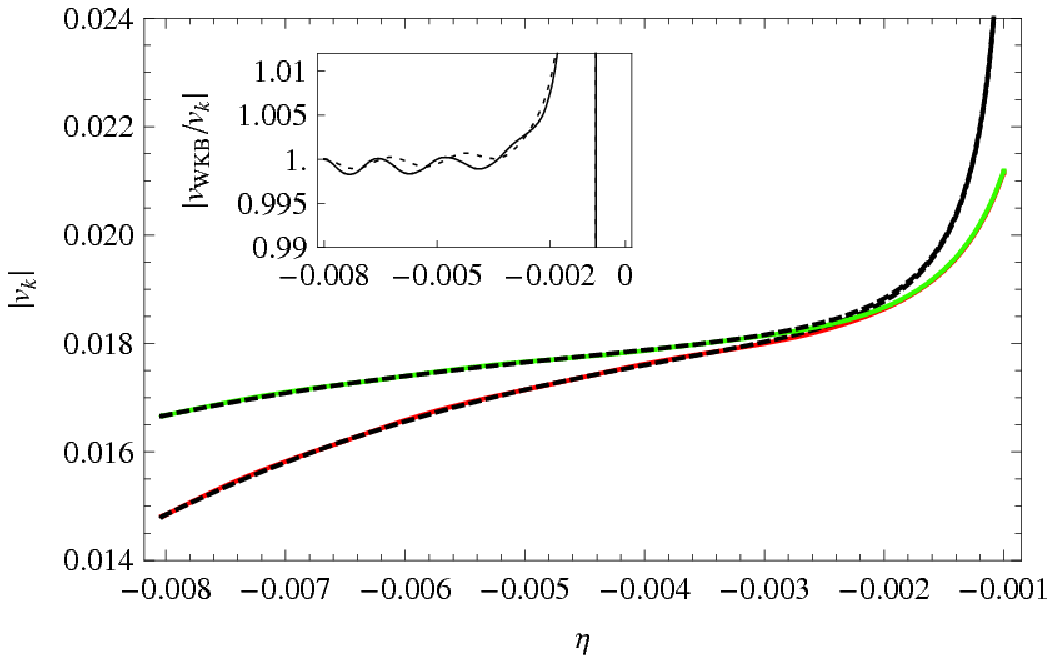}
 \includegraphics[width=8.1cm]{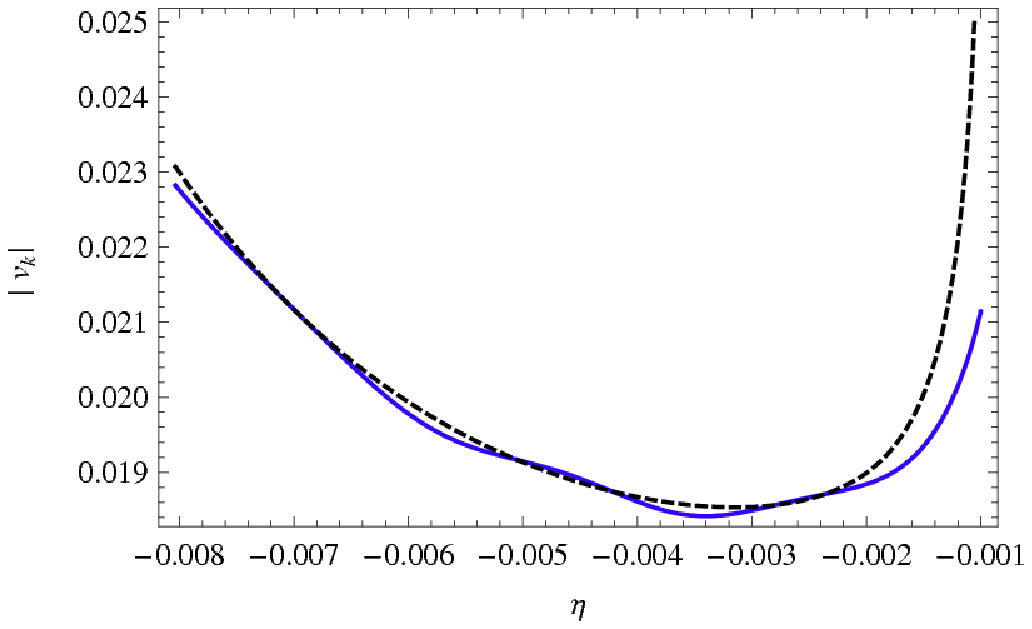}
 \caption{Comparison of the exact (solid lines) solutions and the WKB (dashed lines)
 solutions~(\ref{e:bwkb}) for the scalar modes for three different directions
 with $k=10k_{\rm ref}$ at the end of inflation. We have considered a
 generic Bianchi spacetime with $\alpha=\pi/4$. The figures show two genuine WKB modes
(left panel) and one non-WKB mode (right panel). The inner-left
figure shows the ratio $|v_\bk^{\rm WKB}/v_\bk|$ for the modes
which satisfy the WKB approximation.}\label{fig9a}
\end{figure}

\begin{figure}[htb]
 \includegraphics[width=8.1cm]{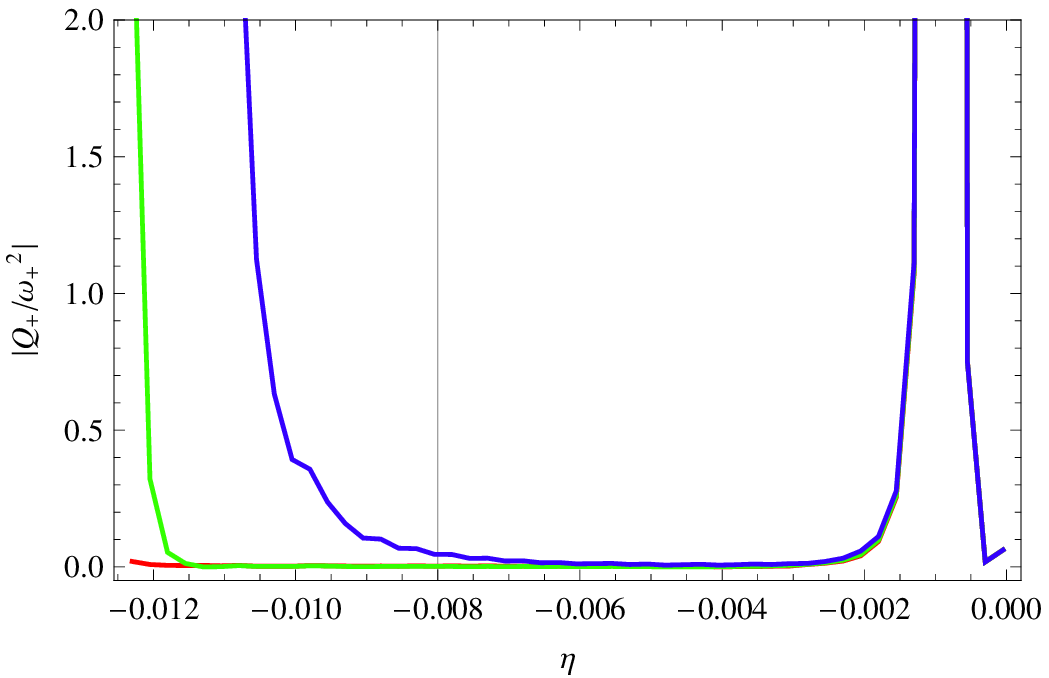}
 \includegraphics[width=8.1cm]{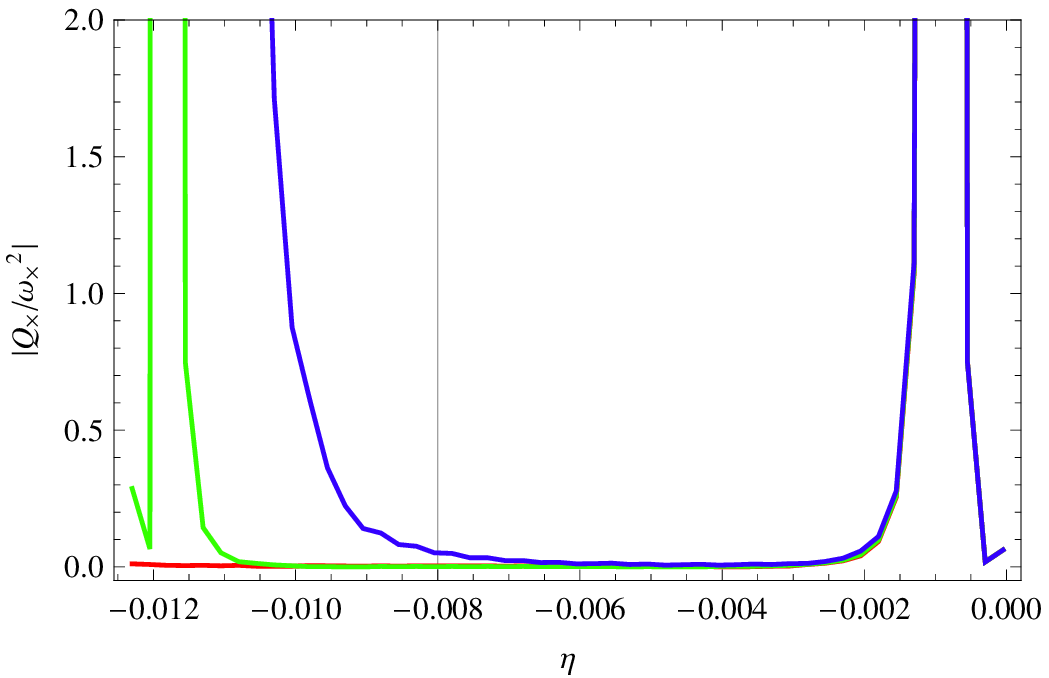}
 \caption{Evolution of $|Q_\lambda^{\rm WKB}/\omega_\lambda^2|$ for the two tensor polarisations
 (left: $\lambda=+$, right: $\lambda=\times$) and for various modes with the same modulus $k=10k_{\rm ref}$
at the end of inflation. The vertical lines represent the time $\eta(\tau_*)$. We have 
considered a generic Bianchi spacetime with $\alpha=\pi/4$.}\label{fig10}
\end{figure}

Fig.~\ref{fig9a} compares the exact (numerical) solution and the
WKB approximation for a mode $k=10\kref$. Fig.~\ref{fig10} is similar to 
Fig.~\ref{fig9} but for the tensor modes. Indeed, we reach the same conclusions.

\subsubsection{Couplings}

The third difference arises from the coupling between scalar and
tensor modes.

As we demonstrated in our previous analysis~\cite{ppu1} (see
\S~IV.D), deep in the sub-Hubble regime (that is when $k/SH$ is
large enough), the three degrees of freedom decouple and behave as
a collection of three independent harmonic oscillators.

However, on larger scales the couplings are a priori
non-negligible. We thus need to evaluate with care the scales for
which this is a good approximation.

\begin{figure}[htb]

 \includegraphics[width=5.4cm]{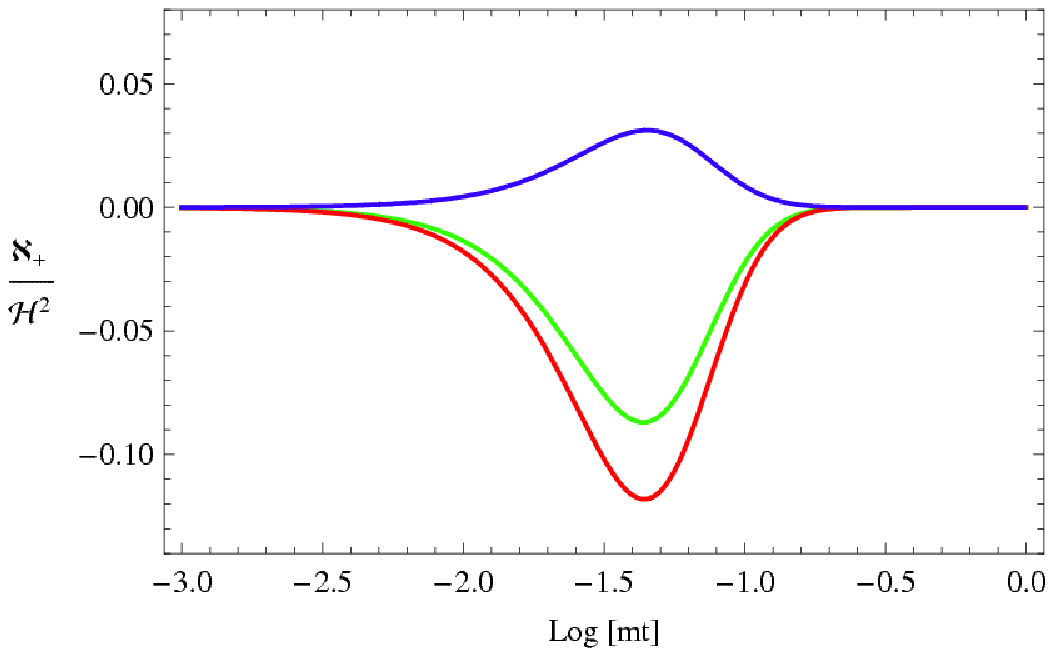}
 \includegraphics[width=5.4cm]{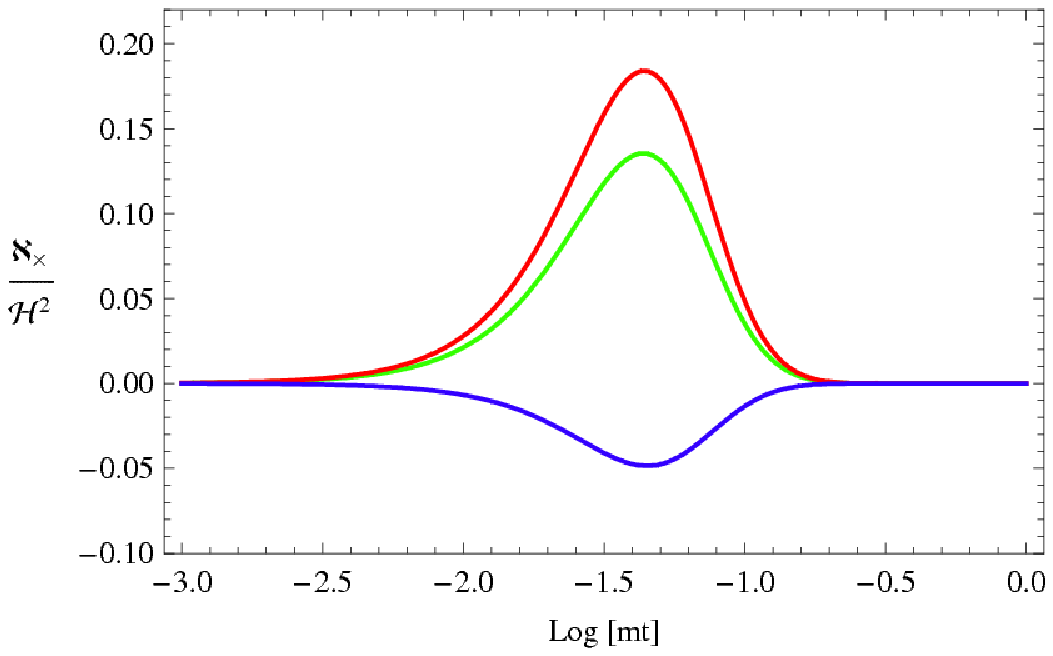}
 \includegraphics[width=5.4cm]{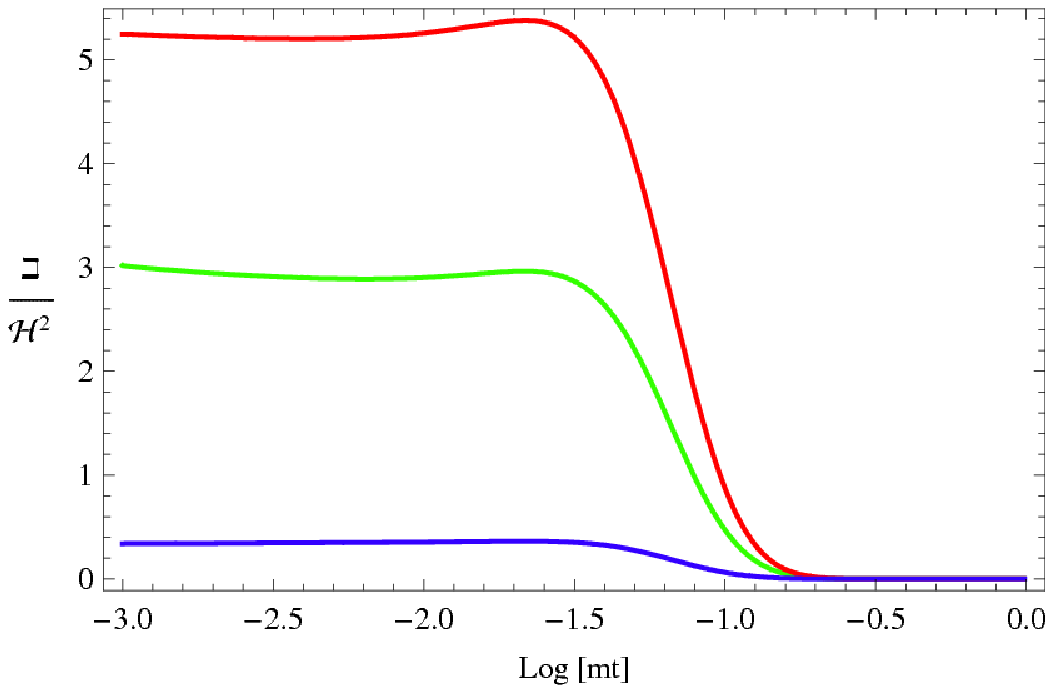}
 \caption{Evolution of $\aleph_\lambda/\HH^2$ (left: $\lambda=+$,
 middle: $\lambda=\times$) and $\beth/\HH^2$
 for three modes each of which is aligned with one of three orthogonal arbitrary directions
 (represented by three different colors). We have considered a generic Bianchi
 spacetime with $\alpha=\pi/4$. Note that these functions depend only on the direction
 of the wavenumber and not on its modulus.}
 \label{fig11}
\end{figure}

Fig.~\ref{fig11} shows that, while the functions $\aleph_\lambda$,
which couple gravity waves and scalar modes can be neglected on
small scales, this is certainly not the case for the coupling
$\beth$ between the two gravity wave polarizations. This coupling
cannot be neglected, even at early time, for modes which are not
sub-Hubble enough. Typically, the modes for which this coupling
cannot be neglected correspond to modes for which the WKB regime
cannot be reached.

Let us compare this situation  with the case of multi-field inflation. 
The equations of evolution for the various scalar field perturbations 
are also usually coupled (see e.g. Ref.~\cite{wands1} for a recent review).
However generally, it is possible to extract independent fields,
at least in the sub-Hubble regime~\cite{bw} so that one can
introduce a set of independent stochastic fields. To our
knowledge, the situation where this is not possible has not been
addressed.

The situation is analogous for us and the long wavelength modes
will be particularly difficult to trustfully deal with because
they can never be considered as independent.

\subsection{Prescription for setting the initial
conditions}\label{sec2.2.c}

\subsubsection{Prescription for Bianchi~$I$ spacetimes}

It is part of the nature of quantum fluctuations that they do not
have initial conditions in the sense that they are continuously
excited. As soon as a mode can oscillate, it will be sourced by
these quantum fluctuations. Given the discussion of the preceding
section, we will thus set the initial conditions at the time where
a mode is the deepest in the WKB regime. This time, $t_{i}(\bk)$
say, depends explicitly on the mode, which is not a problem since
all the modes are independent from each other. In practice, we have fixed
$t_{i}(\bk)$ by minimizing $\omega_v$.

\begin{figure}[htb]
 \includegraphics[width=8cm]{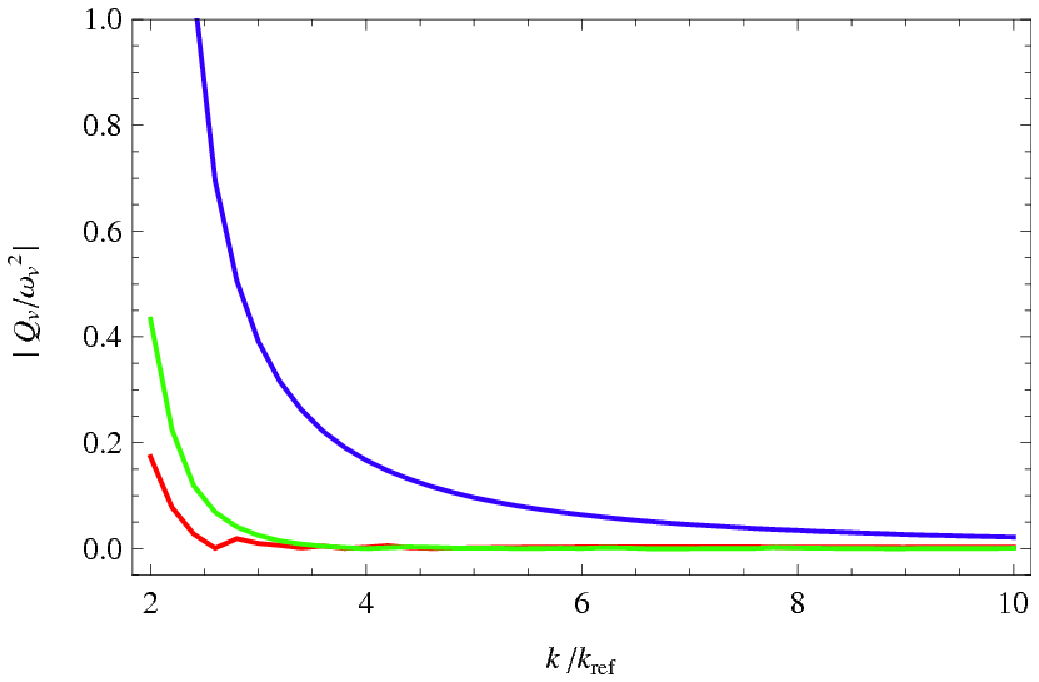}
 \includegraphics[width=8cm]{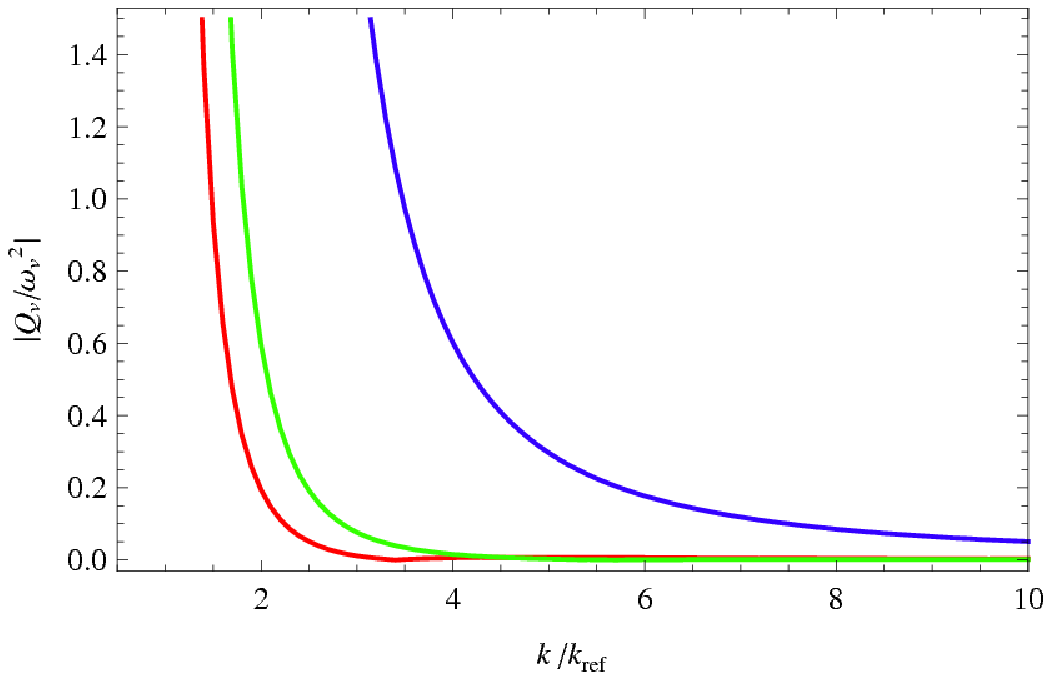}
 \caption{Validity of the WKB approximation at the time we set the
 initial conditions. (left) We set the initial conditions at $t_i(\bk)$
 and (right) at $\tau_*$.
 The quantity $|Q_v^{\rm WKB}/\omega_v^2|$
 is shown for three orthogonal modes and we have considered a
 generic Bianchi spacetime with $\alpha=\pi/4$.}\label{fig12a}
\end{figure}

\begin{figure}[htb]
  \includegraphics[width=6cm]{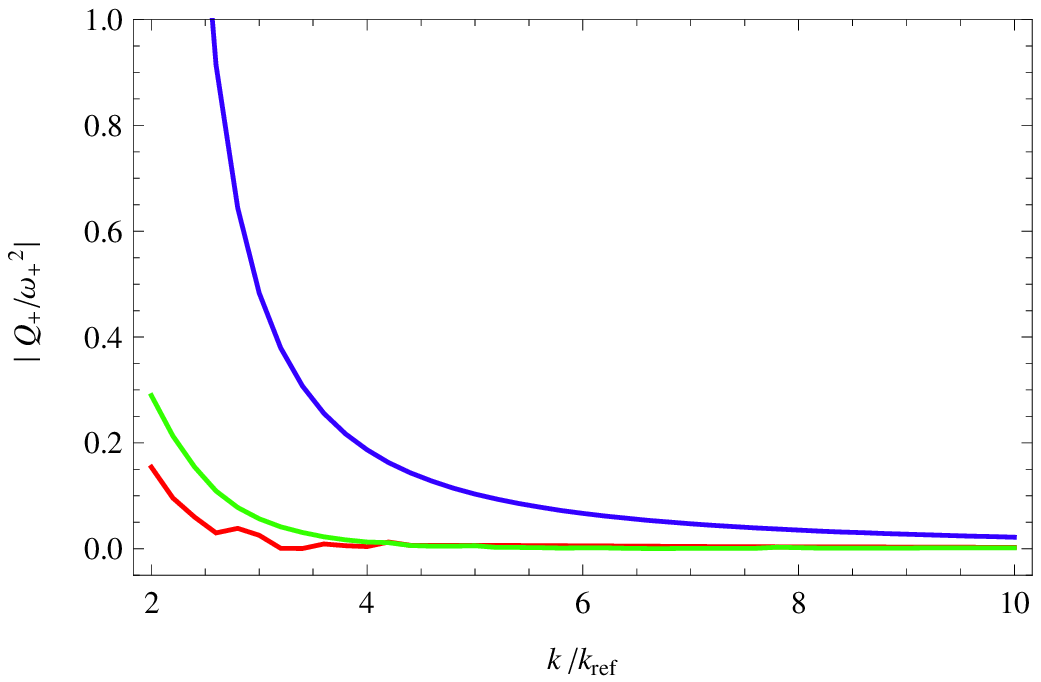}
  \includegraphics[width=6cm]{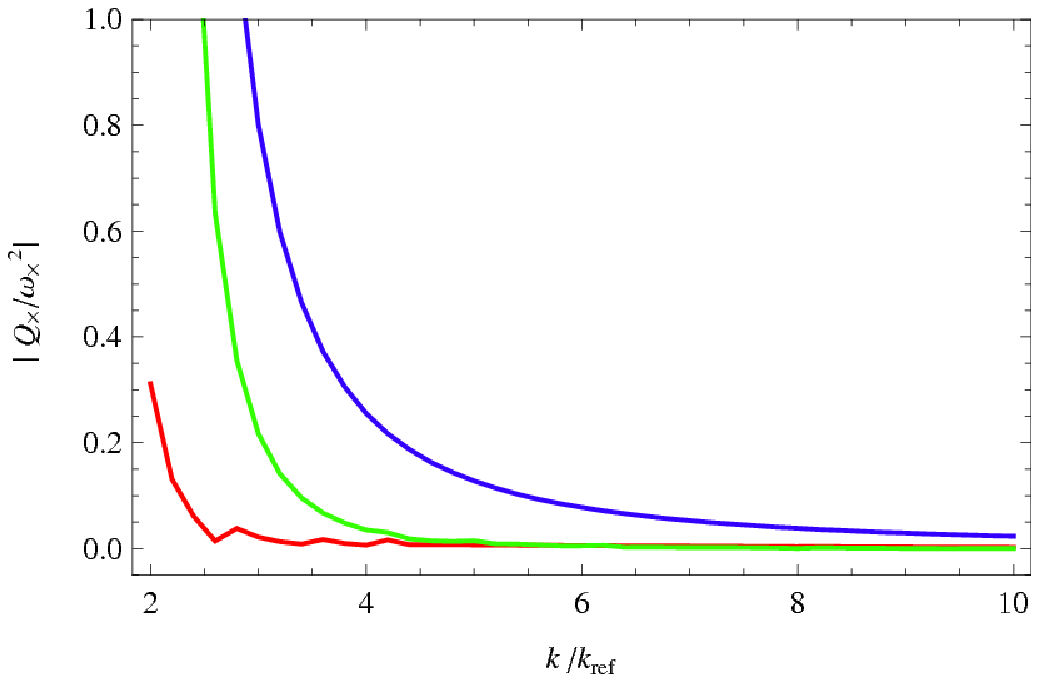}

  \includegraphics[width=6cm]{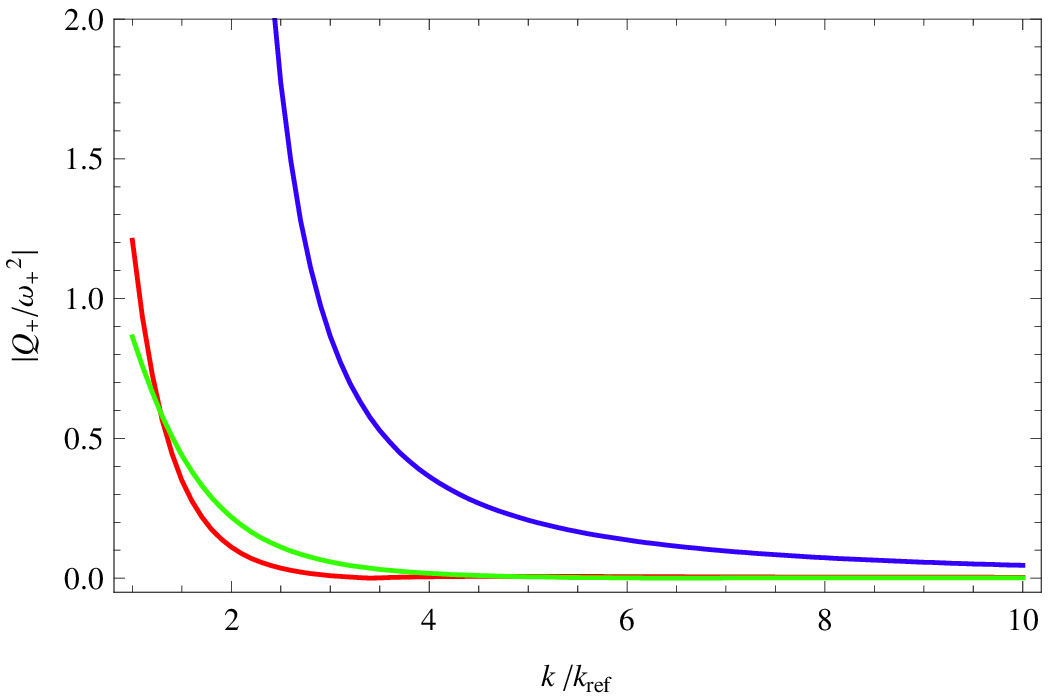}
  \includegraphics[width=6cm]{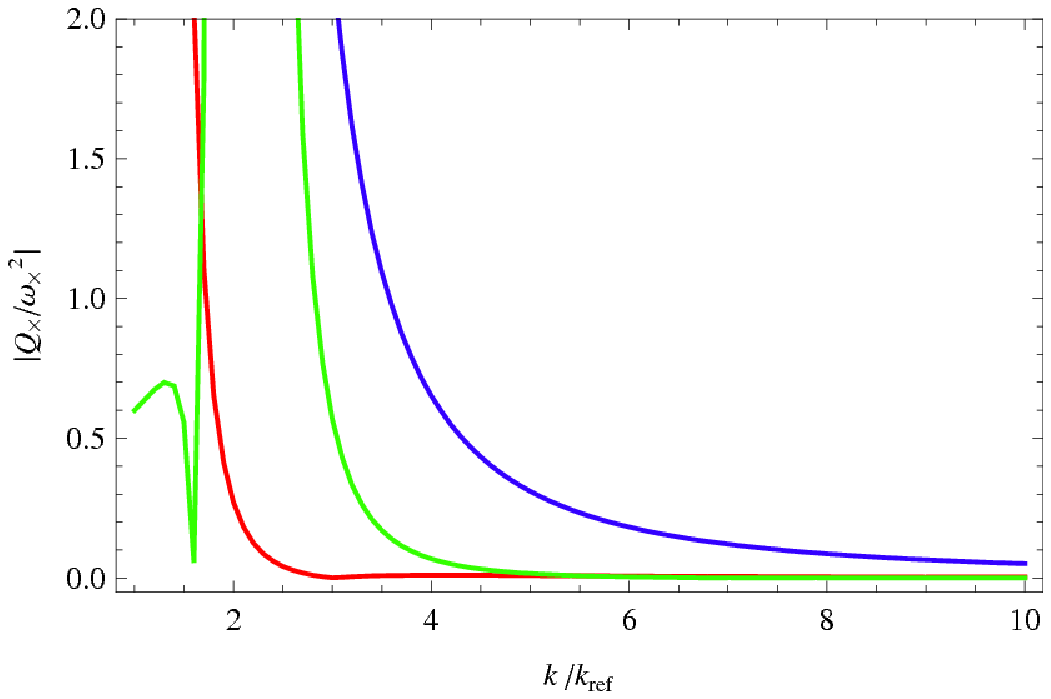}
 \caption{Validity of the WKB approximation for tensor modes. We show the quantity
 $|Q_\lambda^{\rm WKB}/\omega_\lambda^2|$ for $\lambda=+$ (left panel) and $\lambda=\times$ (right panel)
 for three orthogonal modes. We compare
 setting the initial conditions at $t_i(\bk)$ (top)
 and at $\tau_*$ (bottom).
 We have considered a generic Bianchi spacetime with $\alpha=\pi/4$.
}\label{fig12b}
\end{figure}

As we saw, modes with $k<\kref$ never enter a WKB regime. Given
the fact that modes up to approximatively $k_0$ have been excited
and have, at least as a good approximation, a scale invariant
power spectrum, we have to assume that $\kref\lesssim k_0$. Then,
from Figs.~\ref{fig12a} and~\ref{fig12b}, we deduce that the WKB
is reached for all modes with $k>\kref$. Indeed this amounts to a
fine tuning on the shear such that only the largest observable
modes today were affected by the Bianchi phase.

To check the robustness of this procedure, we have varied the time
$t_{i}(\bk)$. In particular, we have also assumed, as a test, that
$t_{i}(\bk)=\tau_*$ for all modes. It can be shown that this does
not affect the predictions for the modes with $k\gtrsim2\kref$
while long wavelength modes are more affected. This is simply due
to the fact that the duration of their WKB phase is smaller (see Fig.~\ref{fspec:iso}). Also
note that the procedure is more robust for the two directions
which are not bouncing.

\begin{figure}[htb]
\includegraphics[width=5.4cm]{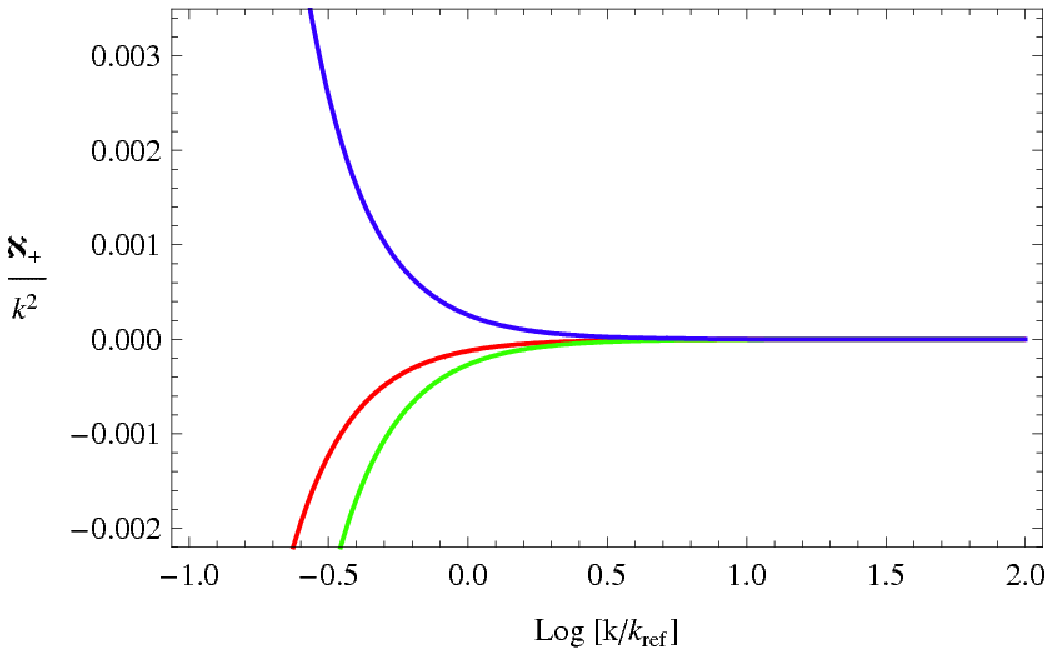}
\includegraphics[width=5.4cm]{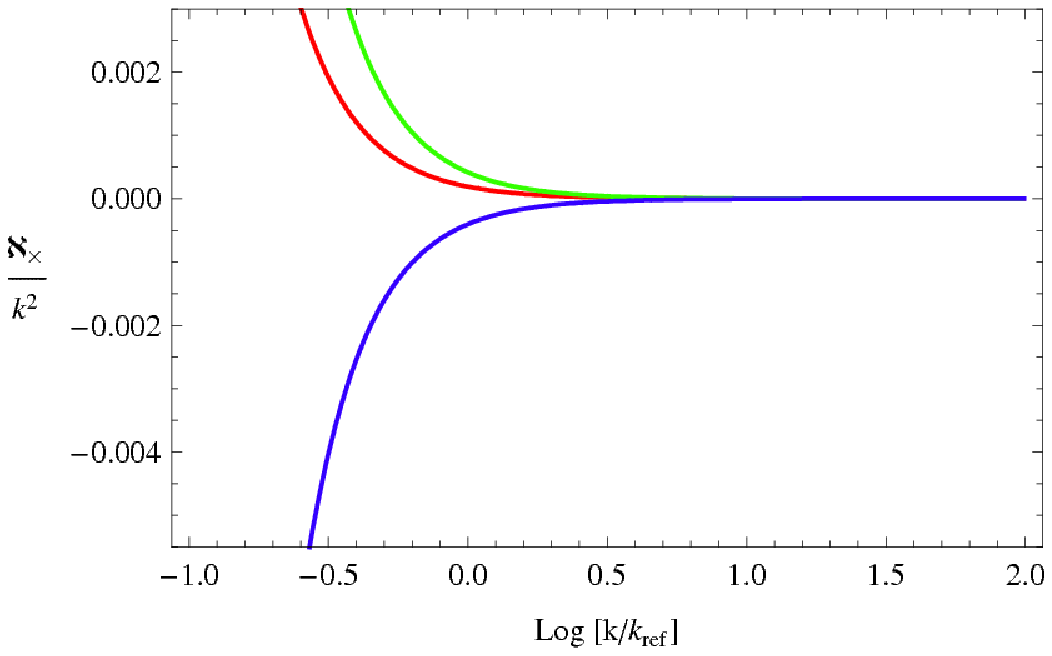}
\includegraphics[width=5.4cm]{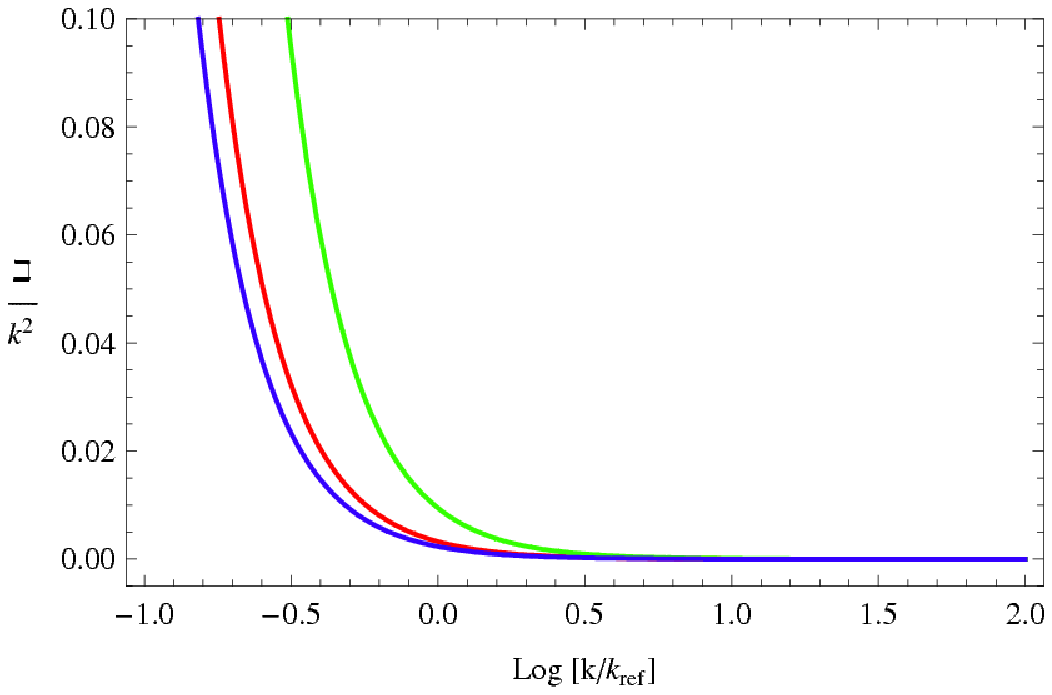}
 \caption{Statistical independence at $\tau_*$. $\aleph/k^2$
 (left) and $\beth/k^2$ (right) as a function of $k$ at $t=\tau_*$.
}\label{fig12c}
\end{figure}

For these modes, and as can be seen from Fig.~\ref{fig12c}, it is
also a good approximation to neglect the coupling terms in
Eqs.~(\ref{b1:v}-\ref{b1:mu}). We emphasize that this hypothesis
breaks down approximatively at the same time when the WKB
approximation also breaks down.

Therefore we will assume that the three modes are
independent at the time when they are excited by the quantum
fluctuations.

Technically, we thus start our computation by setting
\begin{equation}
 v_\bk = \frac{\hbox{e}^{- i \int\omega_v
 \dd\eta}}{\sqrt{2\omega_v(\bk,\tau)}}\, e_v(\bk)\ ,
 \qquad {\rm and}\qquad
 \mu_{\lambda,\bk} = \frac{\hbox{e}^{- i
 \int\omega_\lambda\dd\eta}}{\sqrt{2\omega_\lambda(\bk,\tau)}}\, e_\lambda(\bk)\ ,
\end{equation}
up to an arbitrary relative phase which can be absorbed in the
definition of the unit random variables and where the three random
variables satisfy
$$
 \langle e_X(\bk)e_Y^*(\bk')\rangle
 =\delta_{XY}\delta^{(3)}(\bk-\bk')\ .
$$

\subsubsection{Freedom in a time redefinition}\label{sec2.2.d}


As mentioned in Ref.~\cite{GCP}, the dynamical system of equation for the Mukhanov-Sasaki 
variables admits time reparameterisations that conserve the canonicity of the system, in 
the sense that if $v(\eta)$ is a canonical variable that satisfies
\[
\frac{{\rm d}^{2}v}{{\rm d}\eta^{2}}+\omega_{v}^{2}v=0\quad\textrm{and}\quad 
 v\frac{{\rm d}v^{*}}{{\rm d}\eta}-v^{*}\frac{{\rm d}v}{{\rm d}\eta}=i\,,
\]
then there is a function $f$ and a time $\tau$ defined as
\begin{equation}\label{funcf}
f(\eta)^{2}{\rm d}\tau={\rm d}\eta
\end{equation}
through which we can define a new variable $u=fv$ that satisfies
\[
\frac{{\rm d}^{2}u}{{\rm d}\tau^{2}}+\omega_{u}^{2}u=0\quad\textrm{and}\quad
 u\frac{{\rm d}u^{*}}{{\rm d}\tau}-u^{*}\frac{{\rm d}u}{{\rm d}\tau}=i\,.
\]
If, for example, $v$ is a variable for which the WKB condition does not hold, we
might wonder whether exists a function $f$ that would lead to a different conclusion 
concerning the validity of the WKB condition for the variable $u$.

In the present case, the transformation Eq.~(\ref{funcf}) would lead to the same equations 
of motion satisfied by the new canonical variables
\begin{equation}
\tilde v\equiv f v,\quad\tilde \mu_+\equiv f \mu_+,\quad\tilde
\mu_\times\equiv f \mu_\times\,, 
\end{equation}
where the new pulsations are defined according to
\begin{equation}
\tilde \omega_v^2 \equiv
\frac{\omega_v^2}{f^4}-\frac{1}{f}\frac{\dd^2 f}{\dd \tau^2}\,
\qquad \tilde \omega_\lambda^2 \equiv
\frac{\omega_\lambda^2}{f^4}-\frac{1}{f}\frac{\dd^2 f}{\dd
\tau^2}\ ,
\end{equation}
and the coupling functions become
\begin{equation}
\tilde \aleph_\lambda\equiv \frac{\aleph_\lambda}{f^4}\ ,
\qquad
\tilde \beth \equiv \frac{\beth}{f^4}\ .
\end{equation}

For the sake of simplicity, let us drop $\aleph_\lambda$ and
$\beth$ in the following discussion. It has been shown in
Ref.~\cite{GCP} that, if $\omega_v$ and $\omega_\lambda$ satisfy
the WKB condition, then $f$ is required to satisfy the condition
\begin{equation}\label{Eq_deWKBaWKB}
 \frac{1}{f}\left|\frac{\dd^n f}{\dd \eta^n}\right| \ll
 \omega_v^n,\omega_\lambda^n\ , \qquad \text{with}\quad n=1\ldots4
\end{equation}
in order for the new equations to also satisfy the WKB condition.
Under such conditions, it would lead to the same quantization
procedure. Note that the condition for $n=4$ is required when we
take $|Q^{\rm WKB}/\omega^2| \ll 1$ for the (correct) WKB
condition rather than just $\omega'/\omega^2 \ll 1$, as assumed in
Ref.~\cite{GCP}.\\


In our case the pulsations $\omega_v$ and $\omega_\lambda$ scale like 
$S H$ when $t \rightarrow 0$, or like $1/\eta$ in conformal time. Let us choose
the integration constant $W_0$ of Eq.~(\ref{Eq_Wdet}) such that
the initial singularity corresponds to $\eta=0$. Now, if
$\omega\simeq C/\eta$ then the associated function $Q_{\rm WKB}$
behaves as
$$
\frac{Q_{\rm WKB}}{\omega^2} \simeq -\frac{1}{4C^2}\ .
$$
Let us consider now a time redefinition associated to a function
$f(\eta)=\eta^A$. Then the WKB condition involves
\begin{equation}
 \frac{\tilde Q_{\rm WKB}}{\tilde \omega^2}
  \simeq -\frac{\left(A+\frac{1}{2}\right)^2}{C^2+A(A+1)}\ \ .
\end{equation}
We conclude that it is possible for the WKB condition to be
fulfilled with a new time coordinate by choosing $A=-1/2$,
provided $|C| \neq 1/2$.

Unfortunately, using the expansion~(\ref{Eq_zsprprsurzs}) and the
asymptotic behaviours obtained in
Eqs.~(\ref{Eq_slowroll_asymptotique}), we deduce that when the
leading term of $\omega_v^2$ is $z_s''/z_s$, then $\omega_v^2
\rightarrow -\HH^2$, which implies that for this pulsation,
$|C|=1/2$.

Thus, it is never possible to construct a time redefinition which
would enable the WKB condition to be satisfied for $\tilde
\omega_v$. The only exception arises when the leading term of
$\omega_v^2$ is $k^2$. This happens for $\alpha=\pi/2$ since in
this particular case $k\sim t^{-2/3}\sim \eta^{-1}$ and we recover
the conclusions reached in Ref.~\cite{GCP}.

As a conclusion, though we might naively think that redefining
time could lead to equations satisfying the WKB conditions for the
new canonical variables, it is impossible however to build such a
change of time coordinates. The choice of the canonical variables
is thus unique up to the reparameterisation satisfying the
conditions~(\ref{Eq_deWKBaWKB}), which would not change our
predictions.

\subsection{Discussion}\label{sec2.2.e}

Let us discuss our procedure to set the initial conditions.\\

First for modes smaller than $\kref$, the WKB regime was never
reached. We have no {\it natural} prescription to determine their
amplitude. A solution, that we do not investigate in this article,
would be to fix them by assuming that they minimize their energy,
as proposed in Ref.~\cite{lubo} in the study of some
trans-Planckian models in which the WKB regime is violated.

On the other hand, and probably in a more conservative way, one
could just assume their initial value to be completely random and
introduce a free function to describe the initial conditions on
large scales. Such a function would then need to be measured from
e.g. large angular scale properties of the CMB or predicted by
some processes that arise at the Planck or string scale, and that
indeed cannot be accounted for in our description.

In the former case, we loose the predictive power on large scales,
that actually may just be beyond the actual size of the observable
universe. It may seem that we are back to the (historical)
pre-inflationary times, where the form of the initial power
spectrum of the Harrison-Zel'dovich type had to be postulated in
order to reproduce the observations of the large-scale structures.
This is somehow a very standard approach in physics in which one
learns about the initial conditions of a system by observing its
evolution

Inflationary theories allowed to actually predict this spectrum,
which makes them very predictive. We realize with this study that
these inflationary predictions are very sensitive to the existence
of an (classical) initial shear. To recover such a predictive
power, we have to hope that a theory handling the dynamics of the
universe on this scales~\cite{billiard,kaloper,lcq} or allowing to
generate the shear~\cite{ncg} will also provide a better
understanding of the initial conditions.\\

Indeed, one may wonder whether the arbitrary (pre-WKB era)
conditions can be amplified and compete in amplitude with the ones
of quantum origin seeded during the WKB regime.

To estimate this, note that at early time (apart from the
particular case $\alpha=\pi/2$), $\omega_v^2 $ behaves as
$-z_s''/z_s$ and $\omega_\lambda^2 $ behaves as
$-z_\lambda''/z_\lambda$. 
The expansion~(\ref{Eq_zsprprsurzs}) then leads to the conclusion that, at
early times,
$$
 \omega_v^2 \simeq +\HH^2\,,
$$
whereas at late time, we deduce from our previous analysis that
$$
 \omega_v^2 \simeq -2 \HH^2\ .
$$
Thus, the solutions of Eq.~(\ref{b1:v}) has an oscillatory behaviour until the time when $\omega_v=0$ and the arbitrary initial
conditions are not amplified. For the tensor modes the situation is different because for some configurations of ${\bf k}$,
we can have $\omega_\lambda^2 \sim - \HH^2 $ at early time, and this leads to an
exponential growth. However, since $|\omega_\lambda| \lesssim
C/\eta$ with $C={\cal O}(1)$, this exponential growth is typically at most of order
$$
\exp\left(\frac{{\cal O}(1)}{\eta}\times \eta\right) \sim e\ .
$$
It follows that the arbitrary initial conditions either are not amplified or
do actually grow, but in the latter case they are amplified by no more than a factor of order unity. We thus
expect the transitional tachyonic behaviour and our ignorance of
the initial state of the perturbations before the WKB regime not
to significantly alter the validity of our prescription for the
initial conditions of quantum origin set during the WKB
regime.\\

From a practical point of view, we can estimate the effect of the
coupling functions $\aleph$ and $\beth$ by changing their
amplitude by hand. We have checked that they hardly affect the
predictions that are presented in the following section. In
particular, that teaches us that the main directional dependence
of the power spectra is induced mainly by the directional
dependence of the comoving wavenumbers (see Fig.~\ref{fig2}). We
have also checked that when $k/\kref$ increases our prediction is
similar to the one obtained in
standard inflation.\\

In conclusion, we can trust our prescription for modes larger than
$\kref$ and we have to assume (somehow as an observational input)
that $\kref<k_0$. This sets a tuning on the primordial shear that
we cannot explain with the model at hand. In this regime, we have
checked that it is a robust prescription that is not affected by
the unknown preexisting perturbations that are completely
arbitrary and that fixes the properties of the long wavelength
modes.

\section{Primordial spectra: numerical examples and predictions}\label{sec3}

\subsection{Definition of the spectra}\label{sec3.1}

To set the initial conditions as previously detailed, we need to first 
solve numerically the system~(\ref{b1:v}-\ref{b1:mu}). Because of
the couplings, each of the three variables, $v$ and $\mu_\lambda$,
will have components along the three independent stochastic
directions, even if they were initially independent. For instance,
$$
 v_\bk(\eta) = v_{v}(\bk,\eta) e_v(\bk) + v_{+}(\bk,\eta) e_+(\bk) + v_\times(\bk,\eta) e_\times(\bk)
$$
and
$$
 \mu_{\bk\lambda}(\eta) =  \mu_{\lambda v}(\bk,\eta) e_v(\bk) +  \mu_{\lambda\lambda}(\bk,\eta) e_\lambda(\bk) +  \mu_{\lambda(1-\lambda)}(\bk,\eta) e_{1-\lambda}(\bk)
$$
and we deduce from the properties of the random variables that
\begin{equation}
 \langle v_\bk(\eta) v_{\bk'}^*(\eta)\rangle = \left(
 |v_{v}(\bk,\eta)|^2 + |v_{+}(\bk,\eta)|^2 + |v_{\times}(\bk,\eta)|^2
 \right)\delta^{(3)}(\bk-\bk')\ .
\end{equation}
The power spectrum of the curvature perturbation at the end of
inflation (once the shear has decayed away) is thus
\begin{equation}
 P_{\mathcal R}(\bk) = \frac{2\pi^2}{k^3} {\mathcal P}_{\mathcal R}(\bk) = \frac{1}{z_{\rm S}^2}
 \left(
 |v_{v}(\bk,\eta)|^2 + |v_{+}(\bk,\eta)|^2 + |v_{\times}(\bk,\eta)|^2
 \right)
 \ .
\end{equation}
It can be checked, as expected, that for super-Hubble modes,
$\mathcal{R}$ is conserved once the shear is negligible. We thus
perform our numerical integration in a time interval long enough so that the universe
has been isotropized and all the observable modes have become super-Hubble.

The power spectra of the gravity waves are defined analogously by
\begin{equation}
 P_\lambda(\bk) = \frac{2\pi^2}{k^3} {\mathcal P}_\lambda(\bk) = |\mu_{\lambda v}(\bk,\eta)|^2
 + |\mu_{\lambda \lambda}(\bk,\eta)|^2 +|\mu_{\lambda(1-\lambda)}(\bk,\eta)|^2
 \ .
\end{equation}

At the beginning of the radiation era, the background spacetime
can be described by a Friedmann-Lema\^{\i}tre solution and the
primordial anisotropy is now encoded on the statistical properties
of the perturbations on large scales. It is thus convenient to
decompose the power spectra on spherical harmonics according to
\begin{equation}
 {\mathcal P}_{\mathcal R}(\bk)=f_R(k)\left[1+\sum_{\ell=1}^{\ell=\infty}\sum_{m=-\ell}^{m=+\ell}
   r_{\ell m}(k)\, Y_{\ell m}(\hat\bk)\right]\ ,
\end{equation}
and
\begin{equation}
 {\mathcal P}_\lambda(\bk)=f_\lambda(k)\left[1+\sum_{\ell=1}^{\ell=\infty}\sum_{m=-\ell}^{m=+\ell}
   r^\lambda_{\ell m}(k)\, Y_{\ell m}(\hat\bk)\right]\ .
\end{equation}
The three functions $f_R(k)$ and $f_\lambda(k)$ represent the
power spectra averaged over the spatial directions,
$$
 f_X(k) = \int {\mathcal P}_X(\bk)\frac{\dd^2\hat\bk}{4\pi} \ .
$$
The three series, $r_{\ell m}(k)$ and $r^\lambda_{\ell m}(k)$,
characterise the deviation from statistical isotropy. Indeed, we
expect the anisotropy to be negligible on small scales, that is
$$
 r_{\ell m}(k)\rightarrow 0 \qquad\hbox{when}\qquad k\gg \kref\ ,
$$
and that
$$
 r_{\ell m}(k)\rightarrow 0 \qquad\hbox{when}\qquad \ell\gg 1\ ,
$$
the same being true to $r^\lambda_{\ell m}(k)$. Additionaly, because of the symmetries 
of the spectrum, the only non-vanishing coefficients are obtained for even $\ell$ 
and even $m$. It can also be checked that these coefficients are real and that their 
values do not depend on the sign of $m$. Thus, we conclude
that there are only $1+\ell/2$ independent real coefficients,
$$
 r_{\ell m}\in \mathbb{R}\, \qquad \ell=2\ell'\ ,\, m=2m'\ ,\quad
 m'=0\ldots\ell'\ .
$$
The gravity waves and curvature perturbation are also correlated
so that
\begin{equation}
 \langle v_\bk(\eta) \mu_{\lambda \bk'}^*(\eta)\rangle = \left(
 |v_{v}\mu_{\lambda v}| + |v_{\lambda}\mu_{\lambda\lambda}|
  + |v_{(1-\lambda)}\mu_{\lambda(1-\lambda)}|
 \right)\delta^{(3)}(\bk-\bk')\ .
\end{equation}

\subsection{Predictions}\label{sec3.2}

To illustrate the signatures of a Bianchi~$I$ inflationary era, we
consider a Bianchi spacetime with $\alpha=\pi/4$. We fix the
initial value $\varphi_0/M_p= 3.25 \sqrt{8 \pi}\simeq 16.3$ for
the inflaton field. This implies that the number of $e$-folds of
the accelerating phase is $N[\varphi_i]\simeq 67$. We also set the
reference wavenumber to $\kref\simeq 157 m\simeq 157~ 10^{-6}M_P$. This
corresponds to the definition~(\ref{Eq_defkref}) evaluated with the approximate
cosmological constant solution~(\ref{Eq_S_purelambda}) and (\ref{Eq_H_purelambda}),
with the expression~(\ref{valtau}) for assessing $\tau_*$. 

\begin{figure}[htb]
 \includegraphics[width=6cm]{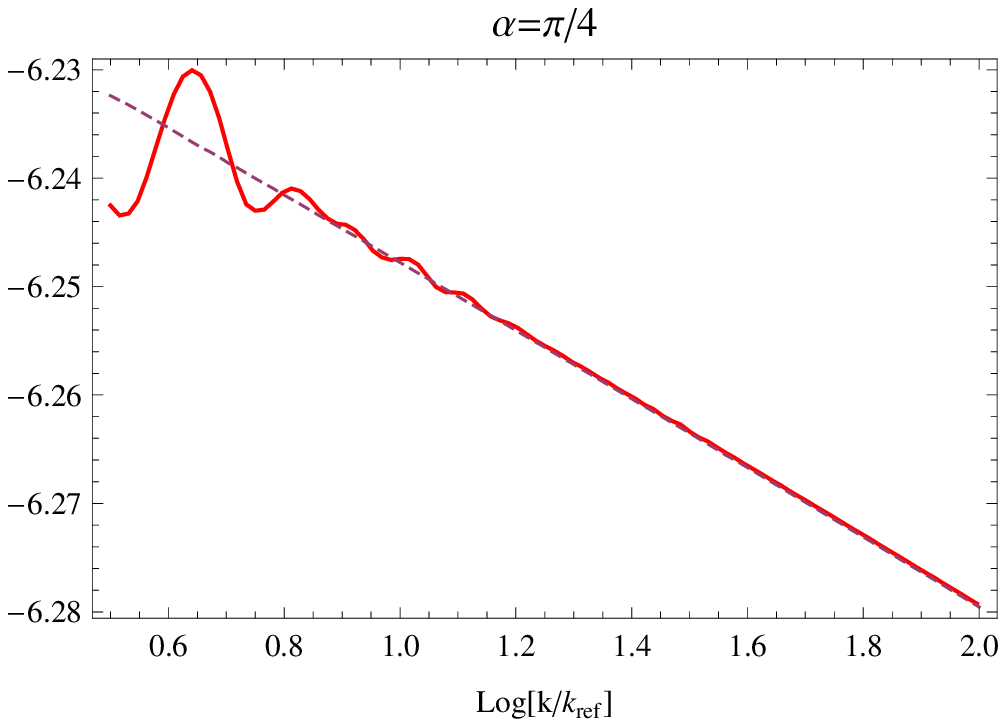}
 \includegraphics[width=6cm]{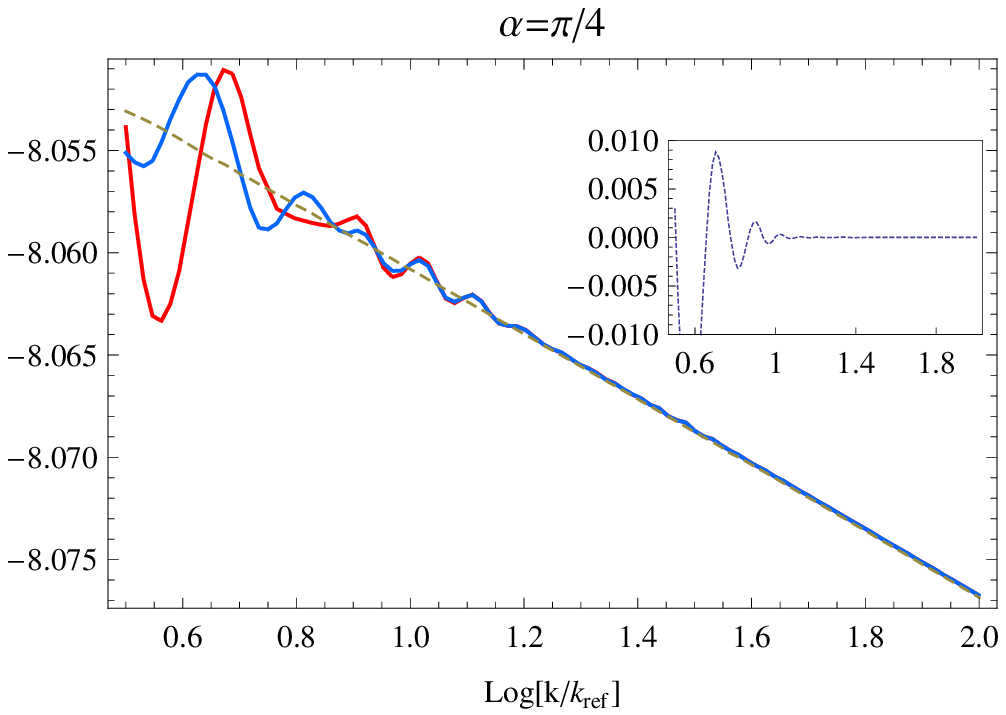}
 \includegraphics[width=6cm]{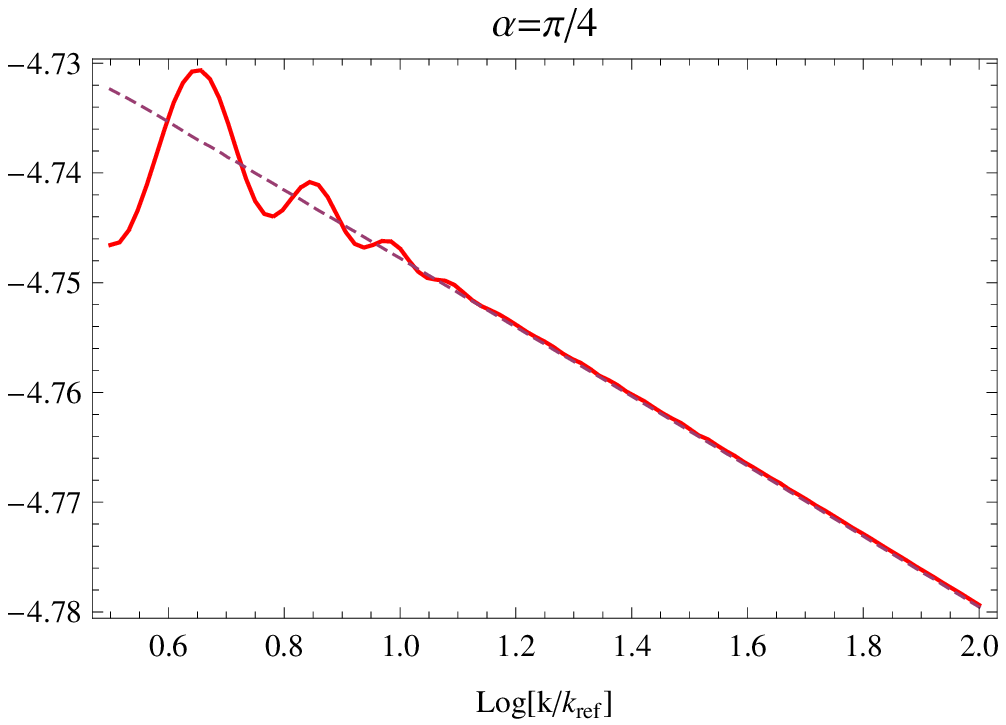}
 \includegraphics[width=6cm]{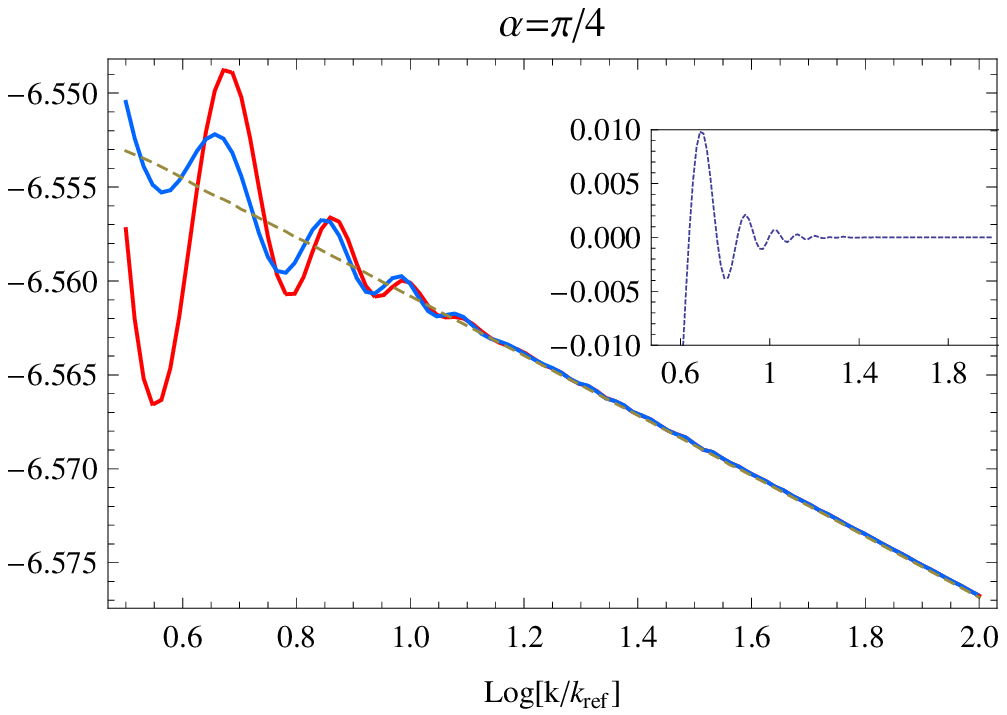}
 \caption{Evolution of $\log[f_R(k)]$ (left) and $\log[f_\lambda(k)]$ for the two
 polarizations(right) as a function
 of $\log[k/\kref]$ for $\varphi_0=16.3 M_p$. The FL case is in dashed
 line. We also depict (in the inner right figure) the relative difference 
 between the two polarisations, which shows that on small scales we recover
 that $P_\times=P_+$, as expected when isotropy is restored. On the upper line
 the initial conditions where fixed at $t_i(\bk)$ whereas in the bottom line
 they where fixed at $\tau_*$. We see that for $\log\left(k/\kref
 \right)\gtrsim 0.8$ the spectra are identical. For smaller wavenumbers, the
 WKB regime is too short to unambiguously fix the initial conditions.
}\label{fspec:iso}
\end{figure}

\begin{figure}[htb]
\includegraphics[width=6cm]{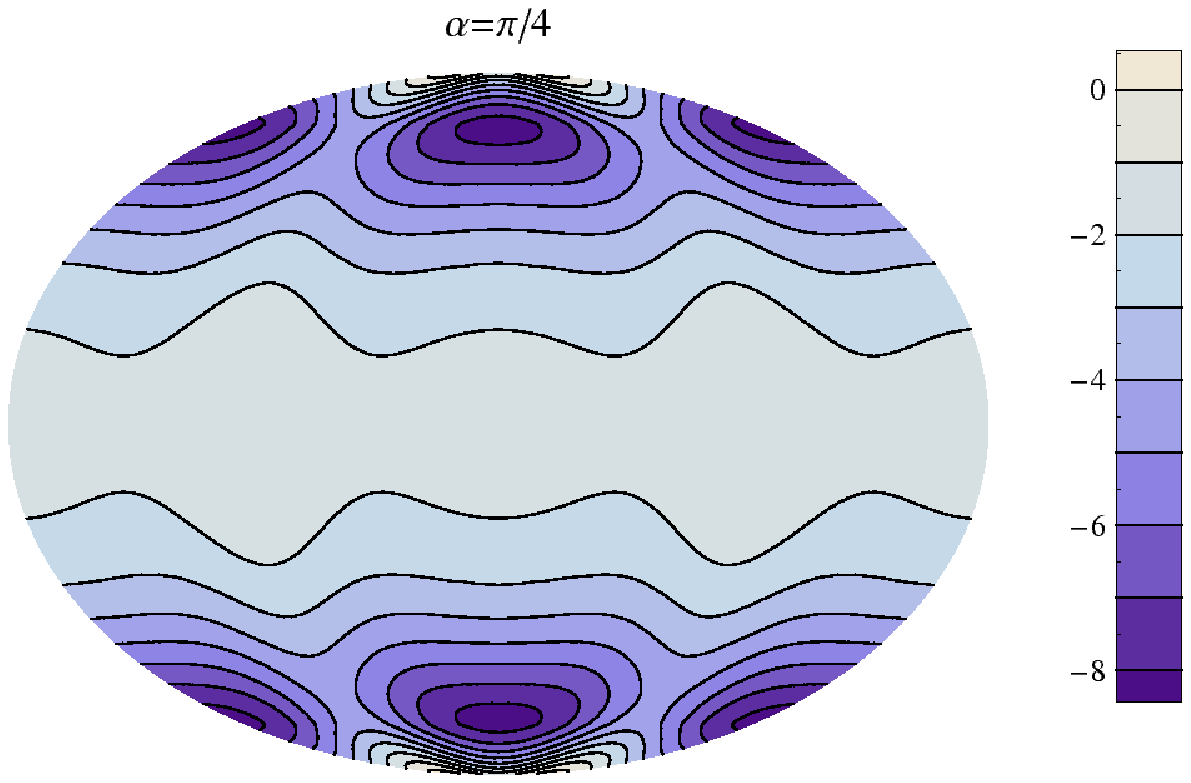}
\includegraphics[width=6cm]{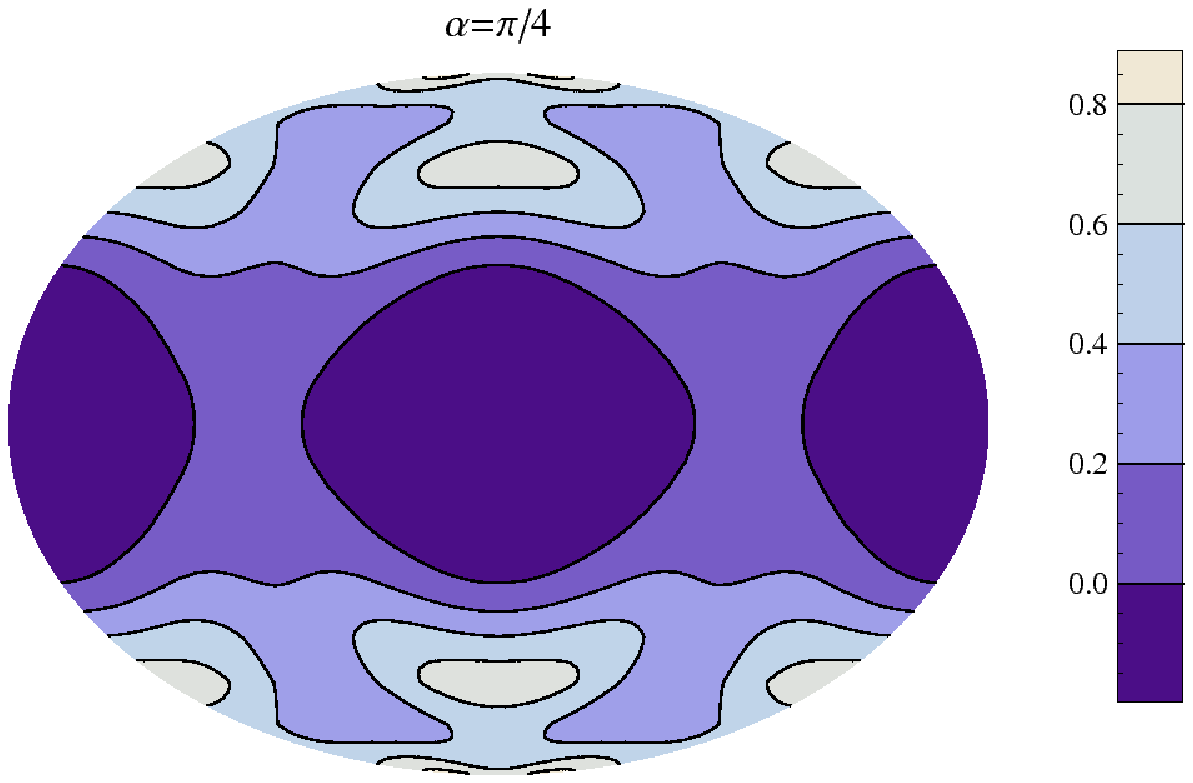}
\includegraphics[width=6cm]{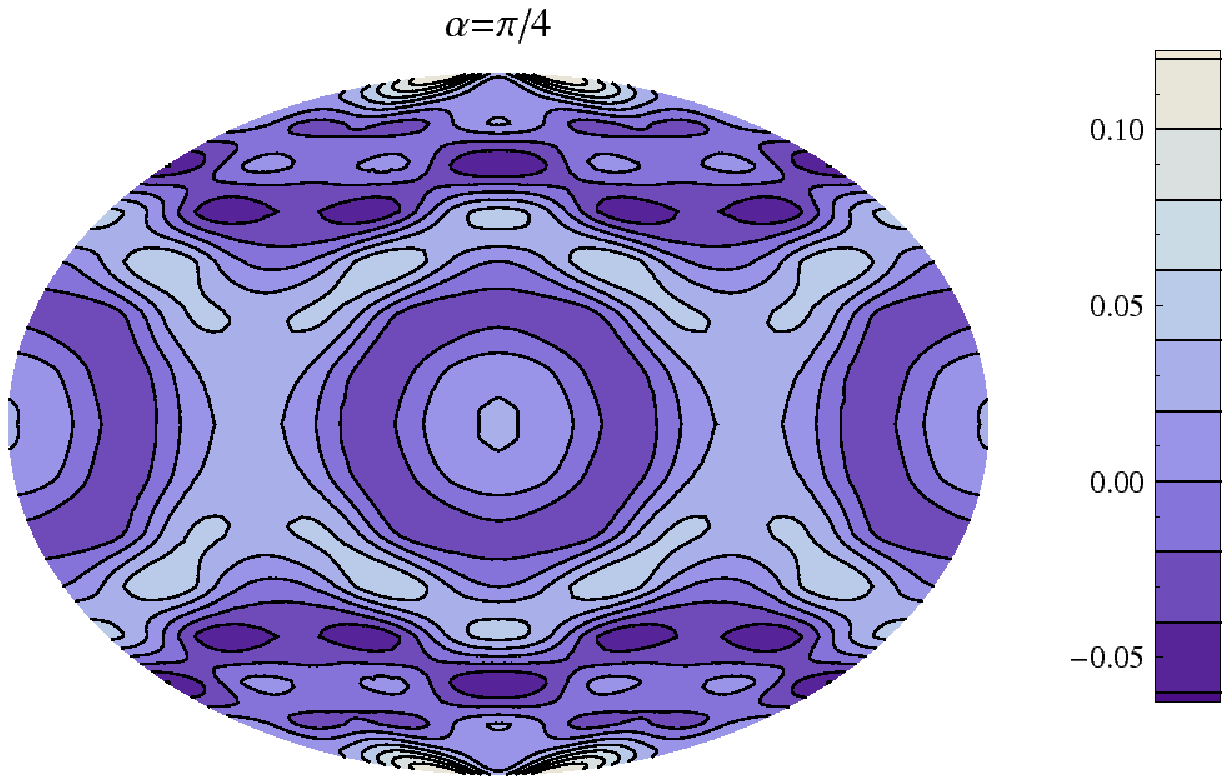}
\includegraphics[width=6cm]{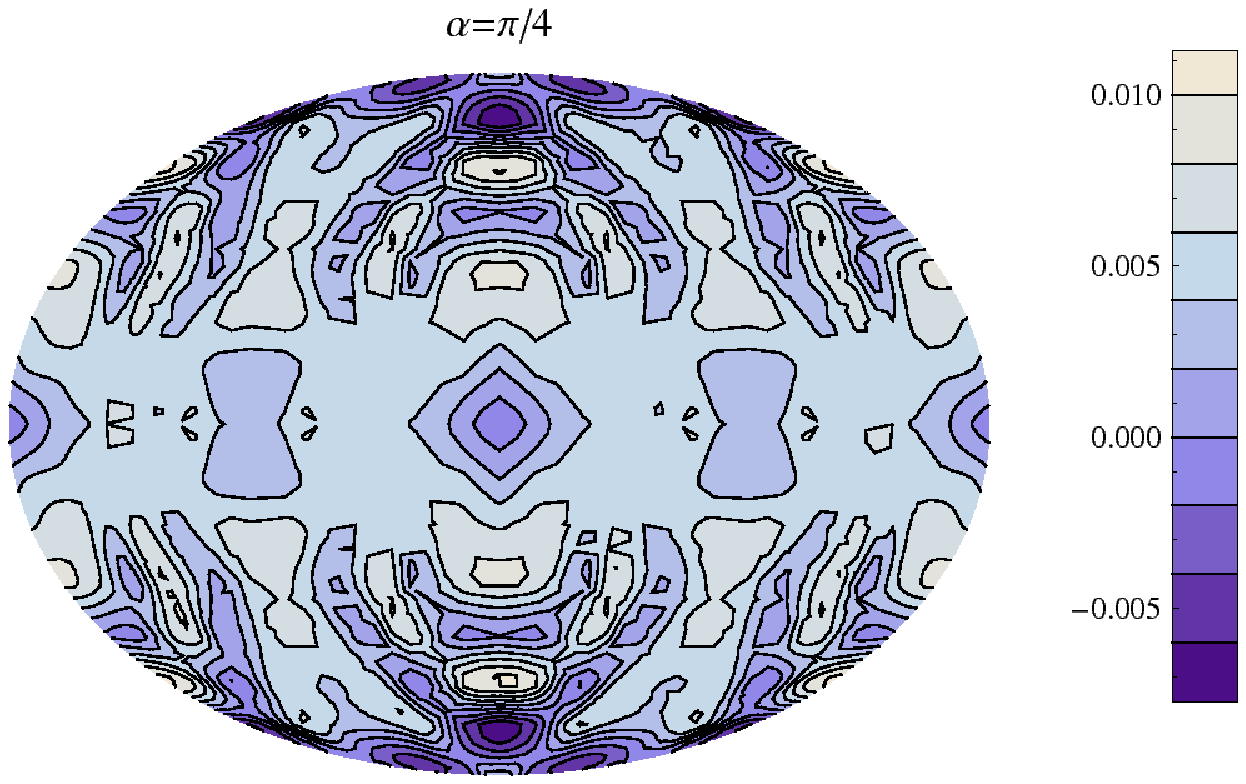}
  \caption{Mollweide projection of the ratio between ${\mathcal P}_{\mathcal R}(\bk)$ and its value
 in the FL case expressed in percentage, for  $\log[k/\kref] = 1/2,\,1,\, 3/2,\, 2$ from left to right
 and top to bottom.}\label{fspec:molw}
\end{figure}

\begin{figure}[htb]
  \includegraphics[width=8cm]{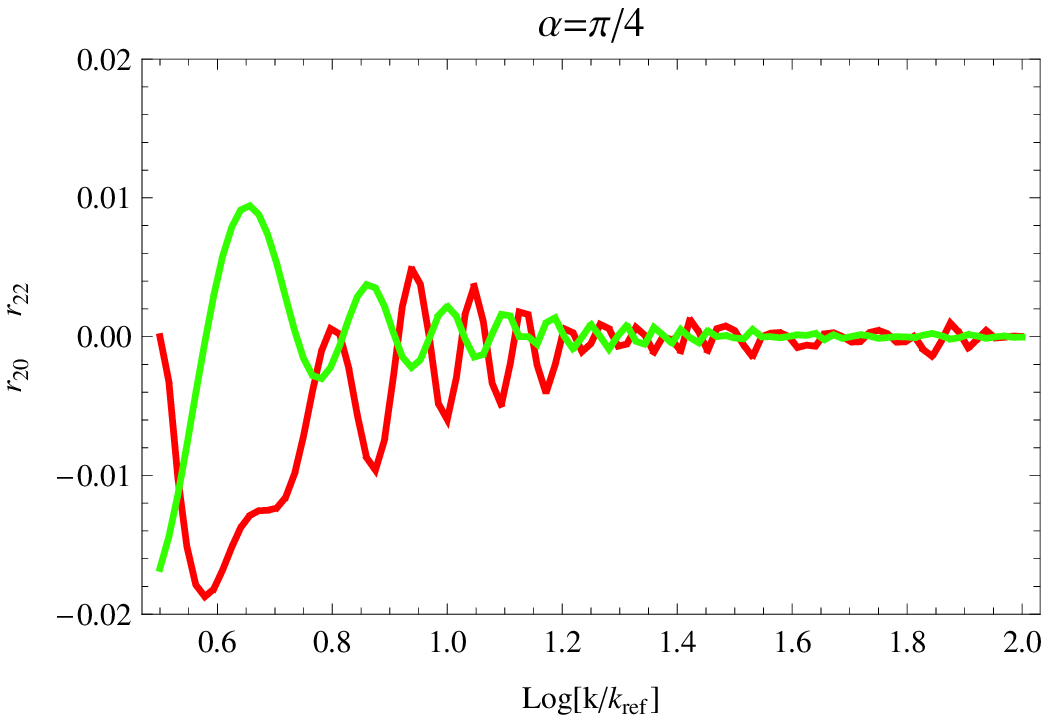}
  \includegraphics[width=8cm]{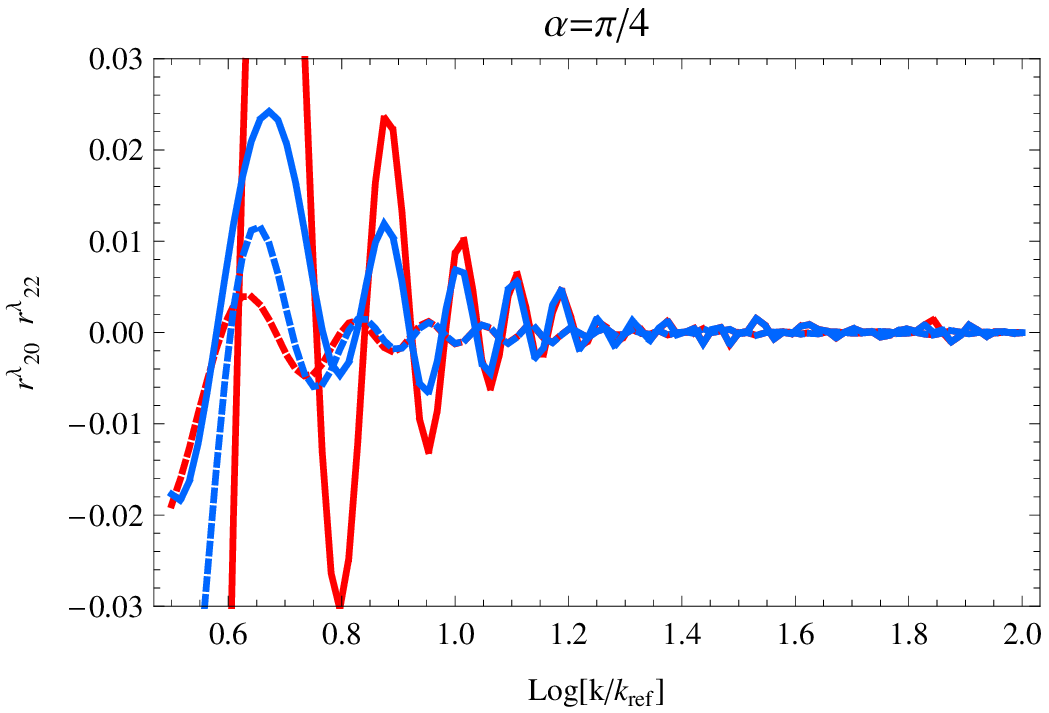}
  \caption{Evolution of $r_{\ell m}(k)$ (left) and $r^\lambda_{\ell
 m}(k)$ (right) as a function of $\log[k/\kref]$ for the lowest $\ell=2$.
$m=0$ and $m=2$ are respectively in red and green on the left. On the right $m=0$ and $m=2$ are respectively in
continuous and dashed line, the red being the $+$ polarisation and the blue
the $\times$ polarisation. 
}\label{fspec:rlm1}
\end{figure}

We solve numerically the dynamics of the background and the
evolution of the perturbations, as detailed in Appendix~A, in
order to compute the spectra defined in \S~\ref{sec3.1}. In order
to understand their behaviour, we present:
\begin{itemize}
 \item Fig.~\ref{fspec:iso}: the evolution of the functions $f_R(k)$ and $f_\lambda(k)$
 as a function of $k$ for modes ranging from $3 \kref$ to $100 \kref$.
 This shows the evolution of the isotropic part, which dominates the small scales.

 We conclude that the curvature perturbation power spectrum has a
 spectral index $n_s-1\simeq-0.032$. For this model,
 $\delta\ll\epsilon$ and the modes depicted become super-Hubble
 when $\epsilon$ is almost constant and $\epsilon\sim0.008$
 (see Fig.~\ref{fig:srpara}). The expected spectral index in
 standard isotropic inflation is thus of order
 $n_s-1=2\delta-4\epsilon\sim-0.032$, in full agreement with our
 numerical computation.

 For tensor modes, we clearly see that $P_+$ and $P_\times$ differ
 on large scales and converge to the spectrum predicted by
 standard inflation on small scales. In particular, it can be
 checked that $n_T=-2\epsilon\sim-0.016$, in full agreement with our
 numerical computation.

 \item Fig.~\ref{fspec:molw}: a Mollweide projection of ${\mathcal
 P}_{\mathcal R}(\bk)$ for different wavenumbers and for the
 scalar modes. This provides a visual intuition of the
 isotropization on small scales.

 \item Fig.~\ref{fspec:rlm1}: $r_{\ell m}(k)$ and $r^\lambda_{\ell m}(k)$
 as a function of $k$ for the lowest multipoles ($\ell=2$). It
 quantifies the isotropisation on small scales, as was observed on
 the previous figure.

 \end{itemize}.

Such predictions can be generated for any Bianchi~$I$ spacetime
and up to an arbitrary multipole. In Appendix~C, we provide
another example, namely of the particular case $\alpha=\pi/2$
considered in Ref.~\cite{GCP}.

\section{Conclusion}\label{sec4}

In this article, we have worked out the predictions of an
anisotropic inflationary era for a generic Bianchi~$I$ spacetime.
We have discussed the isotropization both at the background level
and at the linear order in perturbation theory, both for scalar
modes and gravity waves.

Generically these spacetimes always enjoy a bouncing direction,
apart from the particular case $\alpha=\pi/2$ considered in
Ref.~\cite{GCP}. (Note that the predictions for this case are in
fact singular among the predictions since they do not converge
uniformly when $\alpha\rightarrow\pi/2$).

Since at early time, the modes are not in a WKB regime, we had to
extend the standard procedure to fix the initial conditions (see
\S~\ref{sec2.2.c}). We showed that the modes larger than
$\kref$ always enter a WKB regime before they become super-Hubble
during inflation. As we discussed, our procedure reproduces the
standard one on small scales.

In the particular case where we tune the initial shear so that
these initial conditions can be set unambiguously while still
having an imprint of the CMB anisotropy, we presented the imprint
of the primordial anisotropy in the power spectra of the gravity
waves and curvature perturbation at the end of inflation. Two
examples were studied but we can provide predictions for any
Bianchi~$I$ universe. Note that these predictions were drawn by
assuming that the slow-roll attractors were reached before the time
the modes of observational relevance had exited the horizon. In the case
of inflation with small number of $e$-folds before that time,
it is not clear that this is a realistic hypothesis. If so, one
would have to make the predictions in terms of trajectories, that
is predictions that will depend on the initial conditions of the
scalar field. This is not specific to Bianchi universes but to all
models in which the inflationary period is short~\cite{levprivate}.\\

Concerning the initial conditions, two problems arise (see
discussion in \S~\ref{sec2.2.e}). First, there exists an early
shear-dominated phase where the WKB approximation is violated.
This forbids us to set the initial conditions as in a
Friedmann-Lema\^{\i}tre universe. As we showed, the sub-Hubble
modes at the onset of the accelerating phase can be quantized
because quantum fluctuations act at all times and can always
source oscillatory solutions, when they exist. On the other hand, there 
always exist non-oscillating modes. These modes are
expected not to be much smaller than the Hubble radius today
(since there is no trivial imprint of the anisotropy on the CMB).
This implies that there exists a cut-off scale above which we
cannot predict the spectra from first principles, at least in the
theoretical set-up we are considering. Consequently, we conclude that 
above this scale we can only measure the power spectra or postulate
their functional form, as was actually done to set the initial
conditions before the invention of inflation. We have shown that
even if unknown pre-WKB initial conditions can grow, this growth
is at most of order unity. Therefore we can safely assume that the
perturbations at the end of inflation reflect only those modes that have
been seeded during the WKB regime.

Second, for these modes, one cannot assume that they are
independent. It implies that we have to consider 3 interacting
fields, which also complicates the quantization procedure. Such an
issue was addressed perturbatively in the interaction picture in
the case of self-interacting field in order to estimate the
non-Gaussianity~\cite{interacting} but no general formalism has
been designed when no interaction-free regime can be exhibited.\\

In our analysis we have assumed the validity of general relativity
up to the singularity. Indeed, we do not take this model for more
than what it actually is, e.g. we cannot extrapolate it beyond the
Planck or the string scale. There, more degrees of freedom have to
be included and can change the dynamics. This could introduce an
early chaotic phase~\cite{billiard} since Bianchi~$I$ models
coupled to $p$-form fields have never ending oscillatory behaviour
exhibited by generic string theory. Examples of an early dynamics
have also been given in terms of Kalb-Ramond axion
fields~\cite{kaloper}, non-commutative geometry~\cite{ncg} and
recently loop quantum gravity~\cite{lcq}. Any of these
developments may give a description of both the early phase and a
procedure to fix the initial conditions.\\

Coming back to inflation, our study demonstrates to which extent
its predictions are sensitive to initial (classical) large scale
anisotropies and that in the presence of a non-vanishing shear, it
is impossible to define a Bunch-Davies vacuum in the standard way
(see also Ref.~\cite{picon2} for a discussion of this issue in the
standard picture and the arbitrariness on the choice of the
initial state and its influence of the prediction of inflation).
Indeed, if we tune the initial conditions such that the number of
$e$-folds is large, none of the problems we address here will
affect the observable modes, simply because only modes such
that $k/\kref\gg1$ are observable and we have shown that in this
limit we recover the standard inflationary predictions. If this is the 
case, one would have no observational imprint of the primordial
shear. But our analysis and conclusions may be of some 
relevance for inflationary model building in the framework of string 
theory if the feeling that no large field model (and thus no large number of $e$-folds) 
can be constructed persists, an issue far beyond the scope of this article.

Our analysis also shows the importance (and peculiarity) of the
Friedmann-Lema\^{\i}tre background in our theoretical predictions
and on the quantization procedure, and it gives as well an
explicit construction of the difficulties encountered when these
symmetries do not exist. We showed that if the number of $e$-folds is large
the inflationary predictions converge toward the isotropic predictions, hence
demonstrating that they are robust in that regime. In the case of
a small number of $e$-folds, they are very sensitive to the initial shear
but also the theoretical construction is less under control.

If the indication of the breakdown of statistical isotropy from
the CMB were to be confirmed and related to such an early
anisotropic phase, then a new coincidence will appear in the
cosmological models since one would need to understand why the
characteristic scale is of the order of the Hubble radius today,
$\kref\sim k_0$.

From a more pragmatic attitude, the present work allows us to draw the CMB
signatures of such an anisotropic early phase, for any Bianchi~$I$
universe. We stress that the general form of the initial power
spectra are more general than those heuristically considered in
the analysis performed in Refs.~\cite{acker,kam} and go beyond the
one derived for the singular case $\alpha=\pi/2$ studied in
Ref.~\cite{GCP}. The existence of correlation between gravity
waves and curvature perturbation, as well as the fact that the two
gravity wave polarisations do not share the same spectrum, may
also lead to specific signatures to be investigated.

\section*{Acknowlegements}

TSP thanks the Brazilian research agency Fapesp for financial
support. We thank George Ellis, Lev Kofman, Marco Peloso and Raul Abramo 
for discussions. We also thank Slava Mukhanov and Francis Bernardeau for
reading and commenting the manuscript. The background and perturbed equations have been
checked using the tensor calculus package xAct \cite{JMM}. We
thank Jos\'e Mart\'in-Garc\'ia for his help in mastering this package.

\pagebreak

\pagebreak
\appendix
\section{Integrating the perturbations}

We are interested in both the evolution of a cosmological mode and
the decomposition of the shear according to this mode until it
reaches the Friedmann stage. The covector of components $k_i$ is
constant by definition [see the discussion below Eq.~(\ref{fft})],
but in the Bianchi regime the vector $k^i=\gamma^{ij}k_j$ is not
constant since $\gamma_{ij}$ (and thus $\gamma^{ij}$) is time
dependent. This is why we used these covectors to label a mode.

Once the background equations are solved for the functions
$\beta_i(t)$ and thus $a_i(t)$, $\gamma_{ij}$ is completely
determined so that
\begin{equation}
 k^2 = \sum_i \frac{k_i^2}{a_i^2(t)}
\end{equation}
and the associated normal vector is
\begin{equation}
 \hat k_i = \frac{k_i}{k(t)},\qquad
 \hat k^i  = \frac{k_i}{a_i^2(t)k(t)}\ .
\end{equation}

\subsection{Evolution of the shear components}

The decomposition of the shear with respect to
$\hat\bk$ also requires the construction of an orthonormal base
$\left\{ {\bf e}_{1}(t),{\bf e}_{2}(t),\hat {\bf k}(t)\right\}$
that shall satisfy \cite{ppu1}
\begin{equation}
{\bf e}_{1}'(t).{\bf e}_{2}(t)={\bf e}_{1}(t).{\bf e}_{2}'(t)\ ,
\end{equation}
where the scalar product is meant in terms of the spatial metric
$\gamma_{ij}$.

In principle, it is only necessary to determine this basis at a
given initial time, $t_{\rm init}$ say, in order to extract the
initial value of the shear components through
Eqs.~(\ref{Eq_dec_shear1}-\ref{Eq_dec_shear2})
\begin{equation}
 {\bm\Sigma}(t_{\rm init})=(\spar,\sigma_{_{\rm V}1},\sigma_{_{\rm V}2},\sigma_{_{\rm T}\times}
,\sigma_{_{\rm T}+}).
\end{equation}
Then, to determine the value of ${\bm\Sigma}$ at any time, we use
that Eq.~(\ref{e:fried3Cb}) implies (see Ref.~\cite{ppu1} for
details)
\begin{eqnarray}
 &&\spar'+2\HH\spar = -2\sum_a\sigma_{_{\rm V}a}^2\,, \label{background_sigma}\\
 &&\sva' +2\HH\sva = \frac{3}{2}\sva\spar
  -\sum_{b,\lambda}\svb\stl\mathcal{M}_{ab}^\lambda\,, \\
 &&\stl'+2\HH\stl = 2\sum_{a,b}\mathcal{M}_{ab}^\lambda\sva\svb \label{background_sigma3}\,,
\end{eqnarray}
where the matrix $\mathcal{M}_{ab}^\lambda$ is defined by
\begin{equation}\label{defMab}
 {\cal M}_{ab}^\lambda \equiv \varepsilon^\lambda_{ij}e^i_a e^j_b\,,
\end{equation}
which is manifestly symmetric in $ab$. It is explicitly given by
\begin{equation}
 {\cal M}_{ab}^\lambda = \frac{1}{\sqrt{2}}
      \left(\begin{array}{cc}
             1 &0\\0&-1
             \end{array}\right)\delta^\lambda_+
             +
     \frac{1}{\sqrt{2}}
      \left(\begin{array}{cc}
             0 &1\\1&0
             \end{array}\right)\delta^\lambda_\times\,.
\end{equation}
One interesting aspect of the
system~(\ref{background_sigma}-\ref{background_sigma3}) is that it
does not depend explicitly on $k_i$. Once this system has been
solved, we can deduce easily the functions
$(\omega_v,\omega_\lambda,\aleph,\beth)$ that enters the equations
of evolution of the pertubative modes.

In practice however, in the regime where $\sigma \ll \mathcal{H}$,
small numerical oscillations in the components of the shear are
converted into huge numerical instabilities in the system
(\ref{background_sigma}-\ref{background_sigma3}). In order to avoid 
these numerical instabilities we turn to another method which we now 
describe.

\subsection{Systematic construction}

This method relies on the fact that the solutions of the system
(\ref{background_sigma}-\ref{background_sigma3}) are, in a general
sense, given by
\begin{equation}\label{shearcomponents}
\sigma_{\parallel}=\sigma_{ij}\hat{k}^{i}\hat{k}^{j}\,,
\quad\sigma_{_{{\rm V}}a}=\sigma_{ij}\hat{k}^{i}e_{a}^{j}\,,
\quad\sigma_{_{{\rm
T}}\lambda}=\sigma_{ij}\varepsilon_{\lambda}^{ij}
\end{equation}
as a function of time.

Instead of solving
(\ref{background_sigma}-\ref{background_sigma3}) after having
extracted the components of the shear at a fixed initial time, we
can determine the dynamics of the shear components through the
dynamics of the basis vectors, the polarisation tensor and the
shear. Once the $\beta_i'(t)$ are determined numerically, the
shear is known.

In order to determine the base vectors $\left\{ {\bf
e}_{1}(t),{\bf e}_{2}(t)\right\}$, we first start from a triad
$\left\{{\bf e}_{x}(t),{\bf e}_{y}(t), {\bf e}_{z}(t)\right\}$
aligned with the $xyz$-proper axis. Then we introduce three Euler
angles to rotate this triad to any given direction.

Let us recall how to deal with rotations in spaces endowed with a
general (non orthonormal) metric. Any rotation matrix about a unit
vector $\mathbf{\hat{n}}$ can be constructed by means of
infinitesimal rotations of the form
$$
 R\left(\theta/N\right)=1+\mathbf{J\cdot\hat{n}}\frac{\theta}{N}
=1+\gamma^{ij}J_{j}\hat{n}_{i}\frac{\theta}{N}\ ,
$$
where $\theta/N$ is some infinitesimal angle.  The $J_{j}$ are the
generators of the group of rotations. In particular, if we want to
specify a rotation about the $z$-axis, then the unit vector
aligned with this axis has components
$$
 \hat{n}_{i}=\left(0,0,e^{\beta_{3}}\right)\ ,
$$
and we have $$
 R_{z}\left(\theta\right)=\lim_{N\rightarrow\infty}
\left(1+e^{-\beta_{3}}J_{z}\frac{\theta}{N}\right)^{N}
=\exp\left[\theta e^{-\beta_{3}}J_{z}\right]\,.
$$
It can be shown that in order to conserve the scalar product, the
generators must satisfy $(J_k)_{ij}=-\epsilon_{kij}$, where
$\epsilon_{ijk}$ is totally antisymmetric normalised such that
$\epsilon_{123}=\sqrt{\det(\gamma_{ij})}=1$. A rotation matrix
around the $z$-axis then requires to use $\left(J_{z}\right)_{\,\,
j}^{i}=(\gamma)^{ik}(J_{z})_{kj}$, which is explicitly given by
\begin{equation}\label{genz}
 \left(J_{z}\right)^{i}_{\,\,j}=
 \left(\begin{array}{ccc}
   0 & -e^{-2\beta_{1}} & 0\\
   e^{-2\beta_{2}} & 0 & 0\\
   0 & 0 & 0
 \end{array}\right).
\end{equation}
Consequently, taking into account the constraint~(\ref{beta}),
i.e. $\sum_{i}\beta_{i}=0$, any rotation about the $z$-axis can be
written as
\begin{equation}
 \left[R_{z}(\theta)\right]^i_{\,\,j}=
   \left(\begin{array}{ccc}
     \cos(\theta) & -e^{(\beta_{2}-\beta_{1})}\sin(\theta) & 0\\
     e^{(\beta_{1}-\beta_{2})}\sin(\theta) & \cos(\theta) & 0\\
     0 & 0 & 1\end{array}\right)\ .
\end{equation}

Similarly, for a rotation about the $y$-axis, we use the generator
\begin{equation}\label{geny}
\left(J_{y}\right)^i_{\,\,j}=
    \left(\begin{array}{ccc}
    0 & 0 & e^{-2\beta_{1}}\\
    0 & 0 & 0\\
    -e^{-2\beta_{3}} & 0 & 0
    \end{array}\right)
\end{equation}
to get
\begin{equation}
  \left[R_{y}(\theta)\right]^i_{\,\,j}=
    \left(\begin{array}{ccc}
    \cos(\theta) & 0 & e^{(\beta_{3}-\beta_{1})}\sin(\theta)\\
    0 & 1 & 0\\
    -e^{(\beta_{1}-\beta_{3})}\sin(\theta) & 0 &
    \cos(\theta)\end{array}\right)\ .
\end{equation}
Explicitly, the components of the triad from which
we start are
\begin{equation}
 (e_{x})^{i}\equiv\left(\begin{array}{c}
 \hbox{e}^{-\beta_1}\\0\\0\end{array}\right),
    \quad
 (e_{y})^{i}\equiv\left(\begin{array}{c}
0\\ \hbox{e}^{-\beta_2}\\0\end{array}\right),
    \quad
 (e_{z})^{i}\equiv\left(\begin{array}{c}0\\
0\\ \hbox{e}^{-\beta_3}\end{array}\right)\ .
\end{equation}

The three Euler angles $\left(\alpha,\beta,\gamma\right)$ are then
used to rotate this triad according to, respectively, the $x$, $y$ and $z$-axis. 
Explicitly, the triad after rotation is given by
\begin{eqnarray}\label{generaltriad}
(e_{1})^{i} & \equiv & R_{z}(\gamma)_{\,\,
j}^{i}R_{y}(\beta)_{\,\, l}^{j}R_{z}(\alpha)_{\,\,
p}^{l}(e_{x})^{p}\,,
\nonumber \\
(e_{2})^{i} & \equiv & R_{z}(\gamma)_{\,\,
j}^{i}R_{y}(\beta)_{\,\, l}^{j}R_{z}(\alpha)_{\,\,
p}^{l}(e_{y})^{p}\,.
\nonumber \\
(u_{k})^{i} & \equiv & R_{z}(\gamma)_{\,\,
j}^{i}R_{y}(\beta)_{\,\, l}^{j}R_{z}(\alpha)_{\,\,
p}^{l}(e_{z})^{p}\,,
\end{eqnarray}
which determines the orthonormal basis $\{{\bm e}_1,{\bm e}_2,{\bm
u}_k \}$. This prescription is still incomplete since our
problem also requires the covector ${\bm u}_{k}$ to be equal to the covector
$\hat\bk$ during the whole evolution of the system, i.e.
$$
 (u_k)_i(t)=\hat k_i(t)\ ,
 \qquad
 \forall t\ .
$$

Since $k_{i}$ should not depend on time, we need to determine the
Euler angles as a function of time such that, for any two times
$t$ and $t'$, we have $[u_{k}(t)]_{i}= f(t,t')[u_{k}(t')]_{i}$.
This condition is satisfied provided
\begin{eqnarray}
\tan(\gamma) & = &
\tan(\gamma_{f})\exp\left[(\beta_{1}-\beta_{2})\right]
\label{conditiongamma}\\
\tan(\beta) & = &
\exp\left[(\beta_{3}-\beta_{1})\frac{\cos(\gamma_{f})}{\cos(\gamma)}\tan(\beta_{f})\right]\,,
\label{conditionbeta}
\end{eqnarray}
where $\gamma_{f}$ and $\beta_{f}$ are the angles which give the
final direction that we consider when the shear has vanished.

Additionally, our problem also requires the condition
$$ \left[(e_{2})^{i}\right]^{\prime}\left[(e_{1})_{i}\right]=
\left[(e_{1})^{i}\right]^{\prime}\left[(e_{2})_{i}\right]\ ,
$$
as a prescription for the choice polarisation basis to be
continuous. It is fulfilled if
\begin{equation}
\alpha^{\prime}=-\cos(\beta)\gamma^{\prime}.\label{conditionalpha}
\end{equation}
This equation can be integrated with the help of
Eq.~(\ref{conditiongamma}) and Eq.~(\ref{conditionbeta}).
Consequently, provided we know $\left(\beta_{f},\gamma_{f}\right)$
describing the final orientation of a mode, we can determine
$\left(\alpha(t),\gamma(t),\beta(t)\right)$ which will select the
triad adapted to the mode {\it at any time}.\\

In conclusion, the set of
equations~(\ref{generaltriad}-\ref{conditionalpha}) gives a
complete description of the time-evolving triad necessary to
extract the components of the shear according to
Eq.~(\ref{shearcomponents}) at any time. This is the procedure we
have implemented in our numerical calculations, and is robust to numerical
errors.

\section{Slow-roll expressions}

We give here, for the sake of completeness, the explicit expressions
for $z^{\prime\prime}_{\rm S}/z_{\rm S}$, $z^{\prime\prime}_{\lambda}/z_{\lambda}$, 
$\aleph_\lambda$ and $\beth$ in terms of the slow-roll parameters. Setting 
\begin{equation}
w\equiv \frac{\spar}{2\HH},\quad y_{\lambda}\equiv \frac{\stl}{\sqrt{6}\HH},
\end{equation}
we have 
\begin{eqnarray}\label{Eq_zsprprsurzs}
\frac{z^{\prime\prime}_{\rm S}}{z_{\rm S}} & = & \mathcal{H}^{2}\left\{2+\epsilon\left[5-4
\delta+2\epsilon-\frac{\left(3-\delta\right)^{2}}{3-\epsilon}\right]\right.\\
 &  & +x^{2}\left[ -3-6\delta+\epsilon\left(7+4\delta-4\epsilon\right)
\right] +2x^{4}\left(3-\epsilon\right)^{2}\nonumber\\
&& -6 x^2 \left\{\epsilon-\delta+x^2 (3-\epsilon)\right\}+ 2\frac{\epsilon}{\HH}\left(1-x^{2}\right)\left(\frac{w}{1-w}\right)^{\prime}\nonumber\\
&&\left.+\left(\frac{w}{1-w}\right)\left[6\epsilon+2\epsilon^2-4\epsilon\delta+x^{2}\left(-4\epsilon^{2}+4\epsilon\delta\right)+2x^{4}
\left(\epsilon^{2}-6\epsilon\right)\right]\right\}\,\nonumber\\
\frac{z''_\lambda}{z_\lambda}&=&\mathcal{H}^{2}\left\{2-\epsilon+6w-2\epsilon
\left[w+3\left(\frac{y_{\lambda}^{2}}{1-w}\right)\right]+12y_{(1-\lambda)}^{2}
+18\left(\frac{y_{\lambda}^{2}}{1-w}\right)\right.\nonumber\\
 &  & \left.+\left(\epsilon-3\right)x^{2}\left[1+2w
+6\left(\frac{y_{\lambda}^{2}}{1-w}\right)\right]+2\frac{w^{\prime}}{\mathcal{H}}
+\frac{6}{\mathcal{H}}\left(\frac{y_{\lambda}^{2}}{1-w}\right)^{\prime}\right\}\\
\aleph_\lambda&=&\sqrt{6}\mathcal{H}^{2}
\sqrt{2\epsilon\left(1-x^{2}\right)}\left[\frac{1}{\mathcal{H}}\left(\frac{y_{\lambda}}{1-w}\right)
^{\prime}+\frac{1}{2}\left(6-\epsilon-\delta\right)\left(\frac{y_{\lambda}}{1-w}\right)\right]\
,\\
\beth&=&6\mathcal{H}^{2}\left[\frac{1}{\mathcal{H}}
\left(\frac{y_{+}y_{\times}}{1-w}\right)^{\prime}+6\left(3-\epsilon+\left(3-\epsilon\right)x^{2}\right)
\left(\frac{y_{+}y_{\times}}{1-w}\right)\right]\ .
\end{eqnarray}

\section{Particular case $\alpha=\pi/2$}

As we stressed, the case $\alpha=\pi/2$ is particular in the sense
that this is the only Bianchi~$I$ models which does not have any
bouncing direction.

Since this model enjoys a planar symmetry (conveniently chosen
here to be in the $xy$-plane), the component $\stcross$ of the
shear vanishes for any choice of the Euler angles provided that
$\alpha=0$, which corresponds to the symmetry around the $z$-axis.
This implies that $\aleph_\times\propto\sigma_\times$ also
vanishes at all times. The function $\beth$ behaves as $\sigma_+\sigma_\times$
and for the same reason also vanishes. In conclusion, when (and only when)
$\alpha=\pi/2$, we have
$$
 \aleph_\times = 0\ ,
 \quad
 \beth = 0\ .
$$
This is the reason why it was possible to construct a
``simplified'' perturbation theory in Ref.~\cite{GCP} in splitting
in $2+1$ degrees of freedom from the start of the analysis. As we
saw, this is not generic.

\begin{figure}[htb]
 \includegraphics[width=8cm]{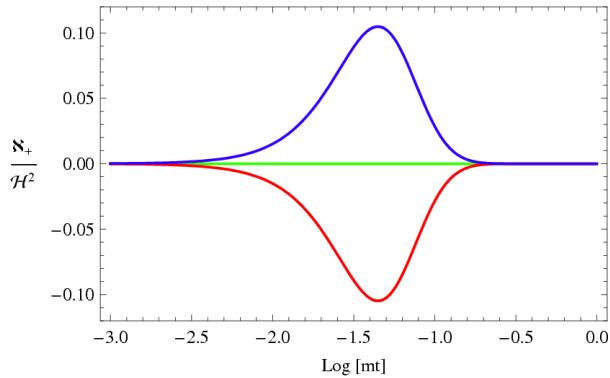}
 \caption{Evolution of $\aleph_+/\HH^2$ for the three orthogonal modes aligned with the $x, y,$
 and $z$-axis (represented by three different colors).
 We have considered here a generic Bianchi spacetime with $\alpha=\pi/2$.}\label{couplingGCP}
\end{figure}

\begin{figure}[htb]
 \includegraphics[width=8.1cm]{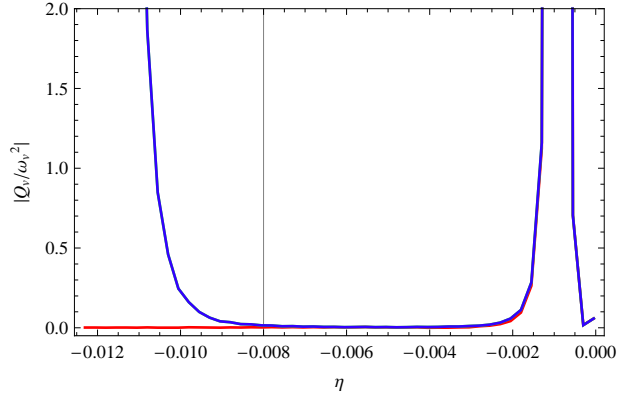}
 \caption{Evolution of $|Q_v^{\rm WKB}/\omega_v^2|$ for three different
modes, each of them aligned with one of the three ortoghonal
directions and with the same modulus $10k_{\rm ref}$ at the end of
inflation. We have considered a generic Bianchi spacetime with
$\alpha=\pi/2$.}\label{figpos1}
\end{figure}

\begin{figure}[htb]
 \includegraphics[width=8cm]{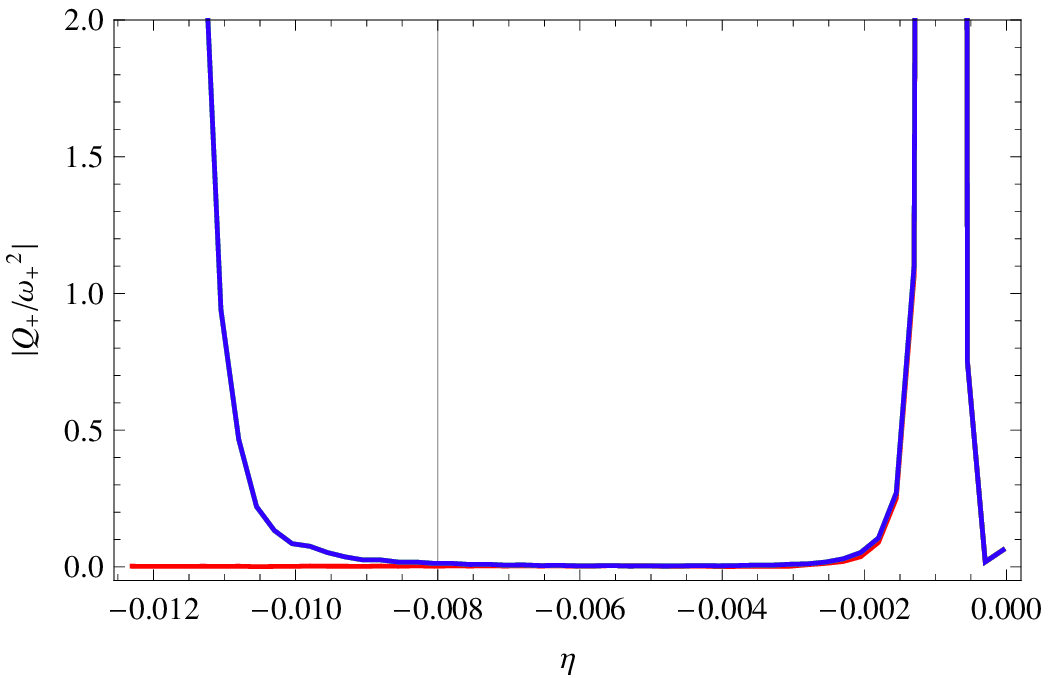}
 \includegraphics[width=8cm]{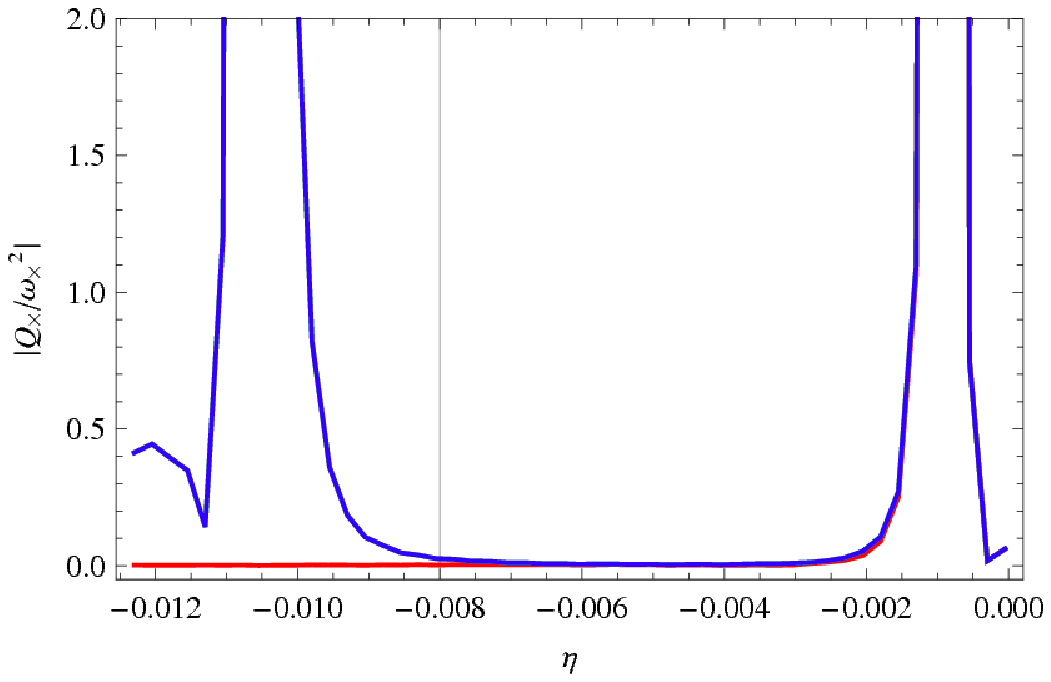}
 \caption{Evolution of $|Q_\lambda^{\rm WKB}/\omega_\lambda^2|$ for $\lambda=+$ (left) and
$\lambda=\times$ (right) for various modes with the same
 modulus $10k_{\rm ref}$ at the end of inflation. We have considered a
 generic Bianchi spacetime with $\alpha=\pi/2$.}\label{figpos2}
\end{figure}

Figure~\ref{couplingGCP} now shows that 
$$
 \frac{\aleph_+}{\HH^2}\rightarrow 0
$$
at early time which implies that the modes were in fact
decoupled. Figs.~\ref{figpos1} and~\ref{figpos2} consider the
validity of the WKB condition. In particular, one can show that on
sub-Hubble scales
$$
 \frac{z_{\rm s}''}{z_{\rm s}}\rightarrow \frac{k_z^2}{a_z^2}\ ,
 \quad
 \frac{z_\lambda''}{z_\lambda}\rightarrow \frac{k_z^2}{a_z^2}\ .
$$
We have checked that there exists only one time redefinition which
leads to adiabatic canonical variables and that agree with all the
conclusions in Ref.~\cite{GCP}.

\begin{figure}[htb]

 \includegraphics[width=8cm]{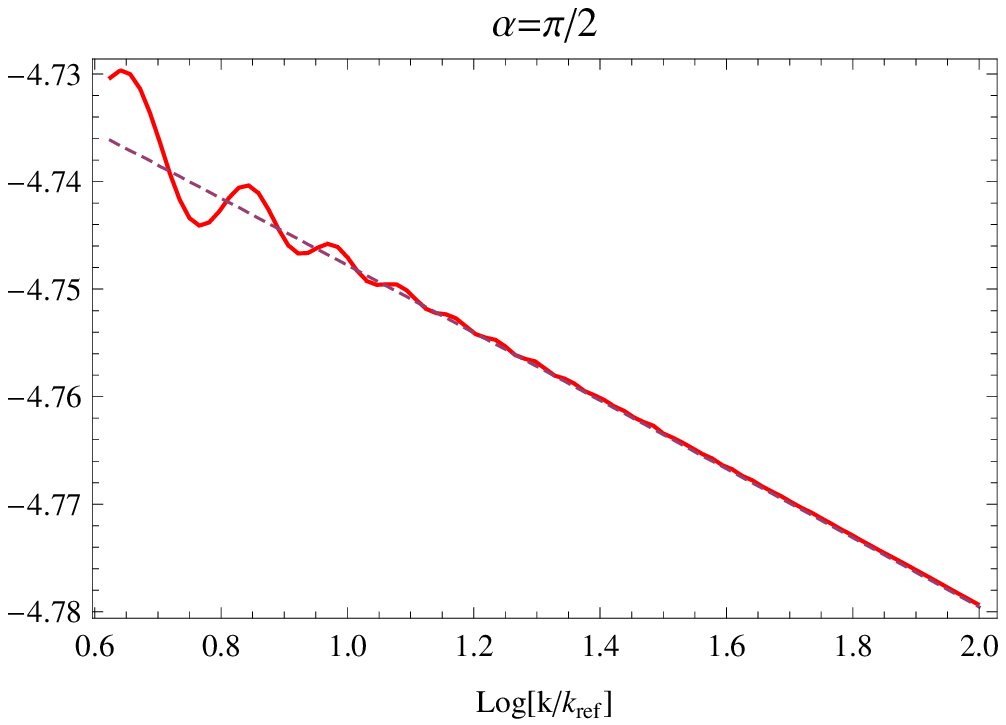}
\includegraphics[width=8cm]{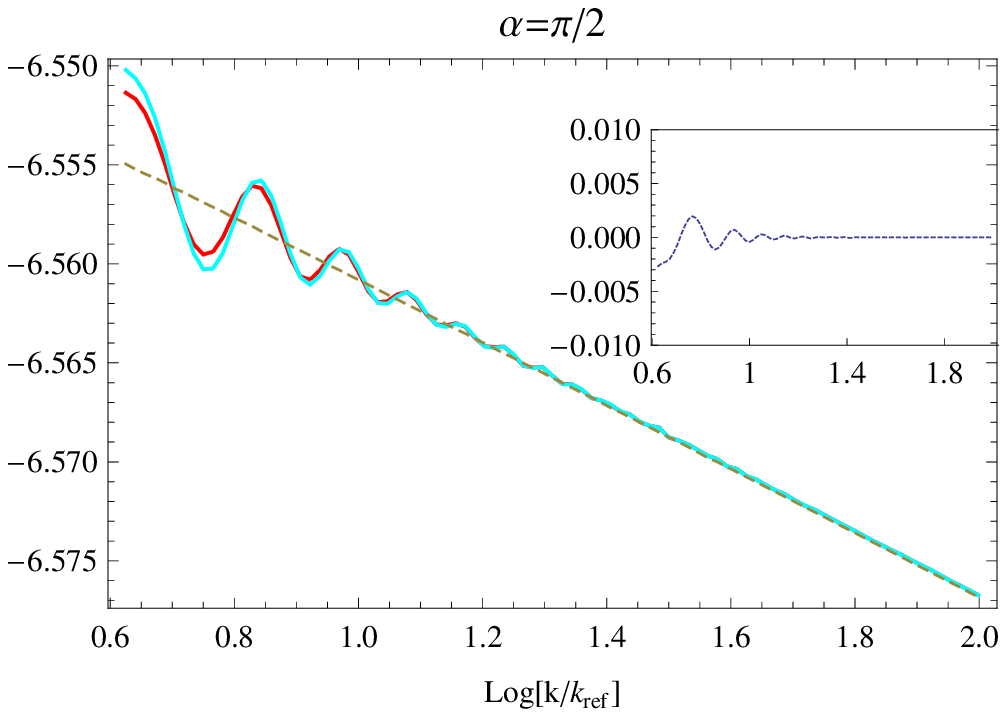}
\caption{Evolution of $f_R(k)$ (left) and $f_\lambda(k)$ (right) as a function
 of $\log[k/\kref]$. The FL case is in dashed line.
}\label{fspec:isoGCP}
\end{figure}

\begin{figure}[htb]
 \includegraphics[width=6cm]{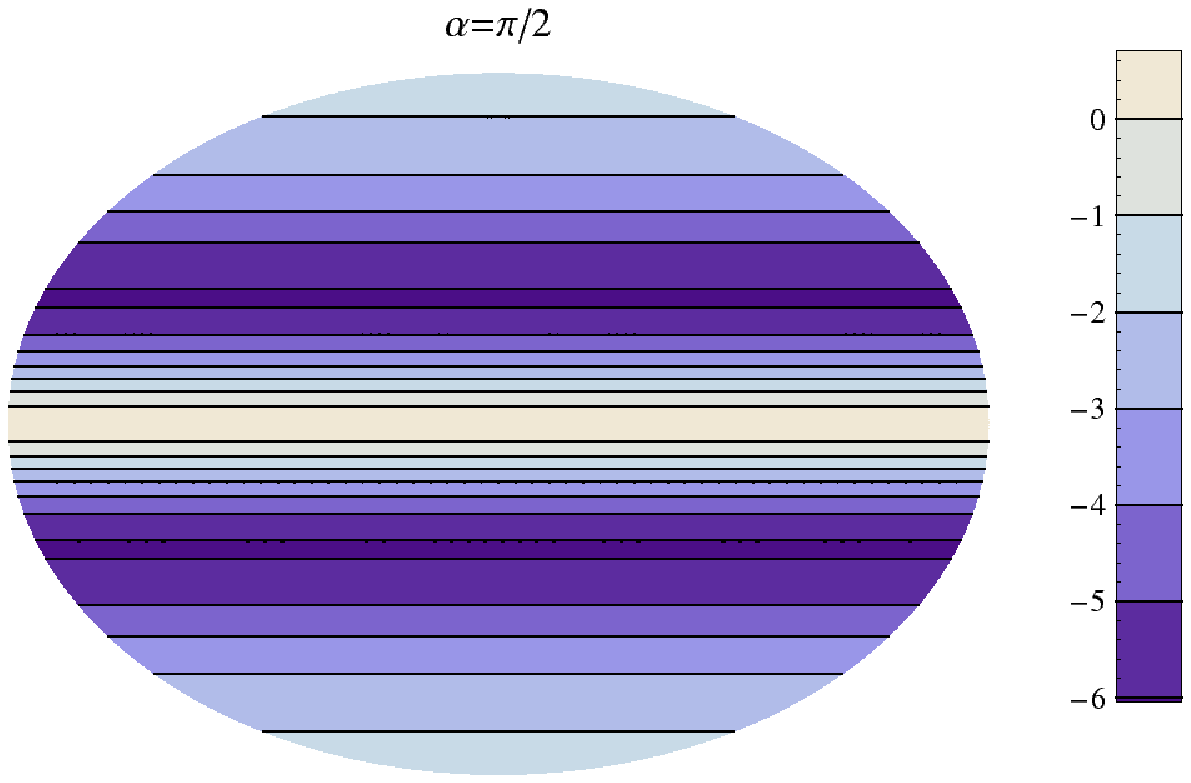}
 \includegraphics[width=6cm]{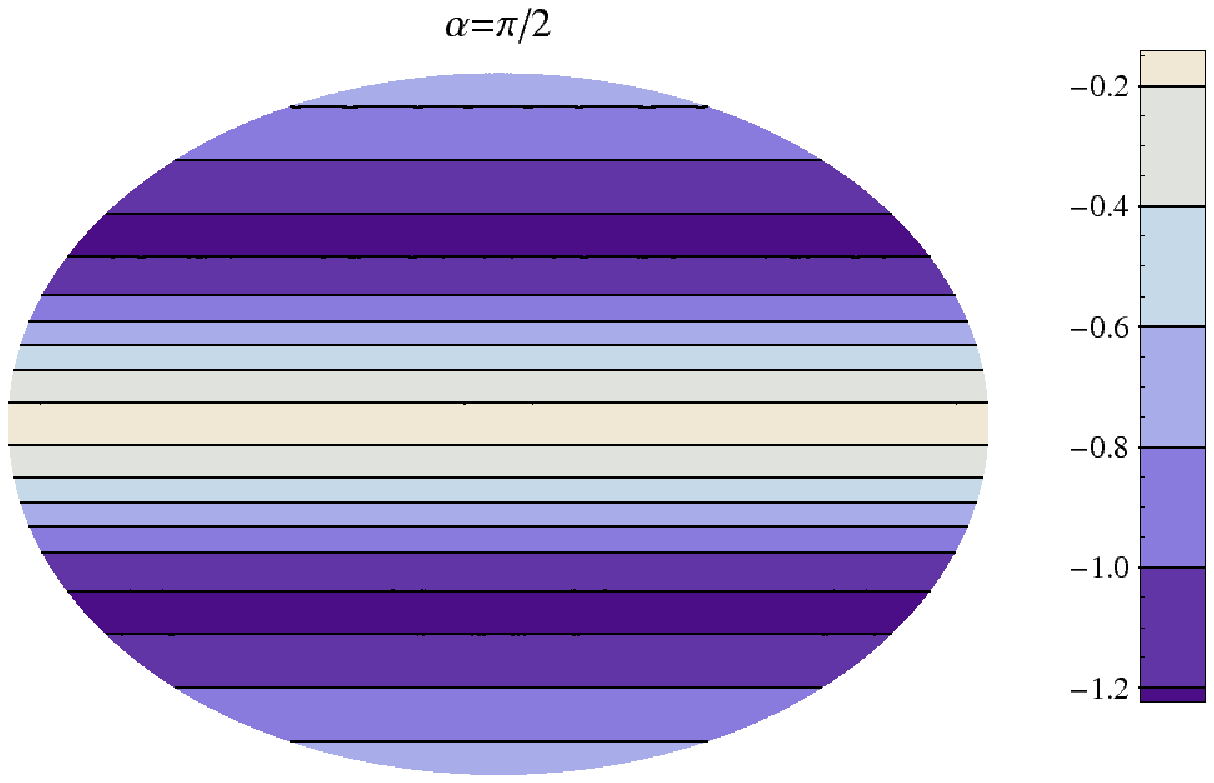}
\includegraphics[width=6cm]{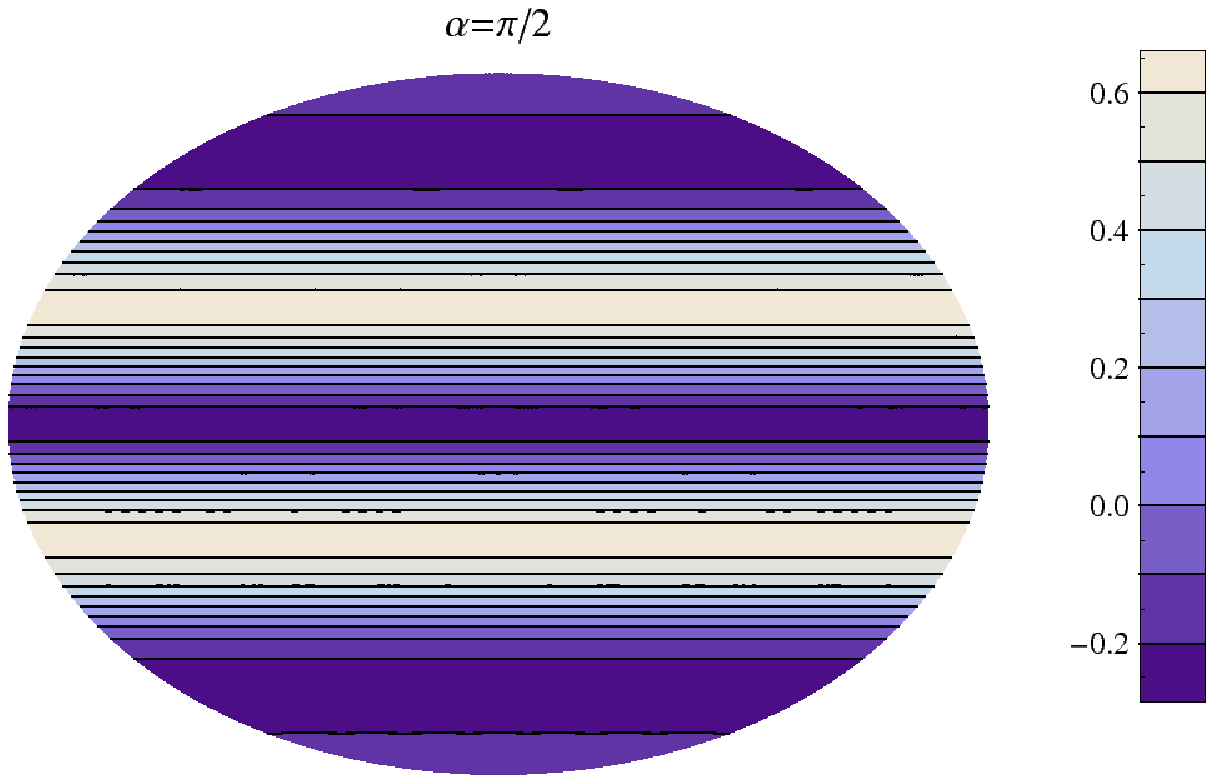}
\includegraphics[width=6cm]{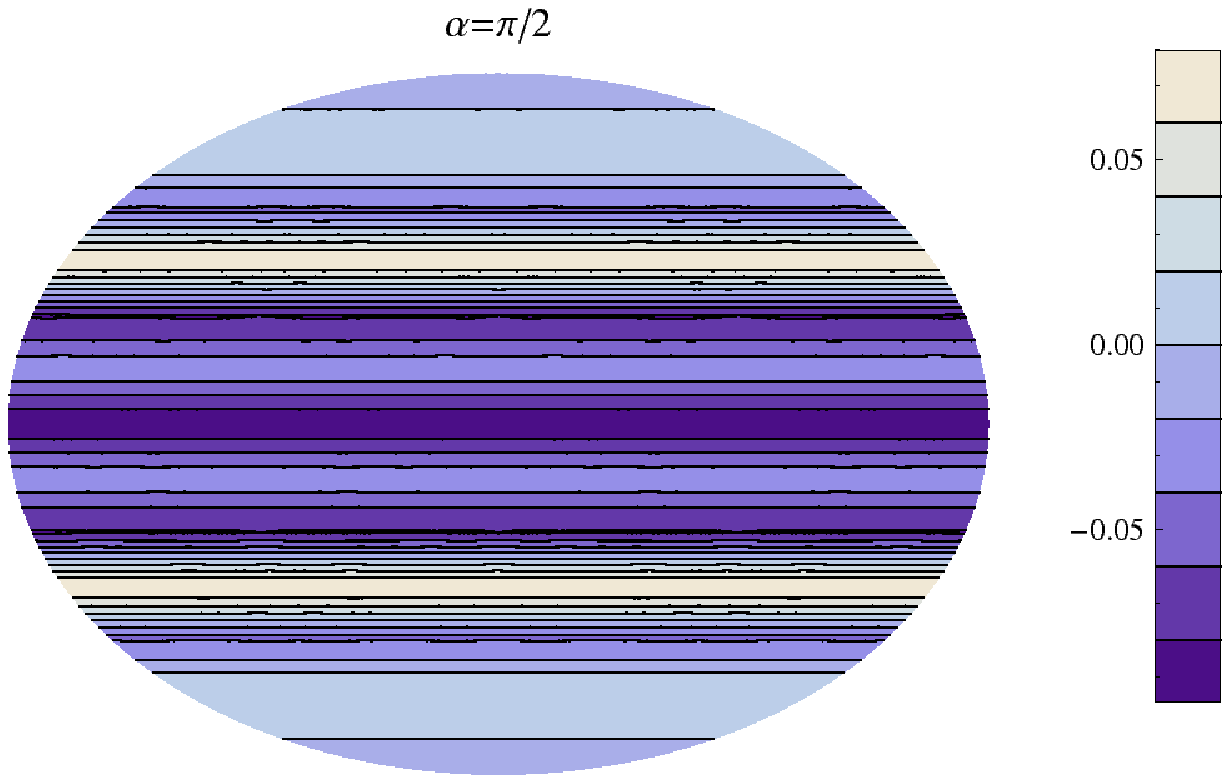}
  \caption{Mollweide projection of the ration between ${\mathcal P}_{\mathcal R}(\bk)$
and its value in the FL case expressed in percentage for  $\log[k/\kref] =1/2,\, 3/4,\, 1,\, 3/2$
from left to right, top to bottom. The $x$-axis as been choosen as the vertical axis}
\label{fspec:molwGCP}
\end{figure}

\begin{figure}[htb]

  \includegraphics[width=8cm]{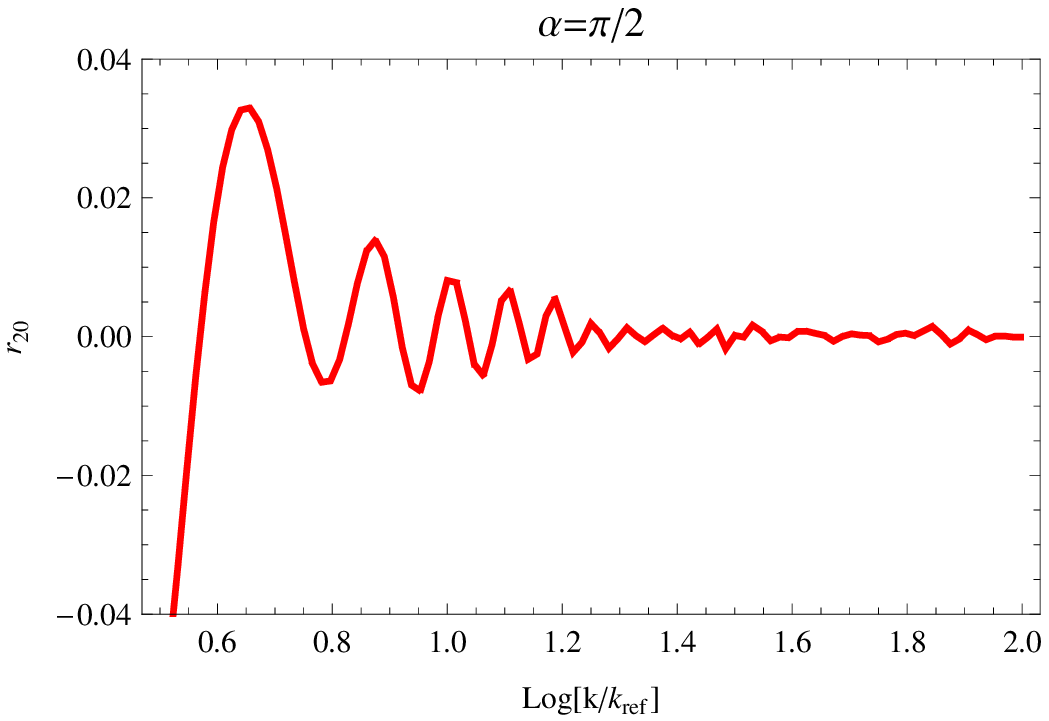}
  \includegraphics[width=8cm]{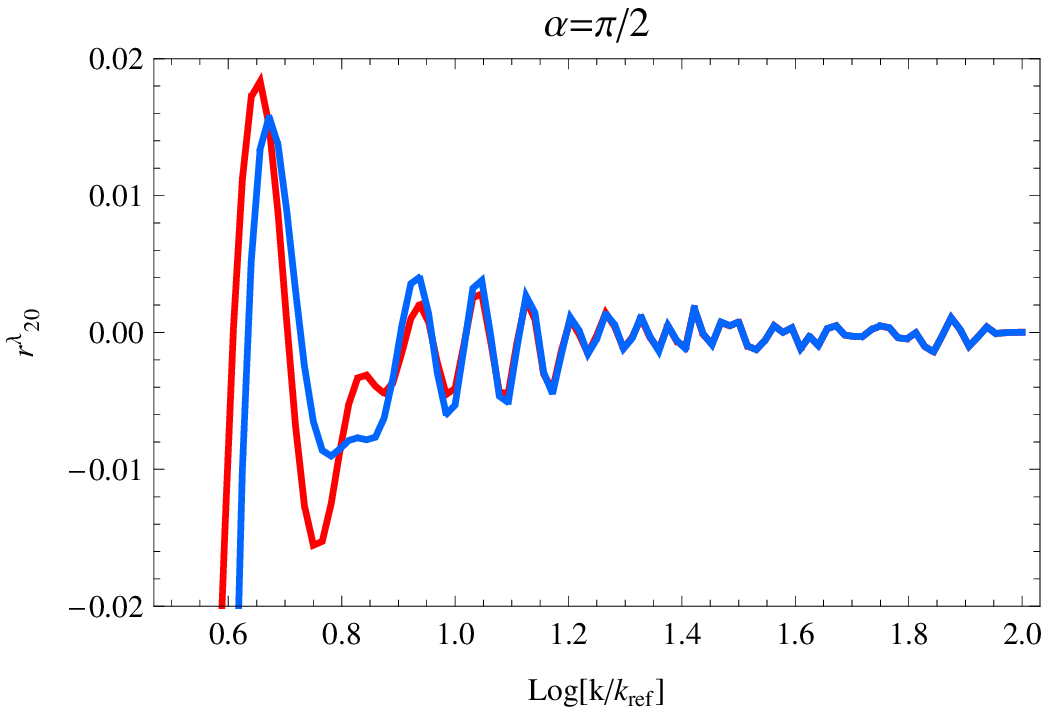}
  \caption{Evolution of $r_{\ell m}(k)$ (left) and $r^\lambda_{\ell
 m}(k)$ (right) as a function of $\log[k/\kref]$ for the lowest multipoles ($\ell=2$). 
The $x$-axis has been used to perform the spherical harmonics decomposition. As a
result of symmetries, only the coefficients with $m=0$ do not vanish. On the
right the red is the $+$ polarisation and the blue is the $\times$ polarization.}
\label{fspec:rlm1GCP}
\end{figure}

Following our quantization procedure, we compute the predictions
for the power spectra (Figs.~\ref{fspec:isoGCP}
to~\ref{fspec:rlm1GCP}). Note that the planar symmetry is obvious
on the Mollweide projection, that shall be compared to the generic
case presented on Fig.~\ref{fspec:molw}. This illustrates how a
detection of anisotropy may permit us to reconstruct the particular
Bianchi~$I$ structure (if any~!) during inflation.

In this particular case, there always exist a WKB regime We can
thus define our initial conditions in the standard way. Following
the procedure described in this article, we can compute the power
spectra of this solution. We present here the same figures as in
\S~\ref{sec3.2}.

\begin{thebibliography}{99}

\bibitem{lindebook}
 A.D. Linde,
 {\it Particle physics and inflationary cosmology},
 Harwood (Chur, Switzerland, 1990);\\
 S. Mukhanov,
 {\it Physical foundations of cosmology},
 Cambridge University Press (Cambrige, UK, 2005).

\bibitem{pubook}
 P. Peter and J.-P. Uzan,
 {\it Cosmologie primordiale},
 Belin (Paris, France, 2005);\\
 F. Bernardeau,
 {\it Cosmologie, des fondement th\'eoriques aux observations},
 CNRS \'Editions (Paris, France, 2007).

\bibitem{linde2007}
  A.~Linde,
  [\urel{arXiv:0705.0164 [hep-th]}].

\bibitem{ref:inf}
 V. Mukhanov and G.V. Chibisov, JETP Lett. {\bf33} (1981) 532;\\
 S.W. Hawking, Phys. Lett. B {\bf115} (1982) 295.

\bibitem{mbf}
 V.F. Mukhanov, F.A. Feldman and R.H. Brandenberger,
 Phys. Rep. {\bf215}, 203 (1992).

\bibitem{iso}
 A.D. Linde,
 Phys. Lett. B {\bf158}, 375 (1985);\\
 J. Garc\'{\i}a-Bellido and D. Wands,
 Phys. Rev. D {\bf 53}, 5437 (1996);\\
 V.F. Mukhanov and P.J. Steinhardt,
 Phys. Lett. B {\bf 422} 52 (1998);\\
 D. Langlois,
 Phys. Rev. D {\bf59}, 123512 (1999).

\bibitem{ng}
 E. Komatsu {\it et al.},
 Phys. Rep. {\bf402}, 103 (2006),
 [\urel{arXiv:astro-ph/0406398}];\\
 A. Linde and V. Mukhanov,
 Phys. Rev. D {\bf 56} (1997) 535,
 [\urel{arXiv:astro-ph/9610219}];\\
 F. Bernardeau and J.-P. Uzan,
 Phys. Rev. D {\bf66}, 103506 (2002),
 [\urel{arXiv:hep-ph/0207295}];\\
 F. Bernardeau and J.-P. Uzan,
 Phys. Rev. D {\bf67}, 121301(R) (2003),
 [\urel{arXiv:astro-ph/0209330}];\\
 B. Osano {\it et al.},
 JCAP {\bf004}, 003 (2007),
 [\urel{arXiv:gr-qc/0612108}];\\
 J. Maldacena,
 JHEP {\bf0305}, 013 (2005),
 [\urel{arXiv:astro-ph/0210603}].

\bibitem{bku}
 F. Bernardeau, L. Kofman, and J.-P. Uzan,
 Phys. Rev. D {\bf70}, 083004 (2004),
 [\urel{arXiv:astro-ph/0403315}].

\bibitem{olive}
 K. Olive,
 Phys. Rept. {\bf190}, 307 (1990).

\bibitem{infanisogen}
 C.B. Collins and S.W. Hawking, Astrophys. J. {\bf180}, 317 (1973);\\
 J.D. Barrow, Quart. J. Roy. Astron. Soc. {\bf23}, 344 (1982);\\
 R.M. Wald, Phys. Rev. D {\bf28}, 2118 (1983);\\
 G. Steigman and M. Turner, Phys. Lett. B {\bf128}, 295 (1983);\\
 O. Gron, Phys. Rev. D {\bf32}, 2522 (1985);\\
 T. Rothman and M.S. Madsen, Phys. Lett. B {\bf159}, 256 (1985);\\
 T. Rothman and G.F.R. Ellis, Phys. Lett. B {\bf180}, 19 (1986);\\
 L.G. Jensen and J.A. Stein-Schabes, Phys. Rev. D {\bf34}, 931 (1986);\\
 L.G. Jensen and J.A. Stein-Schabes, Phys. Rev. D {\bf35}, 1146 (1987);\\
 T. Patcher, Eur. Phys. Lett. {\bf4}, 1211 (1987);\\
 I. Moss and V. Sahni, Phys. Lett. B {\bf178}, 159 (1986);\\
 A.B. Burd and J.D. Barrow, Nuc. Phys. B {\bf308}, 929 (1988);\\
 A. Raychaudhuri and B. Modak, Class. Quant. Grav. {\bf5}, 225
 (1988);\\
 S. Byland and D. Scialom, Phys. Rev. D{\bf57}, 6065 (1998);\\
 J. Aguirregabiria and A. Chamorro, Phys. Rev. D{\bf62}, 084028 (2000);\\
 R.V. Buniy, A. Berera, and T.W. Kephart, [\urel{arXiv:hep-th/0511115}];\\
 J.D. Barrow and S. Hervik, Class. Quant. Grav. {\bf23}, 3053 (2006).

\bibitem{braneinf}
 R. Maartens, V. Sahni, and T.D. Saini,
 Phys. Rev. D {\bf63}, 063509 (2001)
 [\urel{arXiv:gr-qc/0011105}];\\
 M.G. Santos, F. Vernizzi, and P.G. Ferreira,
 Phys. Rev. D {\bf64}, 063506 (2001),
 [\urel{arXiv:hep-ph/0103112}];\\
 B.C. Paul,
 Phys. Rev. D {\bf64}, 124001 (2001),
 [\urel{arXiv:gr-qc/0107005}];\\
 J. Aguirregabiria, L.P. Chimento, and R. Lazkoz,
 Class. Quant. Grav. {\bf21}, 823 (2004),
 [\urel{arXiv:gr-qc/0303096}].

\bibitem{inflK}
 G.F.R. Ellis {\it et al.},
 Gen. Rel. Grav. {\bf34}, 1445 (2002),
 [\urel{arXiv:gr-qc/0109023}];\\
 J.-P. Uzan, U. Kirchner, and G.F.R. Ellis,
 Month. Not. R. Astron. Soc. {\bf344}, L65 (2003),
 [\urel{arXiv:astro-ph/0302597}].

\bibitem{MSvar}
 V.F. Mukhanov,
 JETP Lett. {\bf41}, 493 (1985)
 [Pisma Zh. Eksp. Teor. Fiz. {\bf41}, 402 (1985)];\\
 M. Sasaki,
 Prog. Theor. Phys. {\bf76}, 1036 (1986).

\bibitem{Allister}
 L. McAllister and E. Silverstein,
 [{\tt arXiv:0710.2951}].

\bibitem{bauman}
 D. Bauman and L. McAllister,
 [{\tt hep-th/0610285}].

\bibitem{roulette}
 J.R. Bond \etal.,
 Phys. Rev. D {\bf75}, 123511 (2007),
 [{\tt hep-th/0612197}].

\bibitem{TomitaDen}
 K. Tomita and M. Den,
 Phys. Rev. D {\bf34}, 3570 (1986).

\bibitem{Dunsby}
 P. Dunsby,
 Phys. Rev. D {\bf 48}, 3562 (1993).

\bibitem{pitrou07}
  C.~Pitrou and J.-P.~Uzan,
  Phys. Rev.  D {\bf 75}, 087302 (2007),
  [\urel{arXiv:gr-qc/0701121}].

\bibitem{NohHwang}
 H. Noh and J.-C. Hwang,
 Phys. Rev. D {\bf 52}, 1970 (1995).

\bibitem{Abbotetal}
 R.B. Abbott, B. Bednarz and D. Ellis,
 Phys. Rev. D {\bf33}, 2147 (1986).

\bibitem{Qaniso}
 Y.B. Zeldovich and A.A. Starobinsky,
 Sov. Phys. JETP {\bf34}, 1159 (1972);\\
 B.J. Berger,
 Phys. Rev. D {\bf12}, 368 (1975);\\
 P.K. Suresh,
 [\urel{arXiv:gr-qc/0308080}].

\bibitem{ppu1}
 Thiago S. Pereira, C. Pitrou, and J.-P. Uzan,
 JCAP {\bf0709}, 006 (2007),
 [\urel{arXiv:0707.0736}].

\bibitem{GCP}
  A.E. G\"umr\"uk{\c c}\"uo\v{g}lu, C.R. Contaldi and M. Peloso,
  JCAP {\bf0711}, 005 (2007),
 [\urel{arXiv:0707.4179}].

\bibitem{lowQ}
 A. Oliveira-Costa {\it et al.},
 Phys. Rev. D {\bf69}, 063516 (2004),
 [\urel{arXiv:astro-ph/0307282}];\\
 H.K. Eriksen {\it et al.},
 Astrophys. J. {\bf605}, 14 (2004),
 [\urel{arXiv:astro-ph/0307507}];\\
 D.J. Schwartz {\it et al.},
 Phys. Rev. Lett. {\bf93}, 0403353 (2004),
 [\urel{arXiv:astro-ph/0403353}];\\
 K. Land and J. Magueijo,
 Phys. Rev. Lett. {\bf95}, 071301 (2005),
 [\urel{arXiv:astro-ph/0502237}].

\bibitem{prunet}
 S. Prunet {\it et al}.,
 Phys. Rev. D {\bf71}, 083508 (2005),
 [\urel{arXiv:astro-ph/0406364}].

\bibitem{topologie}
 J.-P. Luminet {\it et al.},
 Nature {\bf425}, 593 (2003),
 [\urel{arXiv:astro-ph/0310253}];\\
 A. Riazuelo {\it et al.},
 Phys. Rev. D {\bf69}, 103518 (2004),
 [\urel{arXiv:astro-ph/0311314}];\\
 A. Riazuelo {\it et al.},
 Phys. Rev. D {\bf69}, 103514 (2004),
 [\urel{arXiv:astro-ph/0212223}];\\
 J.-P. Uzan {\it et al.},
 Phys. Rev. D {\bf69}, 043003 (2004),
 [\urel{arXiv:astro-ph/0303580}].

\bibitem{modetopo}
 R. Lehoucq, J-P. Uzan, and J. Weeks,
 Kodai Math. Journal {\bf26}, 119 (2003),
 [\urel{arXiv:math.SP/0202072}];\\
 E. Gausmann {\it et al.},
 Class. Quant. Grav. {\bf18}, 5155 (2001),
 [\urel{arXiv:gr-qc/0106038}];\\
 R. Lehoucq {\it et al.},
 Class. Quant. Grav. {\bf19}, 4683 (2002),
 [\urel{arXiv:gr-qc/0205009}];\\
 J.D. Barrow and H. Kodama,
 Class. Quant. Grav. {\bf18}, 1753 (2001),
 [\urel{arXiv:gr-qc/0012074}].

\bibitem{picon}
 C. Armendariz-Picon,
 [\urel{arXiv:0705.1167}].

\bibitem{spinor}
 C.G. B\"ohmer and D.F. Mota,
 [\urel{arXiv:0710.2003 [astro-ph]}].

\bibitem{dulaney}
T. Dulaney, M.I. Gresham and M.B. Wise, [\urel{arXiv:0801.2951 [astro-ph]}].

\bibitem{donoghue}
 E.P. Donoghue and J.F. Donoghue, [\urel{arXiv:astro-ph/0411237}];\\
 J.F. Donoghue, K. Dutta, and A. Ross,
 [\urel{arXiv:astro-ph/0703455}].

\bibitem{Jaffe2005}
  T.R.~Jaffe, {\it et al.},
  Astrophys. J. Lett. {\bf629}, L1 (2005),
  [\urel{arXiv:astro-ph/0503213}];\\
  T.R.~Jaffe, {\it et al.},
  [\urel{arXiv:astro-ph/0606046}];\\
 A. Pontzen and A. Challinor,
 [\urel{arXiv:0706.207 [astro-ph]}].

\bibitem{CMB_bianchi}
  A.E. G\"umr\"uk{\c c}\"uo\v{g}lu, C.R. Contaldi and M. Peloso,
  [\urel{arXiv:astro-ph/0608405}].

\bibitem{acker}
 L. Ackerman, S.M. Caroll, and M.B. Wise,
 Phys. Rev. D {\bf75}, 083502 (2007),
 [\urel{arXiv:astro-ph/0701357}];\\
 A.R. Pullen and M. Kamionkowski,
 [\urel{arXiv:0709.1144 [astro-ph]}].

\bibitem{LimitShear}
 R. Marteens, G.F.R. Ellis, and W.R. Stoeger,
 Asron. Astrophys. {\bf309}, L7 (1996),
 [{\tt astro-ph/9501016}];\\
 R. Marteens, G.F.R. Ellis, and W.R. Stoeger,
 Phys. Rev. D {\bf51}, 1525 (1995),
 [{\tt astro-ph/9510126}];\\
 W.R. Stoeger, M.E Araujo, and T. Gebbie,
 Astrophys. J. {\bf476}, 435 (1997),
 [{\tt astro-ph/9904346}];\\
 A. Kogut, G. Hinshaw, and A.J. Banday,
 Phys. Rev. D {\bf55}, 1901 (1997),
 [{\tt astro-ph/9701090}];\\
 E. Martinez-Gonzalez and J.L. Sanz,
 Astron. Astrophys. {\bf300}, 346 (1995).

\bibitem{bbn}
 K.S. Thorne,
 Astrophys. J. {\bf148}, 51 (1967).

\bibitem{bianchimath}
 M.P. Ryan,
 {\it Homogeneous relativistic cosmologies}
 (Princeton University Press, Princeton, 1975);\\
 G.F.R. Ellis and M.A.H. MacCallum,
 Comm. Math. Phys. {\bf12}, 108 (1969).

\bibitem{lidsey}
 J.E. Lidsey, Class. Quant. Grav. {\bf9}, 1239 (1992).

\bibitem{levprivate}
 L. Kofman, results presented in seminars and conferences.

\bibitem{borde}
 A. Borde, A.H. Guth, and A. Vilenkin,
 Phys. Rev. Lett. {\bf90}, 151301 (2003),
 [\urel{arXiv:gr-qc/0110012}].

\bibitem{JeromeSchwartz}
 J. Martin, D.J. Schwartz, Phys. Rev. D {\bf67}, 083512 (2003), [\urel{arXiv:astro-ph/0210090}].

\bibitem{llN}
 J.E. Lidsey, {\em et al.}, Rev. Mod. Phys. {\bf69}, 373 (1997);\\
 D. Lyth and A. Liddle, Phys. Rept. {\bf 231}, 1 (1993).

\bibitem{sarkar}
 S. Sarkar, Rep. Prog. Phys. {\bf 59}, 1493 (1996).

\bibitem{wands1}
 D. Wands,
 [\urel{arXiv:astro-ph/0702187}].

\bibitem{bw}
 C.T. Byrnes and D. Wands,
 [\urel{arXiv:astro-ph/0605679}].

\bibitem{lubo}
 M. Lemoine {\it et al.}, Phys. Rev. D {\bf65}, 023510 (2002);
 [\urel{arXiv:hep-th/0109128}].

\bibitem{billiard}
 T. Damour and M. Henneaux,
 Phys. Rev. Lett. {\bf85}, 920 (2000),
 [\urel{arXiv:hep-th/0003139}];\\
 T. Damour and M. Henneaux,
 Phys. Lett. B {\bf488}, 108 (2000),
 [\urel{arXiv:hep-th/0006171}].

\bibitem{kaloper}
 N. Kaloper, Phys. Rev. D {\bf44}, 2380 (1991).

\bibitem{lcq}
 D.-W. Chiou,
 [{\tt arXiv:0710.0416}]

\bibitem{ncg}
 E. Di Grezia \etal.,
 Phys. Rev. D {\bf68}, 105012 (2003),
 [\urel{gr-qc/0305050}].

\bibitem{interacting}
 J. Maldacena,
 JHEP {\bf0305}, 013 (2005),
 [\urel{arXiv:astro-ph/0210603}];\\
 T. Brunier, F. Bernardeau, and J.-P. Uzan,
 Phys. Rev. D {\bf 71} (2005) 063529,
 [{\tt hep-th/0412186}];\\
 F. Bernardeau, T. Brunier, and J.-P. Uzan,
 Phys. Rev. D {\bf 69} (2004) 063520,
 [{\tt astro-ph/0311422}];\\
 S. Weinberg,
 Phys. Rev. D {\bf 72} (2005) 043514,
 [{\tt hep-th/0506236}];\\
 S. Weinberg,
 Phys. Rev. D {\bf 74} (2006) 023508,
 [{\tt hep-th/0605244}].

\bibitem{picon2}
  C. Armendariz-Picon,
 [\urel{arXiv:astro-ph/0612288}].

\bibitem{kam}
 A.R. Pullen and M. Kamionkowsky,
 Phys. Rev. D {\bf76}, 103529 (2007).

\bibitem{JMM}
 J. Mart\'in-Garc\'ia, ``xPerm and xAct'',\\
 http://metric.iem.csic.es/Martin-Garcia/xAct/index.html


\end{thebibliography}
\end{document}